\documentclass[a4paper,fleqn,usenatbib]{mnras}
\usepackage[T1]{fontenc}
\usepackage{ae,aecompl}

\usepackage{graphicx}	
\usepackage{amsmath}	
\usepackage{amssymb}	
\usepackage[utf8]{inputenc}
\usepackage{epsfig}
\usepackage{url}
\usepackage{capt-of}
\usepackage{multicol}
\usepackage{epstopdf}
\usepackage{caption}
\usepackage{textcomp}
\usepackage[usenames,dvipsnames]{color}
\usepackage{float,lscape}
\usepackage{pdfpages}
\usepackage{pdflscape}
\usepackage{afterpage}
\usepackage{rotating}
\usepackage{arydshln}

\title[KROSS-SAMI: The TFR Since $\mathit{z \approx 1}$]{KROSS-SAMI: A Direct IFS Comparison of the Tully-Fisher Relation Across 8 Gyr Since $z \approx 1$}

\author[Tiley et al.]{A. L.\ Tiley,$^{1,2\dagger}$ M. Bureau,$^{2}$ L. Cortese,$^{3,4}$ C. M. Harrison,$^{5,1}$ H. L. Johnson,$^{1}$ 
\newauthor J. P. Stott,$^{6}$ A. M. Swinbank,$^{1}$ I. Smail,$^{1}$  D. Sobral,$^{6}$ A. J. Bunker,$^{2,7}$
\newauthor K. Glazebrook,$^{8}$ R. G. Bower,$^{9,1}$ D. Obreschkow,$^{3}$ J. J. Bryant,$^{10,11,4}$ M. J. Jarvis,$^{2,12}$  
\newauthor J. Bland-Hawthorn,$^{10,4,11,13}$ G. Magdis,$^{14,15}$  A. M. Medling,$^{16,17}$ S. M. Sweet,$^{8}$ 
\newauthor C. Tonini,$^{18}$ O. J. Turner,$^{19}$ R. M. Sharples,$^{1,20}$ S. M. Croom,$^{10,4}$ M. Goodwin,$^{21}$ 
\newauthor  I. S. Konstantopoulos,$^{22}$ N. P. F. Lorente,$^{21}$ J. S. Lawrence,$^{21}$  J. Mould,$^{8}$ 
\newauthor  M. S. Owers,$^{23,21}$ S. N. Richards$^{24}$
\\
$^{1}$Centre for Extragalactic Astronomy, Department of Physics, Durham University, South Road, Durham, DH1 3LE, U.K.\\
$^{2}$Sub-dept. of Astrophysics, Department of Physics, University of Oxford, Denys Wilkinson Building, Keble Road, Oxford, OX1 3RH, U.K.\\
$^{3}$International Centre for Radio Astronomy Research, The University of Western Australia, 35 Stirling Hw, 6009 Crawley, WA, Australia\\
$^{4}$ARC Centre of Excellence for All Sky Astrophysics in 3 Dimensions (ASTRO 3D)\\
$^{5}$European Southern Observatory, Karl-Schwarzchild-Str. 2, 85748 Garching b. M{\"u}nchen, Germany\\
$^{6}$Department of Physics, Lancaster University, Lancaster, LA1 4YB, U.K.\\
$^{7}$Affiliate Member, Kavli Institute for the Physics and Mathematics of the Universe, 5-1-5 Kashiwanoha, Kashiwa, 277-8583, Japan\\
$^{8}$Centre for Astrophysics and Supercomputing, Swinburne University of Technology, P.O. Box 218, Hawthorn, VIC 3122, Australia\\
$^{9}$Institute for Computational Cosmology, Durham University, South Road, Durham, DH1 3LE, U.K.\\
$^{10}$Sydney Institute for Astronomy (SIfA), School of Physics, The University of Sydney, NSW 2006, Australia\\
$^{11}$Australian Astronomical Optics, AAO-USydney, School of Physics, University of Sydney, NSW 2006, Australia\\
$^{12}$Department of Physics, University of the Western Cape, Bellville 7535, South Africa\\
$^{13}$Institute of Photonics and Optical Science (IPOS), School of Physics, The University of Sydney, NSW 2006, Australia\\
$^{14}$Dark Cosmology Centre, Niels Bohr Institute, University of Copenhagen, Juliane Mariesvej 30, DK-2100 Copenhagen, Denmark\\
$^{15}$Institute for Astronomy, Astrophysics, Space Applications and Remote Sensing, National Observatory of Athens, GR-15236 Athens, Greece\\
$^{16}$Research School for Astronomy \& Astrophysics Australian National University Canberra, ACT 2611, Australia\\
$^{17}$Cahill Center for Astronomy and Astrophysics California Institute of Technology, MS 249-17 Pasadena, CA 91125, USA\\
$^{18}$Melbourne University, School of Physics Parkville, 3010 Australia\\
$^{19}$Scottish Universities Physics Alliance, Institute for Astronomy, University of Edinburgh, Royal Observatory, Edinburgh EH9 3HJ\\
$^{20}$Centre for Advanced Instrumentation, Department of Physics, Durham University, South Road, Durham, DH1 3LE, U.K.\\
$^{21}$Australian Astronomical Optics, AAO-Macquarie, Faculty of Science and Engineering, Macquarie University, 105 Delhi Rd, North Ryde, \\
NSW 2113, Australia\\
$^{22}$Atlassian 341 George St Sydney, NSW 2000\\
$^{23}$Department of Physics and Astronomy, Macquarie University, NSW 2109, Australia\\
$^{24}$SOFIA Operations Center, USRA, NASA Armstrong Flight Research Center, 2825 East Avenue P, Palmdale, CA 93550, USA\\
$^{\dagger}$E-mail: alfred.l.tiley@durham.ac.uk
}

\date{Accepted XXX. Received YYY; in original form ZZZ}

\pubyear{2018}

\begin{document}
\label{firstpage}
\pagerange{\pageref{firstpage}--\pageref{lastpage}}
\maketitle

\begin{abstract}
We construct Tully-Fisher relations (TFRs), from large samples of galaxies with spatially-resolved H$\alpha$ emission maps from the K-band Multi-Object Spectrograph (KMOS) Redshift One Spectroscopic Survey (KROSS) at $z\approx1$. We compare these to data from the Sydney-Australian-Astronomical-Observatory Multi-object Integral-Field Spectrograph (SAMI) Galaxy Survey at $z\approx0$. We stringently match the data quality of the latter to the former, and apply identical analysis methods and sub-sample selection criteria to both to conduct a direct comparison of the absolute $K$-band magnitude and stellar mass TFRs at $z\approx1$ and $z\approx0$. We find that matching the quality of the SAMI data to that of KROSS results in TFRs that differ significantly in slope, zero-point and (sometimes) scatter in comparison to the corresponding original SAMI relations. These differences are in every case as large or larger than the differences between the KROSS $z\approx1$ and matched SAMI $z\approx0$ relations. Accounting for these differences, we compare the TFRs at $z\approx1$ and $z\approx0$. For disk-like, star-forming galaxies we find no significant difference in the TFR zero-points between the two epochs. This suggests the growth of stellar mass and dark matter in these types of galaxies is intimately linked over this $\approx8$ Gyr period.
\end{abstract}




\section{Introduction}
\label{sec:SAMImotivation}

The Tully-Fisher relation \citep[TFR;][]{Tully:1977aa} describes the correlation between a galaxy's rotation speed and its luminosity. The relation demonstrates an underlying link between the stellar mass of galaxies and their total masses (including both baryonic and dark matter). The relation may be derived from the simple assumption of spherical, circular motion and states that the total luminosity of the system is a function of the galaxy luminosity ($L$), its rotation velocity ($v$), mass surface density ($\Sigma$), and total mass-to-light ratio ($M/L$) such that $L\propto v^{4}/(\Sigma M/L)$. The relation is therefore a useful tool to measure the relative difference in the mass-to-light ratios and surface densities of different populations of galaxies, given a measure of their rotation and luminosity. Over time the TFR has become an effective tool in this regard. 

The TFR in the local Universe is well studied \citep[e.g.][]{TullyPierce2000,Bell:2001aa,Masters:2008aa,Lagattuta:2013aa}. Recent works have studied the TFR at much higher redshift, with a particular focus on the epoch of peak cosmic star formation rate density, $z\approx1$--$3$ \citep[e.g][]{Lilly:1996,Madau:1996,Hopkins:2006,Sobral:2013a,Madau:2014}. At these redshifts, typical star-forming galaxies are found to be much more turbulent than those in the local Universe, with an average ratio of intrinsic rotation velocity-to-intrinsic (gas) velocity dispersion $v/\sigma\sim2$--$3$ \citep[e.g.][]{Stott:2016,Johnson:2018} - lower than late-type disk galaxies at $z\approx0$ ($v/\sigma\sim5$--$20$; \citealt{Epinat:2010}). At this epoch, the extent to which star-forming galaxies obey the assumption of circular motion required for the TFR to hold strictly true varies on an individual basis from system to system. The observed slope, zero-point, and scatter of the TFR in this regime are thus indicators of the $M/L$ and $\Sigma$ of galaxies, but also of the relative dominance of rotational motions in their dynamics.

Approximately $50$ percent of the stellar mass in the Universe was already assembled by $z\approx1$ \citep[e.g.][]{PerezGonzalez:2008}, with massive galaxies at $z\approx1$--$3$ prolifically star-forming in comparison to those in the present day \citep{Smit:2012}. This epoch is one of the key periods in galaxy evolution, and is likely a time in which many key properties of galaxies were defined. It is therefore vital to compare the stellar mass, gas and dark matter content in galaxies at this epoch to those in the present day, and to determine whether this is easily reconciled with the evolving global star formation rate density over the intervening $\approx$8 Gyrs. The TFR provides a simple tool with which to do this.

Thus far, TFR studies that employ slit spectroscopy to measure galaxy kinematics suggest little-to-no evolution in the relation between $z\approx1$ and $z\approx0$ \citep[e.g.][]{Conselice:2005b,Kassin:2007a,Miller:2011,Miller:2012}. Some integral field spectroscopy (IFS) studies also report no change to the TFR over the same period \citep[e.g.][]{Flores:2006}. However, the majority report the opposite, tending to measure significant differences between the TFR zero-point at high redshift and the zero-point at $z\approx0$ \citep[e.g.][]{Puech:2008,Cresci:2009,Gnerucci:2011,Tiley:2016,Ubler:2017} that suggest that, at fixed rotation velocity, galaxies had less stellar mass in the past than in the present day. 

These studies and many others \citep[e.g.][]{Maiolino:2008a,ForsterSchreiber:2009,Mannucci:2009,Contini:2012,Swinbank:2012b,Sobral:2013} fail to reach a robust consensus on whether the TFR has significantly changed over cosmic time, particularly during the period between $z\approx1$ and $z\approx0$. Recently \citet{Turner:2017b} showed that many of these discrepancies can be accounted for by controlling for different sample selections used in each study. However, this study relied on compiling catalogues of values from the literature and was unable to fully account for the different data quality and analyses methods used throughout the studies. Given the implications that any measured evolution would have for galaxy evolution, there is a clear need for a systematic study of the TFR between $z\approx1$ and $z\approx0$.

In \citet{Tiley:2016} we constructed TFRs for $\sim$600 galaxies with resolved dynamics from the KMOS Redshift One Spectroscopic Survey \citep[KROSS;][]{Stott:2016,Harrison:2017}. For ``strictly'' rotation-dominated KROSS galaxies ($V_{80}/\sigma > 3$ where $V_{80}$ is the rotation velocity of the galaxy at a radius equal to the semi-major axis of the ellipse containing 80\% of the galaxy H$\alpha$ flux), we found no evolution of the absolute $K$-band ($M_{K}$) TFR zero-point, but a significant evolution of the stellar mass ($M_{*}$) TFR zero-point ($+0.41 \pm 0.08$ dex from $z\approx1$ to $z\approx0$). Assuming a constant surface mass density, this implies a reduction, by a factor of $\approx2.6$, of the dynamical mass-to-stellar mass ratio for this type of galaxy over the last $\approx 8$ Gyr, and it suggests substantial stellar mass growth in galaxies since the epoch of peak star formation.

In this work we aim to improve on our previous analysis by obtaining a measure of the evolution of the TFR between $z\approx1$ and $z\approx0$ that is unaffected by potential biases that may arise as a result of differences in the sample selection, analysis methods and data quality between TFR studies at different epochs. Our goal is thus to construct TFRs at both $z\approx1$ and $z\approx0$ using the same methodology, with uniform measurements of galaxy properties for samples constructed using the same selection criteria and taken from data matched in spatial and spectral resolution and sampling, and typical signal-to-noise ratios at both redshifts. Any differences in the TFRs between epochs can then be attributed to real differences between the physical properties of the observed galaxies at each redshift. 

In this paper we draw on samples from KROSS and the Sydney-Australian-Astronomical-Observatory Multi-object Integral-Field Spectrograph \citep[SAMI;][]{Croom:2012} Galaxy Survey \citep[e.g.][]{Bryant:2015} to construct TFRs at $z\approx1$ and $z\approx0$, respectively. The SAMI Galaxy Survey provides a convenient comparison sample with which to compare to KROSS, well matched in its sample size, restframe optical bandpass and that it targets star-forming galaxies with star-formation rates typical for their epoch.

This paper is divided in to several sections. In \S~\ref{sec:data} we provide details on the SAMI and KROSS data, as well as describing the process employed to transform the original SAMI data so that it is matched to KROSS in terms of spatial and spectral resolution and sampling, as well as in the typical signal-to-noise ratio of galaxies' nebular emission. Throughout this work, we refer to the transformed SAMI data as the matched SAMI sample (or data). For clarity we refer to the original, unmatched SAMI data as the original SAMI sample (or data). In \S~\ref{sec:measurements} we detail the measurements of galaxy properties made from the KROSS, original SAMI, and matched SAMI data. To construct TFRs we extract sub-samples from each data set using uniform selection criteria. These criteria are detailed in \S~\ref{sec:SAMIvsKROSSsample}. In \S~\ref{sec:KROSSvsSAMIresults} we present the TFRs for each data set and examine the differences between the relations. In \S~\ref{sec:SAMIvsKROSSdiscussion} we discuss the implications of our results for galaxy evolution. Concluding remarks and an outline of future work are provided in \S~\ref{sec:SAMIvsKROSSconclusions}.  

A Nine-Year Wilkinson Microwave Anisotropy Probe \citep[WMAP9;][]{Hinshaw:2013} cosmology  is used throughout this work. All magnitudes are quoted in the Vega system. All stellar masses assume a Chabrier \citep{Chabrier:2003aa} initial mass function. 

\section{Data}
\label{sec:data}

In this section we provide details of the SAMI and KROSS data we use to construct the TFRs. We also describe the process by which we transform the original SAMI data to match the typical quality of the KROSS data.

\subsection{KROSS}
\label{subsec:KROSS}

The $z\approx1$ TFRs presented in this work are constructed from samples drawn from KROSS. For detailed descriptions of the KROSS sample selection, observations, and data reduction see \citet{Stott:2016}. Here we provide only a brief summary. 

KROSS comprises integral field unit observations of 795 galaxies at $0.6 \lesssim z \lesssim 1$, that target H$\alpha$, [N {\sc ii}]6548 and [N {\sc ii}]6583 emission from warm ionised gas that falls in the $YJ$-band ($\approx1.02$--$1.36 \mu$m) of KMOS. Target galaxies were selected to be primarily blue ($r-z<1.5$) and bright ($K_{\rm{AB}} < 22.5$), including H$\alpha$-selected galaxies from HiZELS \citep{Sobral:2013a,Sobral:2015}, and from well-known, deep extragalactic fields: the Extended {\it Chandra} Deep Field South (ECDFS), the Ultra Deep Survey (UDS), the COSMOlogical evolution Survey (COSMOS), and the Special Selected Area 22 field (SA22). ECDFS, COSMOS and sections of UDS all benefit from extensive {\it Hubble Space Telescope} ({\it HST}) coverage. 

All KROSS observations were carried out with KMOS on UT1 of the Very Large Telescope, Cerro Paranal, Chile. The core KROSS observations were undertaken during ESO observing periods P92--P95 (with programme IDs 092.B-0538, 093.B-0106, 094.B-0061, and 095.B-0035). The full sample also includes science verification data \citep[60.A-9460;][]{Sobral:2013,Stott:2014}. KMOS consists of 24 individual integral field units (IFUs), each with a $2\farcs8 \times 2\farcs8$ field-of-view, deployable in a $7'$ diameter circular field-of-view. The resolving power of KMOS in the $YJ$-band ranges from $R\approx3000$--$4000$. The median seeing in the $YJ$-band for KROSS observations was $0\farcs7$. Reduced KMOS data results in a ``standard'' data cube for each target with $14 \times 14$ $0\farcs2$ square spaxels. Each of these cubes is then resampled on to a spaxel scale of $0\farcs1$ before analysis.  

A careful re-analysis of the KROSS sample by \citet{Harrison:2017}, that combines the extraction of weak continuum emission from the KROSS data cube with newly-collated high quality broadband imaging (predominantly from {\it HST} observations) provided improved cube centering and measures of galaxy sizes and inclinations. 

\subsection{SAMI Galaxy Survey}
\label{subsec:SAMIHQ}

The $z\approx0$ TFRs presented in this work are constructed from samples drawn from the SAMI Galaxy Survey  \citep{Bryant:2015}. Using the SAMI spectrograph \citep{Croom:2012} on the 3.9-meter Anglo-Australian Telescope at Siding Spring Observatory. The SAMI Galaxy Survey has observed the spatially-resolved stellar and gas kinematics of $\approx3000$ galaxies in the redshift range $0.004 < z <0.095$, over a large range of local environments. This work uses SAMI observations of 824 galaxies with mapped kinematics out to or beyond one effective radius. 

The Sydney-AAO Multi-obect Integral field spectrograph (SAMI; Croom et al. 2012) is mounted at the prime focus on the Anglo-Australian Telescope that provides a 1 degree diameter field of view. SAMI uses 13 fused fibre bundles \citep[Hexabundles,][]{BlandHawthorn:2011,Bryant:2014} with a high (75 percent) fill factor. Each bundle contains 61 fibres of 1$\farcs$6 diameter resulting in each IFU having a diameter of 15$''$. The IFUs, as well as 26 sky fibres, are plugged into pre-drilled plates using magnetic connectors. SAMI fibres are fed to the double-beam AAOmega spectrograph \citep{Sharp:2015}. AAOmega allows a range of different resolutions and wavelength ranges. The SAMI Galaxy survey uses the 570V grating at $\approx3700$--$5700$ \AA\ giving a resolution of $R\approx1730$ (sigma=74 km s$^{-1}$), and the R1000 grating from $\approx6300$--$7400$ \AA\ giving a resolution of $R\approx4500$ (sigma=29 km s$^{-1}$). Observations were carried out with natural seeing, with a typical range $0\farcs 9$--$3\farcs 0$.The resulting data were reduced via version v0.8 of the SAMI reduction pipeline \citep{Sharp:2015,Allen:2015} and underwent flux calibration and telluric correction. The resultant data cubes have $0\farcs 5 \times 0\farcs 5$ spaxels. 

The work presented in this paper draws upon the internal SAMI data release v0.9 (kindly provided by the SAMI team ahead of its public release), comprising 824 galaxies. It does not include those $\approx600$ SAMI galaxies specifically targeted as being members of clusters \citep[see][]{Bryant:2015}. Since our goal is to compare the restframe ionised gas kinematics (H$\alpha$ and [N{\small II}] lines) of both the KROSS and SAMI samples, we utilise here only those cubes observed in the red SAMI bandpass, yielding a reasonable match in wavelength coverage to the restframe optical bandpass of the KMOS $YJ$ filter at $z\approx1$.  We note that the average star formation rate of the SAMI galaxies considered in this work (being typical of star-forming galaxies at $z\approx0$) is at least an order of magnitude less than that of the KROSS galaxies \citep{Johnson:2018}.

\subsection{SAMI-KROSS Data Quality Match}
\label{subsec:SAMImatching}

In this work, we take steps to remove the potential for systematic biases between TFRs constructed at different redshifts by implementing a novel data ``matching'' process, applied to the SAMI data to transform them so that they match the quality of KROSS observations. As stated in \S~\ref{sec:SAMImotivation}, we refer to these transformed data as the matched SAMI sample (or data). We refer to the original, unmatched SAMI data as the original SAMI sample (or data). 

The data matching process provides two important benefits. Firstly, it removes the potential for bias in our measure of TFR evolution as a result of differing data quality between the KROSS and SAMI samples; any systematic bias resulting from the data quality should be equally present in both the $z\approx1$ and $z\approx0$ TFRs. Secondly, matching the data allows us to identify and quantify any bias (and associated selection function) that is introduced in the $z\approx1$ IFS observations as a result of its lower quality. 

To match the SAMI data we ensure that the spatial resolution and sampling (relative to the size of the galaxy, i.e.\ in physical rather than angular scale), spectral resolution and sampling, and H$\alpha$ signal-to-noise ratio ($S/N$) of the SAMI cubes match those of KROSS observations. We also require that the spatial extent of the matched SAMI data is comparable to that of KROSS -- more specifically, we only require that the field-of-view or spatial extent of the H$\alpha$ emission (whichever is smaller) is enough to extract the rotation velocity measure in the outer (i.e.\ flat) parts of the galaxies' velocity fields, as for KROSS. The radius at which we take our velocity measure is thus a delicate choice and is discussed in \S~\ref{subsec:SAMIrotvels}.

It should be stressed that transforming the SAMI cubes to match the typical H$\alpha$ $S/N$ of KROSS observations does nothing to address the question of how the observed H$\alpha$ $S/N$ of SAMI galaxies would be affected, were they observed with KMOS at similar distances and in the same manner as KROSS galaxies. I.e.\ in this work we do not adjust the fluxes to mimic the effects of ``redshifting" a galaxy to $z\approx1$. We focus only on how IFS observations of galaxies of differing qualities at any epoch bias the resultant galaxy sample and measurements. We thus degrade the SAMI data to match the quality of KROSS purely to negate potential observational biases. 

Each of the fully-reduced, flux-calibrated, telluric-corrected red data cubes from the SAMI internal data release v0.9 was thus transformed in a sequence of steps, outlined in order of their application in \S~\ref{subsubsec:spatialseeing},  \S~\ref{subsubsec:spectralseeing} and  \S~\ref{subsubsec:adjustSN}.

\subsubsection{Spatial Resolution and Sampling}
\label{subsubsec:spatialseeing}

First, the PSF full-width-at-half-maximum of each original SAMI cube ($\text{FWHM}_{\text{0}}$) was calculated by fitting a two-dimensional circular Gaussian to the image of the corresponding reference star. The median seeing of KROSS observations is $\approx 0\farcs7$, corresponding to a scale of $\approx5.3$ kpc at $z=0.8$ (the median redshift of KROSS galaxies). To match the SAMI seeing in physical scale to KROSS, we thus require a $\text{FWHM}_{1}=5.3$ kpc$/S_{\text{D}}$, where $S_{\text{D}}$ is the angular scale at the redshift of the SAMI galaxy.

We therefore simply convolved each spectral slice (i.e.\ each plane of the datacube in the wavelength direction) with a two-dimensional circular Gaussian (normalised so that its integral is unity) of width 

\begin{equation}
\text{FWHM}_{\delta} = \sqrt{\text{FWHM}_{1}^{2}-\text{FWHM}_{0}^{2}}\,\,\,.
\end{equation}
\label{eq:deltasig}

\noindent The width of the SAMI PSF was on (median) average enlarged by a factor of $3 \pm 1$ during this process.

Next, using a third-degree bivariate spline approximation\footnote{The bivariate spline interpolation is similar to a polynomial interpolation and is a standard way to smoothly interpolate in two dimensions. Its use avoids the problem of oscillations occuring between data points when interpolating with higher order polynomials since it minimises bending between points.} from {\sc Scipy} in {\sc Python}, we regrided each of the spatially-convolved spectral slices of the SAMI cube so that the physical size of the spaxels matches that of KROSS. We calculated the number of spaxels, $N_{\text{1}}$, required across the width of each square SAMI slice as 

\begin{equation}
N_{\text{1}}=0\farcs 5\ S_{\text{D}} N_{0} /\ l_{\text{K}}\,\,,
\end{equation}

\noindent where $N_{0}$ is the original number of spaxels across the width of the SAMI slice and $l_{\text{K}} = 0.8$ kpc is the physical width of a $0\farcs 1$ wide KROSS spaxel at the median redshift of KROSS galaxies ($z = 0.8$). 

\subsubsection{Spectral Resolution and Sampling}
\label{subsubsec:spectralseeing}

We then require to match the SAMI resolving power to that of KROSS. We thus convolve the spectrum of each spaxel with a Gaussian of width 

\begin{equation}
\text{FWHM}_{\text{d}} = \sqrt{\text{FWHM}_{\text{K}}^{2}-\text{FWHM}_{\text{S}}^{2}}\,\,,
\end{equation}
\label{eq:deltasigspec}
 
\noindent where $\text{FWHM}_{\text{K}} = \lambda_{\text{central}}/R_{\text{K}}$ is the width of the Gaussian required to match the resolving power of KROSS ($R_{\text{K}}\approx3580$), $\lambda_{\text{central}}$ is the central wavelength of the SAMI red filter and $\text{FWHM}_{\text{S}}$ is the width of the Gaussian corresponding to the original spectral resolution of SAMI ($\text{FWHM}_{\text{S}}= \lambda_{\text{central}}/R_{\text{S}}$, where the resolving power of SAMI $R_{\text{S}}=4500$). This in effect increases the width of the instrumental broadening by a factor of $\approx1.3$. To match the spectral sampling to that of KROSS, we then re-bin the smoothed spectra using a linear interpolation to calculate the flux in each bin.

\subsubsection{H$\alpha$ $S/N$}
\label{subsubsec:adjustSN}

Lastly, we match the median of the distribution of H$\alpha$ $S/N$ in the spaxels of each SAMI cube to that of typical KROSS observations. As discussed in \S~\ref{subsubsec:HQcontmaps}, given that we typically detect only weak spatially-extended continuum emission in KROSS and that we are primarily interested in a comparison of the ionised gas kinematics of SAMI and KROSS galaxies, we first subtract from each SAMI cube the corresponding best fit model continuum cube as computed from {\sc lzifu} (see Section~\ref{subsubsec:HQcontmaps}). Using these continuum-subtracted cubes, we then simultaneously fit the H$\alpha$, [N{\small II}]6548 and [N{\small II}]6583 emission lines of each spectrum with three single Gaussians, in the exact same manner as described in \citet{Stott:2016} and \citet{Tiley:2016}. The emission line fits are performed using the routine {\sc mpfit}. We then take the $S/N$ of the H$\alpha$ emission in each spaxel as the square root of the difference between the $\chi^{2}$ value of the best-fit Gaussian to the H$\alpha$ emission line ($\chi^{2}_{\text{mod.}}$) and that of a straight line equal to the baseline value ($\chi^{2}_{\text{line}}$), avoiding regions of sky emission i.e.\ $ S/N = \sqrt{\chi^{2}_{\text{line}}-\chi^{2}_{\text{mod.}}}$ \citep[e.g.][]{Neyman:1933,Bollen:1989,Labatie:2012}. This approach relies on the assumption that the noise is Gaussian and constant with a single variance (which we verify as true for the KROSS and SAMI cubes).

We define the typical $S/N$ of the H$\alpha$ emission in KROSS observations as the mean of the distribution of median H$\alpha$ $S/N$ values across all KROSS maps (that is flat as a function of stellar mass). We use this as a ``target'' H$\alpha$ $S/N$ for each SAMI galaxy, adding Gaussian noise uniformly to each cube such that the median $S/N$ matches this value. 

The median H$\alpha$ $S/N$ of the original SAMI galaxies is on median average $1.6\pm0.6$ times larger than that for the corresponding matched SAMI galaxies, where the uncertainty is the median absolute deviation from the median itself.

\section{Measurements}
\label{sec:measurements}

In this section we detail our measurements of key properties for the original SAMI, matched SAMI and KROSS galaxies that we use for our analysis.

\subsection{Stellar Masses and Absolute Magnitudes}

\subsubsection{KROSS}
\label{subsubsec:KROSSSED}

Stellar masses ($M_{*}$) and K-corrected absolute $K$-band magnitudes for each KROSS galaxy were derived using {\sc Le Phare} \citep{Arnouts:1999,Ilbert:2006} to compare a suite of model spectral energy distributions (SEDs) to the observed SED of the target. The latter were constructed using integrated broadband photometry spanning the optical to the near-infrared ($u$-, $B$-, $V$-, $R$-, $I$-, $J$-, $H$-, and $K$-bands). Where available we also included mid-infared photometry from {\it IRAC} ($ch1$--$ch4$). The model SEDs were generated using the population synthesis models of \citet{Bruzual:2003aa}. The {\sc Le Phare} routine fits for extinction, metallicity, age, star formation and stellar mass, and allows for single burst, exponential decline, and constant star formation histories. We note the stellar masses used in this work are different to those presented in \citet{Harrison:2017}. The latter are interpolated from the absolute $H$-band magnitudes of the KROSS galaxies, assuming a fixed mass-to-light ratio. Since a galaxy's position in the TFR-plane is itself dependent on the mass-to-light ratio of the galaxy, we prefer to allow the possibility of variation in the ratio between galaxies, rather than assume a constant value, when determining the stellar masses. We note, however, that the two measures are generally consistent with a median difference of $0.0 \pm 0.2$ dex. The stellar masses in this work, calculated with {\sc Le Phare}, also differ from those presented in \citet{Stott:2016} and \citet{Tiley:2016}, calculated with the {\sc hyperz} SED fitting routine \citep{Bolzonella:2000}. We prefer the use of {\sc Le Phare} in this work since it allows for calculation of galaxy stellar mass and absolute magnitudes from a single routine. We note that anyway the two measures of stellar mass generally agree with a median difference of $0.0\pm0.2$ dex. In keeping with \citet{Tiley:2016}, and as commonly employed in studies of high-redshift star-forming galaxies, we adopt a uniform stellar mass uncertainty of $\pm 0.2$ dex throughout this work \citep[e.g.][]{Mobasher:2015} that should conservatively account for the typical deviations in stellar mass values resulting from the use of different, commonly employed SED fitting codes, and the possibility for low photometric signal-to-noise or high photometric uncertainty.

\subsubsection{SAMI}
\label{subsubsec:SAMIphot}

Each of the SAMI galaxies has associated integrated broadband photometry ranging (where available) from the far-ultra-violet ($FUV$ and $NUV$ from the {\it Galaxy Evolution Explorer} \citep[{\it GALEX};][]{Martin:2005}), through the optical ($u$, $g$, $r$, $i$ and $z$ from the Sloan Digitised Sky Survey (SDSS) e.g. \citealt{Stoughton:2002}; \citealt{Abazajian:2003}), near-infrared ($Z$, $J$, $H$ and $K$ from the Visible and Infrared Survey Telescope for Astronomy (VISTA), \citealt{Sutherland:2015}), mid-infrared ($W1$, $W2$, $W3$ and $W4$ from the {\it Wide-Field Infrared Survey Explorer} ({\it WISE}), \citealt{Wright:2010}), to the far-infrared (Photoconductor Array Camera and Spectrometer (PACS) {\it green} ($\approx98$ $\mu$m) and {\it red} ($\approx154$ $\mu$m) and Spectral and Photometric Imaging Receiver (SPIRE) {\it PSW} ($\approx243$ $\mu$m), {\it PMW}  ($\approx341$ $\mu$m) and {\it PLW}  ($\approx482$ $\mu$m) from {\it Herschel}, \citealt{Pilbratt:2010}). The 21-band photometric data sets are taken from the GAMA Panchromatic Data Release \citep[PDR;][]{Driver:2016}. For each galaxy's photometry, the Lambda Adaptive Multi-Band Deblending Algorithm in R \citep[LAMBDAR;][]{Wright:2016} was applied. LAMBDAR is designed to calculate matched aperture photometry across a range of non-homogeneous images with differing PSFs and pixel scales, given prior aperture information from high resolution imaging in the visible regime. \citeauthor{Wright:2016} define an initial aperture for each SAMI galaxy using a combination of Source Extractor \citep[SExtractor;][]{Bertin:1996} and visual inspection of the $r$-band image from SDSS and $Z$-band image from the VISTA Kilo-degree Infrared Galaxy Survey \citep[VIKING;][]{Edge:2013}. 

As with the KROSS galaxies in \citet{Tiley:2016}, absolute rest-frame magnitudes and stellar masses were computed for each SAMI galaxy using the SED fitting routine {\sc Le Phare} \citep[][]{Arnouts:1999,Ilbert:2006}. We note that above $\approx10^{8.8}M_{\odot}$, our masses agree (a median offset $0.0 \pm 0.2$ dex) with the SAMI stellar masses described in \citet{Bryant:2015}, estimated from $g-i$ colours and $i$-band magnitudes following \citet{Taylor:2011}. Below this mass, however, the two measures deviate (a median offset of $0.3 \pm 0.1$ dex).

\subsubsection{Matched SAMI}
\label{subsubsec:SAMILQSED}

Since the purpose of matching the SAMI data cubes is to make a fair and direct comparison of the KROSS $z\approx1$ TFR and the SAMI $z \approx 0$ TFR, we apply the same philosophy to the SED fits as we applied to the data cube matching. Specifically, we must restrict the available SAMI photometry to only include those bandpasses that are available in the restframe for each KROSS galaxies. In practice, this means truncating the full SAMI photometry range to only span the FUV to $K$ band. Absolute $K$-band magnitudes and stellar masses are then derived from the truncated SEDs using {\sc Le Phare} in the exact same manner as for the KROSS and original SAMI photometry.

\subsection{Emission Line Fitting and Maps}
\label{subsec:linemaps}

\subsubsection{KROSS}
\label{subsubsec:KROSSvelfields}

H$\alpha$ imaging and kinematic maps were extracted from the KROSS data cubes by \citet{Stott:2016}. The maps were extracted via a simultaneous triple Gaussian fit to the H$\alpha$, [N {\sc ii}]6548 and [N {\sc ii}]6583 emission lines in each (continuum-subtracted) spectrum of each spaxel for each cube. The central velocity and width of the H$\alpha$ and [{N{\sc ii}] lines are coupled so that they vary in unison. If the H$\alpha$ $S/N < 5$ for a given $0\farcs1$ spaxel, a larger area of  $3 \times 3$ spaxels was considered, and $5 \times 5$ spaxels, as required. If at this point the $S/N$ was still less than 5, that spaxel is excluded from the final maps. H$\alpha$ intensity maps were constructed by plotting the integral of the model H$\alpha$ flux in each spaxel. Line-of-sight velocity maps were constructed by plotting in each spaxel the  best fit central velocity of the H$\alpha$ emission, with respect to the rest frame velocity of the galaxy, as determined from the spectroscopic redshift measurements of \citet{Harrison:2017} (themselves measured from the position of H$\alpha$ and [N{\sc ii}] emission in the integrated KMOS spectrum within a $1\farcs2$ diameter aperture\footnote{the diameter of the aperture was chosen as a compromise between maximising the flux and the signal-to-noise ratio}). Similarly, observed velocity dispersion maps were constructed by plotting for each spaxel the width of the best fit Gaussian to the H$\alpha$ emission, in velocity space and correcting in quadrature for the instrumental broadening of KMOS. 

Building on the original analysis, \citet{Harrison:2017} extracted rotation curves (i.e.\ one dimensional velocity profiles) from the velocity map of each KROSS galaxy within a $0\farcs7$ ``slit'' along the galaxy's major kinematic axis. As a means to reduce the effects of noise in these curves, they find the best fit exponential disk model to the data of each curve, where the model velocity ($v$) as a function of radius ($r$) takes the form

\begin{equation} \label{eq:mod} 
(v(r)-v_{\rm{off}})^{2}=\frac{r^2 \pi G \mu_{0}}{h}(I_{0}K_{0}-I_{1}K_{1}) \,\,,
\end{equation}

\noindent where $G$ is the gravitational constant, $\mu_{0}$ is the peak mass surface density, $h$ is the disk scale radius, $v_{\rm{off}}$ is the velocity at $r=0$, and $I_{\rm{n}}K_{\rm{n}}$ are Bessel functions evaluated at $0.5r/h$. During the fitting process the radial centre is also free to vary. The velocity offset $v_{\rm{off}}$ is subtracted from the KROSS velocity maps before taking any further measurements. 

\subsubsection{SAMI}
\label{subsubsec:HIGHQvelfields}

Following continuum subtraction, we apply the same line-fitting methodology as for KROSS to each of the original SAMI cubes. If the H$\alpha$ $S/N < 5$ for a given 0$\farcs5 \times 0\farcs $5 spaxel, we consider a larger area of  1$\farcs5 \times 1\farcs $5, and 2$\farcs5 \times 2\farcs $5, as required. Once again, if at this point the $S/N$ is still too low, the spaxel is excluded from the resultant maps. We correct the observed velocity dispersion this time for the instrumental broadening of the SAMI spectrograph. We also extract a rotation curve from each original SAMI velocity map as for KROSS along the major kinematic axis, taking the weighted mean of the velocity values in pixel-wide steps, within a ``slit'' of width equal to three spaxels (approximately the FWHM of the PSF) \footnote{We quantified the effect of the slit width on the final measure of velocity finding a median maximum fractional difference of $-1.7^{+0.6}_{-8.1}$ percent in the measured rotation velocity when using a slit of width equal to $0.5$ or $1$ times the FWHM PSF. This translates to an average shift in log-space of $-0.007$ dex, with a range from $+0.003$ dex to $-0.03$ dex. The width of the slit therefore has minimal impact on our measure of galaxy rotation velocity and thus on our final TFRs.}. The major kinematic axis we find by rotating the same slit in 1$^{\circ}$ steps about the continuum centre, taking the position angle that maximises the velocity gradient along the slit. To describe the trend of the rotation curve and to reduce the effects of noise we also find the best fit model rotation velocity to the data of each curve according to Equation~\ref{eq:mod}, using {\sc mpfit}\footnote{translated into {\sc Python} by Mark River and updated by Sergey Koposov)} \citep{Markwardt:2009}, itself employing a Levenberg-Marquardt minimisation algorithm.

\subsubsection{Matched SAMI}
\label{subsubsec:SAMIvelfields}

Once the SAMI cubes are transformed to match the quality of KROSS observations, we extract H$\alpha$ intensity, velocity dispersion, and line-of-sight velocity maps and rotation curves in the same manner as for the original SAMI and KROSS cubes, as described in \S~\ref{subsubsec:KROSSvelfields}. Examples of matched SAMI kinematic maps, along with the same maps extracted from the corresponding original SAMI cube are shown in Figure~\ref{fig:SAMIOGgoodvels}. Here we also include the extracted rotation curve for each galaxy and the corresponding best fit model.

\begin{figure*}
\begin{minipage}[]{1.\textwidth}
\centering
\label{fig:AOGmap}
\includegraphics[width=0.84\textwidth,trim = 5 50 15 15, clip=True]{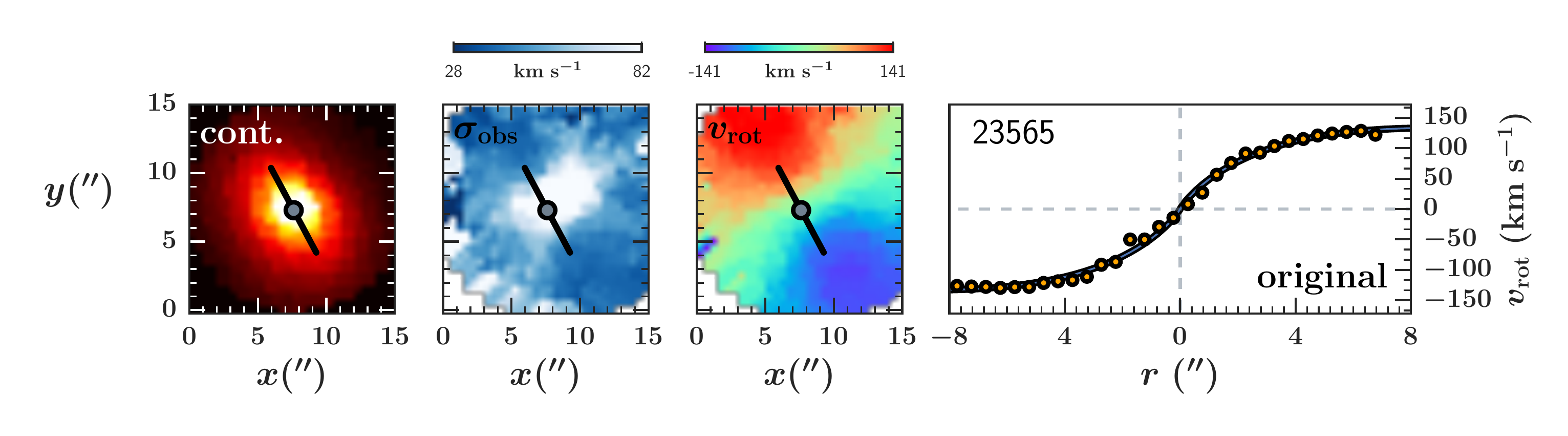}
\end{minipage}
\begin{minipage}[]{1.\textwidth}
\centering
\label{fig:BOGmap}
\includegraphics[width=0.84\textwidth,trim = 5 50 15 15, clip=True]{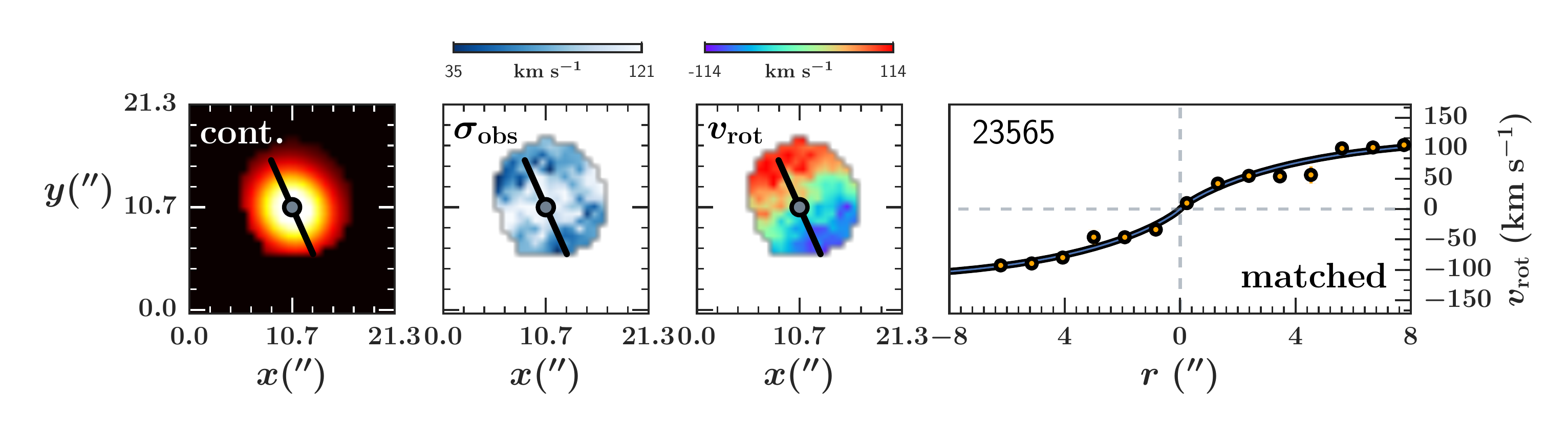}
\end{minipage}\vspace{.35cm}
\begin{minipage}[]{1.\textwidth}
\centering
\label{fig:DOGmap}
\includegraphics[width=0.84\textwidth,trim = 5 50 15 15, clip=True]{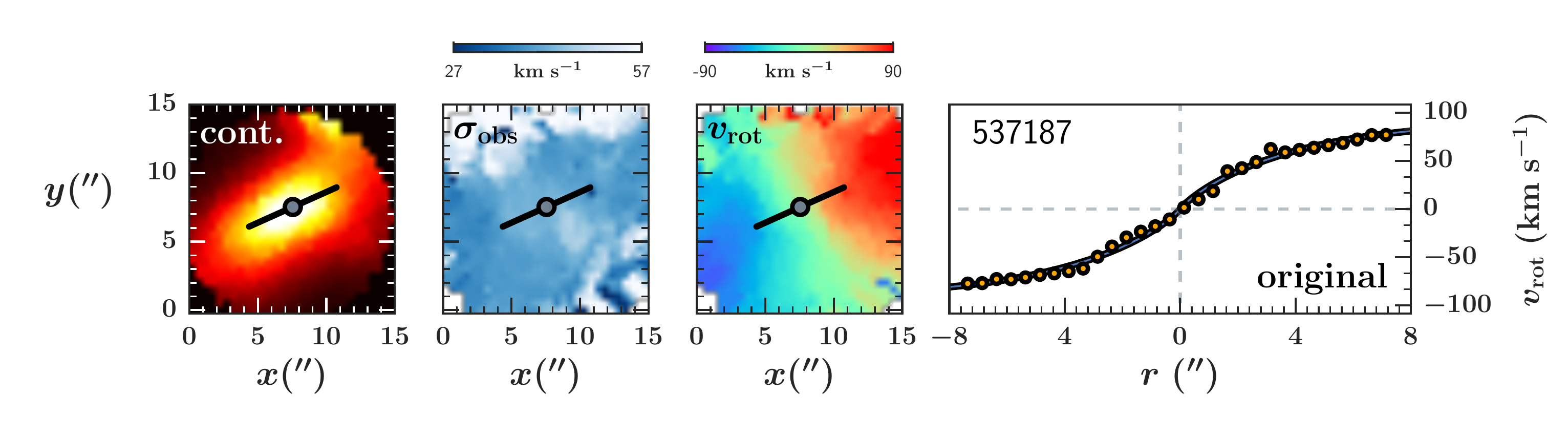}
\end{minipage}
\begin{minipage}[]{1.\textwidth}
\centering
\label{fig:DOGmap}
\includegraphics[width=0.84\textwidth,trim = 5 50 15 15, clip=True]{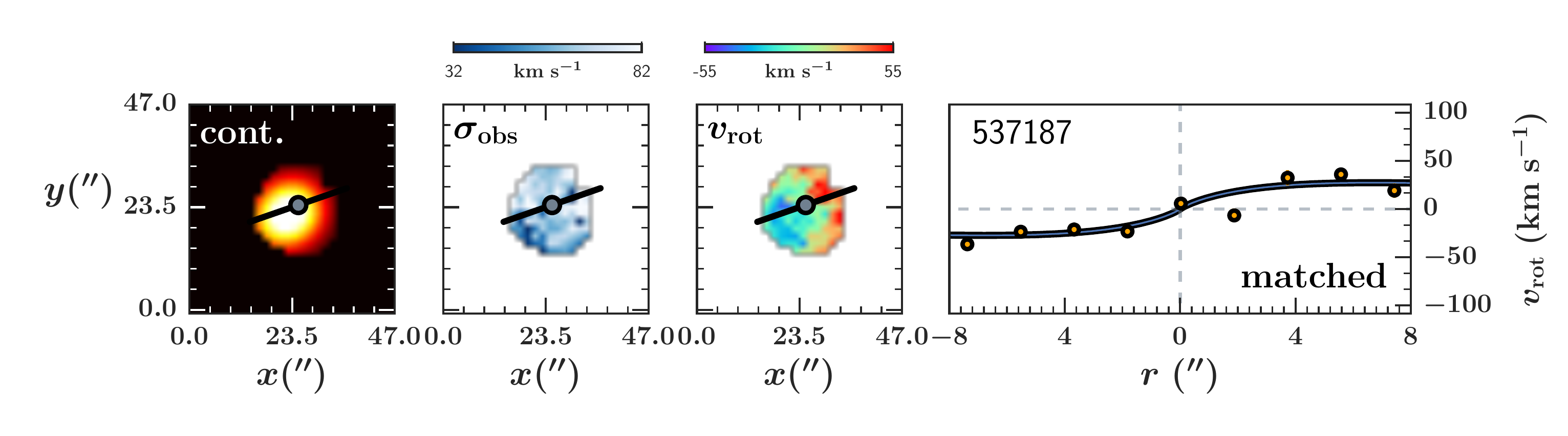}
\end{minipage}\vspace{.35cm}
\begin{minipage}[]{1.\textwidth}
\centering
\label{fig:DOGmap}
\includegraphics[width=0.84\textwidth,trim = 5 50 15 15, clip=True]{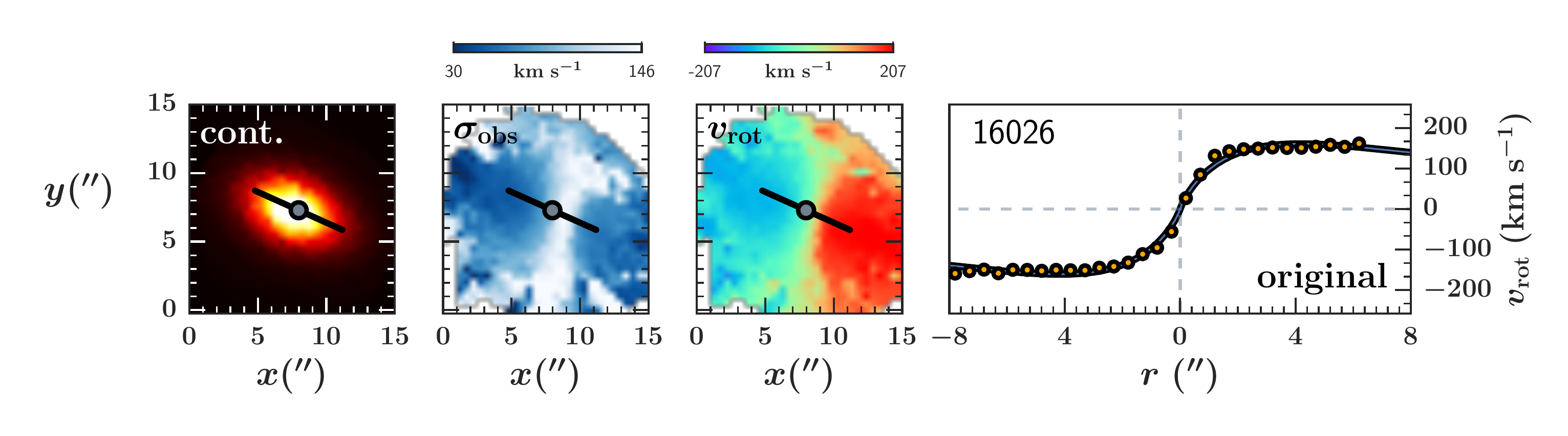}
\end{minipage}
\begin{minipage}[]{1.\textwidth}
\centering
\label{fig:DOGmap}
\includegraphics[width=0.84\textwidth,trim = 10 0 15 15, clip=True]{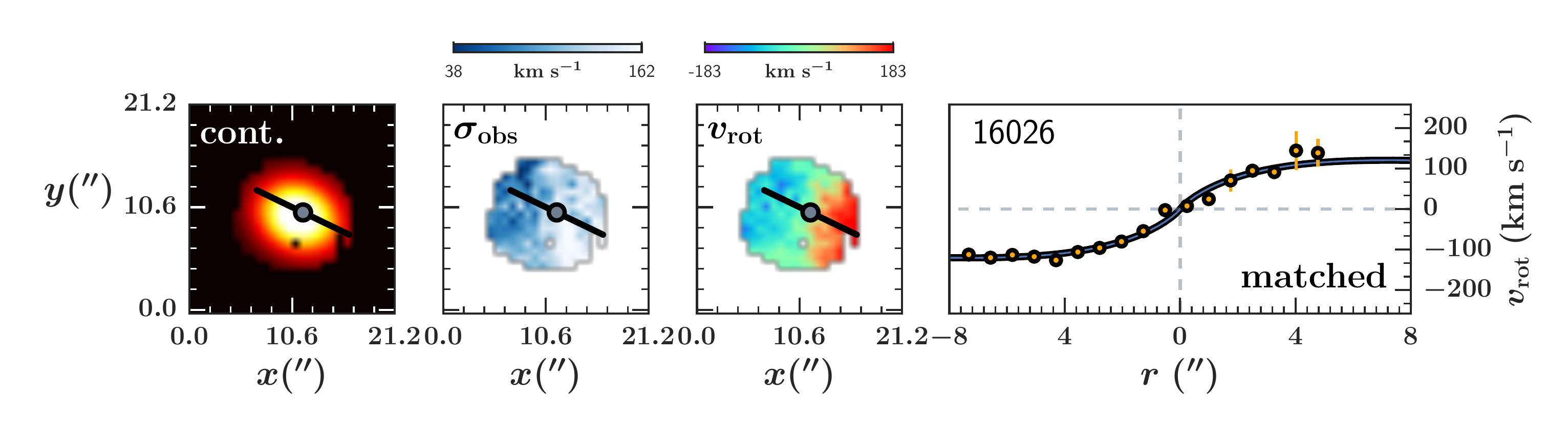}
\end{minipage}
\caption{Examples of original SAMI H$\alpha$ maps with the corresponding matched maps. The original and matched maps are in alternating rows. For each galaxy the (model) continuum intensity (far-left), velocity dispersion (centre-left) and line-of-sight velocity maps (centre-right) are displayed, centred on the position of peak continuum intensity. The galaxy rotation curve is shown on the far-right (orange points). We plot the best fit model curve (blue line) to the data (Equation~\ref{eq:mod}). The SAMI galaxy ID is shown at the top left of this panel. The kinematic centre (adjusted, where required, from the continuum centre using the best fit to the extract rotation curve) and major axis are indicated on each map as respectively a grey circle and black line. Model continuum maps for the matched SAMI galaxies were generated by spatially degrading the corresponding original SAMI model continuum maps in the manner described in \S~\ref{subsubsec:spatialseeing}.}%
\label{fig:SAMIOGgoodvels}
\end{figure*}

\subsection{Continuum and Broadband Imaging}

\subsubsection{KROSS}

To further improve the analysis, \citet{Harrison:2017} constructed $YJ$-band continuum maps for the KROSS galaxies by plotting the median flux of each spectrum of each cube in its corresponding spaxel, after masking any line emission in the cube and performing a $2$-$\sigma$ clip to the spectrum to exclude significant sky emission. In this manner they were able to robustly identify the continuum centroids for 85\% of the detected KROSS galaxies. For the remaining 15\%, centroids were identified from maps of the integrated continuum {\em and} spectral line emission. These centres were then used to align the KMOS data cubes to the centres of the best quality corresponding broadband image for each galaxy.

After aligning the KROSS cubes, \citet{Harrison:2017} measured the half light (or effective) radius ($r_{e}$) and inclinations for each KROSS galaxy from the broadband photometry. We refer the reader to \citeauthor{Harrison:2017} for a full description, but here we provide a brief summary. For a large fraction of KROSS targets {\it HST} imaging was available, of which the longest wavelength data was employed in each case. For 36\% of those targets with {\it HST} imaging, WFC3-$H$-band was the reddest band available, with a point spread function (PSF) full-width-half-maximum of $\approx0\farcs2$. For a further 57\% ACS-$I$ imaging was employed, and for the remaining 7\% ACS-$z'$ imaging (both with PSF FWHMs of $\approx0.1$). $K$-band UK Infrared Telescope (UKIRT) observations from the UKIRT Deep Sky Survey (UKIDSS), with a PSF FWHM $\approx0\farcs65$, were used for those galaxies in UDS without {\it HST} observations. For those galaxies in SA22, $K$-band UKIDSS imaging with a PSF FWHM of $0\farcs85$ was used. For each KROSS galaxy the measure of $r_{e}$ is corrected for the PSF broadening. Inclinations were derived from the axial ratio ($b/a$) measured from the best broadband image, assuming

\begin{equation}\label{eq:inc}
\cos^{2} i = \frac{(b/a)^{2}-q_{0}^{2}}{1-q_{0}^{2}}\,\,,
\end{equation}

\noindent where the intrinsic axial ratio $q_{0}$ is fixed at 0.2, appropriate for a thick disk \citep[e.g.][]{Guthrie:1992,Law:2012,Weijmans:2014}.

\subsubsection{SAMI}
\label{subsubsec:HQcontmaps}

Unlike the KROSS galaxies, the lower redshift of the SAMI galaxies allows for significant detections of the stellar continuum in many of the SAMI spaxels within typical exposure times. As such, we require a detailed fit to the stellar continuum. Before fitting the H$\alpha$ and [N {\sc ii}] emission lines (see \S~\ref{subsubsec:HIGHQvelfields}), we first fit and subtract the stellar continuum in each spaxel using the {\sc lzifu} {\sc idl} routine \citep{Ho:2016}, an emission line fitting ``toolkit'' designed specifically for use with IFS data. The {\sc lzifu} routine itself draws on the penalised pixel fitting routine \citep[{\sc pPXF};][]{Cappellari:2004,Cappellari:2017} to fit a library of model SEDs to the spectrum of each spaxel in order to build a corresponding cube of best fit model stellar continuum emission. We then subtract this from the original SAMI data cube before fitting the gas emission lines. We construct a model continuum map for each original SAMI galaxy cube, by integrating the model continuum emission along each spectrum in each spaxel. It should be noted that the primary purpose for {\sc lzifu} is to model the line emission in IFS datacubes, with the continuum emission modelling an intermediate step in this process. However, in this work we prefer to use it only as a convenient tool to rapidly model (and subtract) the continuum in the SAMI cubes. To maintain homogeneity with the KROSS analysis, we model the line emission in the cubes using the methods of \citet{Swinbank:2006} and \citet{Stott:2016}, described in \S~\ref{subsubsec:HIGHQvelfields}.   

Similar to KROSS, the original SAMI inclinations are estimated from the $r$-band axial ratio \citep{Cortese:2016}, according to Equation~\ref{eq:inc}.

\subsubsection{Matched SAMI}

The original SAMI data cubes are continuum-subtracted and centred before they are degraded to match the quality of KROSS observations. We therefore do not produce continuum maps from the matched SAMI (i.e.\ degraded) cubes.

For the matched SAMI galaxies, we adopt the same inclinations as for the original SAMI galaxies.
   
\subsection{Rotation Velocity and Velocity Dispersion}
\label{subsec:SAMIrotvels}

In this work, for a unique measure of the intrinsic circular velocity for each galaxy we adopt the $v_{2.2}$ parameter of \citet{Harrison:2017} derived from $v_{2.2,\rm{obs}}$, the line-of-sight velocity measured from the best fit model to the rotation curve at 1.3 times the effective (half-light) radius $r_{\text{e}}$ (convolved to the ``native" seeing of the velocity maps)\footnote{We note that $1.3r_{\rm{e}}$ is smaller than the $r_{80}$ used in \citet{Tiley:2016}, that corresponds to $\approx1.8r_{\text{e}}$ ($\approx3$ times the exponential disk scale length).}. This radius corresponds to 2.2 times the scale length of a purely exponential disk (i.e.\ 2.2$h$). The {\em intrinsic} rotation velocity at the same radius, $v_{2.2}$  is then retrieved by correcting for the effects of inclination ($i$) and beam smearing as

\begin{equation}\label{eq:veldef}
v_{2.2}=\frac{\epsilon_{\rm{R,PSF}}}{\sin i}v_{2.2,\rm{obs}}\,\,,
\end{equation}

\noindent where $\epsilon_{\rm{R,PSF}}$ is a beam smearing correction factor that depends on the ratio of the galaxy size to the width of the seeing PSF, and the rotation speed of the galaxy. This correction factor is detailed in \citet{Johnson:2018}, and is based on the creation of thousands of mock IFS observations of disk galaxies, with distributions of key galaxy properties designed to match the KROSS sample. We estimate the uncertainty in the velocity measurement by propagating the bootstrapped uncertainties from the best fit to the rotation curve, also accounting for the uncertainty in the inclination correction.

Extracting a rotation velocity at $1.3r_{\rm{e}}$ is to a certain extent physically motivated as it corresponds to the peak of the rotation for a purely exponential disk \citep[][]{Freeman:1970,Courteau:1997,Miller:2011}. Of course, we do not expect our galaxy sample to comprise only pure exponential disks. Given the limitations of the data however, it at least allows us to extract our velocity measure at the same scale radius across the large majority of galaxies in each of the original SAMI, matched SAMI and KROSS data sets and should recover close to the maximum rotation velocity for each system. Figure~\ref{fig:SAMIvsKROSShists}, panel (a), shows that the distributions of $r_{\text{e}}$ for the SAMI and KROSS galaxies are similar. Similarly, Figure~\ref{fig:SAMIvsKROSShists}, panel (j), shows that the majority of the original SAMI, matched SAMI, and KROSS galaxies have H$\alpha$ emission with sufficient radial extent as to sample the rotation curve at $1.3r_{\rm{e}}$. 

We note that for the KROSS galaxies, for which we are typically able to trace each galaxy's rotation curve out to $\approx2r_{\rm{e}}$ or further (firmly into the ``flat'', outer region of the curve), we find the median ratio of the rotation velocity at $2r_{\rm{e}}$ to that at $1.3r_{\rm{e}}$ is $1.1\pm0.1$, where the uncertainty is the standard deviation. 

We also note here that we obtain consistent measures of rotation velocity, for the KROSS, original SAMI, and matched SAMI data, if we instead fit the rotation curve data with the well-known and commonly employed arctangent model \citep{Courteau:1997aa}. We find a median fractional difference ($v_{2.2,\text{arc}}-v_{2.2,\text{exp}}$)/$v_{2.2,\text{exp}}$ of $-0.1^{+7.3}_{-3.8}$ and $0.4^{+4.0}_{-2.2}$ for respectively the original SAMI and matched SAMI samples. \citet{Harrison:2017} find similarly small differences for the KROSS galaxies.

For a unique measure of the intrinsic velocity dispersion of each original SAMI and matched SAMI galaxy, we follow the same methods employed by \citet{Johnson:2018}, and as measured by \citet{Harrison:2017} for the KROSS galaxies. For a measure of the velocity dispersion of each galaxy, we take the median of each galaxy's {\em observed} line-of-sight velocity dispersion map, (corrected for the instrumental broadening). We define the uncertainty in this measurement as half the difference between the 84$^{\rm{th}}$ and 16$^{\rm{th}}$ percentile of the map. To find the {\em intrinsic} velocity dispersion ($\sigma$) we apply a systematic correction to account for beam smearing, derived in \citet{Johnson:2018}, that again is a function of the ratio of the PSF to the galaxy size, as well as its rotation velocity. We note that, when available (i.e.\ for 48 percent of the sample), \citet{Harrison:2017} prefer to measure the observed velocity dispersion from the outer regions ($>2r_{\rm{e}}$) of the KROSS dispersion maps, rather than take the median of the map, before correcting to an intrinsic value. However, where it is possible to take both measurements, the resultant intrinsic velocity dispersions agree within 4 percent (albeit with a 50 percent scatter). 

\section{Sample Selection}
\label{sec:SAMIvsKROSSsample}

The final step to construct the TFRs is to carefully and uniformly select those galaxies from each data set (original SAMI, matched SAMI, and KROSS) that are suitable for inclusion in the relations. The sample selection criteria are similar to those described in \citet{Tiley:2016} but with some differences. We therefore detail each criterion here. We note that we restrict our criteria to those that are readily applicable to both our high- and low-redshift data sets. Here the limiting factor is the information available on the properties of the KROSS galaxies. For TFRs constructed for sub-samples selected on the basis of additional information available for the $z\approx0$ SAMI galaxies, see \citet{Bloom:2017}. As in the work of \citet{Tiley:2016}, we now also sort the galaxies into three categories. We refer to these as the \textit{parent}, \textit{rot-dom} and \textit{disky} sub-samples, in order of decreasing sub-sample size. 

\subsection{The \textit{parent} sub-samples}
\label{subsec:parentsample}

We define the \textit{parent} sub-samples as those galaxies that are detected and resolved (i.e.\ the radial extent of the H$\alpha$ emission is at least equal to the radius of the PSF) in H$\alpha$ and for which we are able to measure a rotation velocity ($v_{2.2,\rm{obs}}$), even if the H$\alpha$ emission does not extend out to 1.3$r_{\text{e}}$. At this step, we carry out an inspection of the velocity field for each galaxy by eye, to ensure the velocity field extraction has been successful, i.e.\ that the velocity field contains a sufficient number of spaxels to measure a velocity. Additionally, each galaxy must have $M_{K}$ and $M_{*}$ values from SED fitting. 

\begin{table}
\begin{tabular}{ c|c|cccc}
\cline{1-6}
&&&&\\
Sub- & Criterion & SAMI & SAMI & KROSS  \\
sample &  & original & matched &  \\
\cline{1-6}
&&&&\\
& Detected & 824 & 824 & 719 \\
& in H$\alpha$ & & & \\
&&&&\\
\textit{parent} & Resolved & 752 & 586 & 552 \\
& in H$\alpha$ & & & \\
&&&&\\
& $M_{K}$ and $M_{*}$ \\
& from & 751 & 585 & 537 \\
& SED fitting \\
&&&&\\
& $v_{2.2}$ & 669 & 490 & 530 \\
&&&&\\
\cline{1-6}
&&&&\\
& $\frac{\Delta_{v_{2.2}}}{v_{2.2}} \leq 0.3$ & 625 & 355 & 467 \\
&&&&\\
& $\frac{r_{\rm{H}\alpha,\rm{max}}}{r_{\text{e}}} \ge 1.3$ & 527 & 313 & 456 \\
\textit{rot-dom}  &&&&\\
& $45^\circ < i < 85^\circ$ & 367 & 216 & 311 \\
&&&&\\
& $\frac{v_{2.2}}{\sigma} + \Delta\frac{v_{2.2}}{\sigma} > 1$ & 309 & 186 & 259 \\
&&&&\\
\cline{1-6}
&&&&\\
& $\frac{v_{2.2}}{\sigma} + \Delta\frac{v_{2.2}}{\sigma} > 3$ & 151 & 76 & 127 \\
\textit{disky} &&&&\\
& $R^{2} > 80$\% & 134 & 70 & 112 \\

&&&&\\
\cline{1-6}
\end{tabular}
\caption{Summary of the selection criteria and size of the sub-samples defined in \S~\ref{sec:SAMIvsKROSSsample}. Criteria are applied step by step from top to bottom. The numbers represent the size of each sample after each successive cut is applied.}
\label{tab:SAMIKROSSsamples}
\end{table}

\subsection{The \textit{rot-dom} sub-samples}
\label{subsec:rotdomsample}

We then define a \textit{rotation-dominated} (\textit{rot-dom}) sub-sample (which bear similarities to the \textit{all} sub-sample of \citealt{Tiley:2016}) for each data set that comprises all those galaxies that are member of the \textit{parent} sub-sample and with a ratio of intrinsic rotation velocity to intrinsic line-of-sight velocity dispersion that is greater than unity. This criterion is commonly applied in the literature for studies of galaxy kinematics at $z\approx1$--$2$ \citep[e.g.][]{Genzel:2006,ForsterSchreiber:2009,Epinat:2012,Stott:2016,Tiley:2016,Harrison:2017,Johnson:2018} as a crude deliniation between ``rotation-dominated'' systems that obey the criterion and ``dispersion-dominated'' systems that violate it. Accounting for uncertainties in the ratio of rotation-to-dispersion, we require that $v_{2.2}/\sigma + \Delta v_{2.2}/\sigma >1$. We stress that we do not expect this criterion to effectively select systems that obey the assumption of circular motion inherent in the TFR. Rather we employ it as a bare minimum for a galaxy's inclusion in our analysis and for ease of comparison with previous studies. In \S~\ref{subsec:diskysample} we describe a stricter cut in $v_{2.2}/\sigma$ chosen to select galaxies that more closely obey the circular motion assumption.

We apply further selection criteria so that each galaxy in the \textit{rot-dom} sub-samples has a fractional error in the velocity measurement less than or equal to thirty percent, and has sufficient H$\alpha$ radial extent to empirically constrain the velocity measure adopted here. We thus require a maximum H$\alpha$ radius $r_{\text{H}\alpha,\text{max}} \ge 1.3r_{\text{e}}$ (where an error margin of 2 spaxels and 1 spaxel is allowed for respectively the KROSS and SAMI matched data, and the SAMI original data to account for the pixelisation of the H$\alpha$ intensity map). We also require that $45^\circ < i < 85^\circ$. The lower limit is based on the findings presented in \citet{Tiley:2016b} and is designed to exclude those systems that require large inclination correction factors (and are thus very sensitive to inaccuracy in the measure of inclination). The upper limit is imposed in order to exclude very edge-on systems with an increased probability of suffering from significant dust obscuration. 

\subsection{The \textit{disky} sub-samples}
\label{subsec:diskysample}

Finally, as alluded to in \S~\ref{subsec:rotdomsample}, for each data set we also define a more strictly rotation-dominated, \textit{disky} sub-sample of galaxies that appear disk-like in their kinematic properties. This comprises all those members of the \textit{rot-dom} sub-sample for which the ratio of rotation velocity to intrinsic velocity dispersion $v_{2.2}/\sigma + \Delta v_{2.2}/\sigma > 3$. This lower limit is chosen in light of the results of \citet{Tiley:2016} that showed that the TFR zero-point does not change as a function of increasing $v/\sigma$ above this value. More formally, it also ensures that the rotation velocity term in the first velocity moment of the collisionless Boltzmann equation accounts for at least 90\% of the dynamical mass (adopting the reasonable assumption that the velocity anisotropy factor is less than unity e.g. \citealt{Kormendy:2001}). Therefore whilst $v_{2.2}/\sigma=3$ is lower than the $v/\sigma\sim5$--$20$ \citep[][]{Epinat:2010} measured for disk galaxies in the local Universe, these systems should nevertheless effectively obey the assumption of circular motion required for the TFR.    

To exclude potentially merging systems and to further select for systems with regular, disk-like kinematics, we also extend the exponential disk model of Equation~\ref{eq:mod} to two dimensions, fitting it to the observed velocity map of each galaxy using a genetic algorithm \citep{Charb:1995} in the manner of \citet{Swinbank:2012b} and \citet{Stott:2016}. The resultant best-fit model velocity map for each galaxy in the \textit{disky} sub-samples must have an associated goodness of fit parameter $R^{2} > 80$\% i.e.\ more than $80$\% of the total variation in the observed velocity map must be explained by the best fit model map (\citealt{Bloom:2017} show that the most kinematically asymmetric galaxies scatter low off the TFR). In Figure~\ref{fig:SAMIOGbadvels} we show two example original SAMI velocity fields and their corresponding best fit disk model, one with $R^{2} > 80$\%, and the other a poor fit.

\begin{figure}
\begin{minipage}[]{1.\textwidth}
\label{fig:AOGmap}
\includegraphics[width=0.485\textwidth,trim = 5 68 0 20, clip=True]{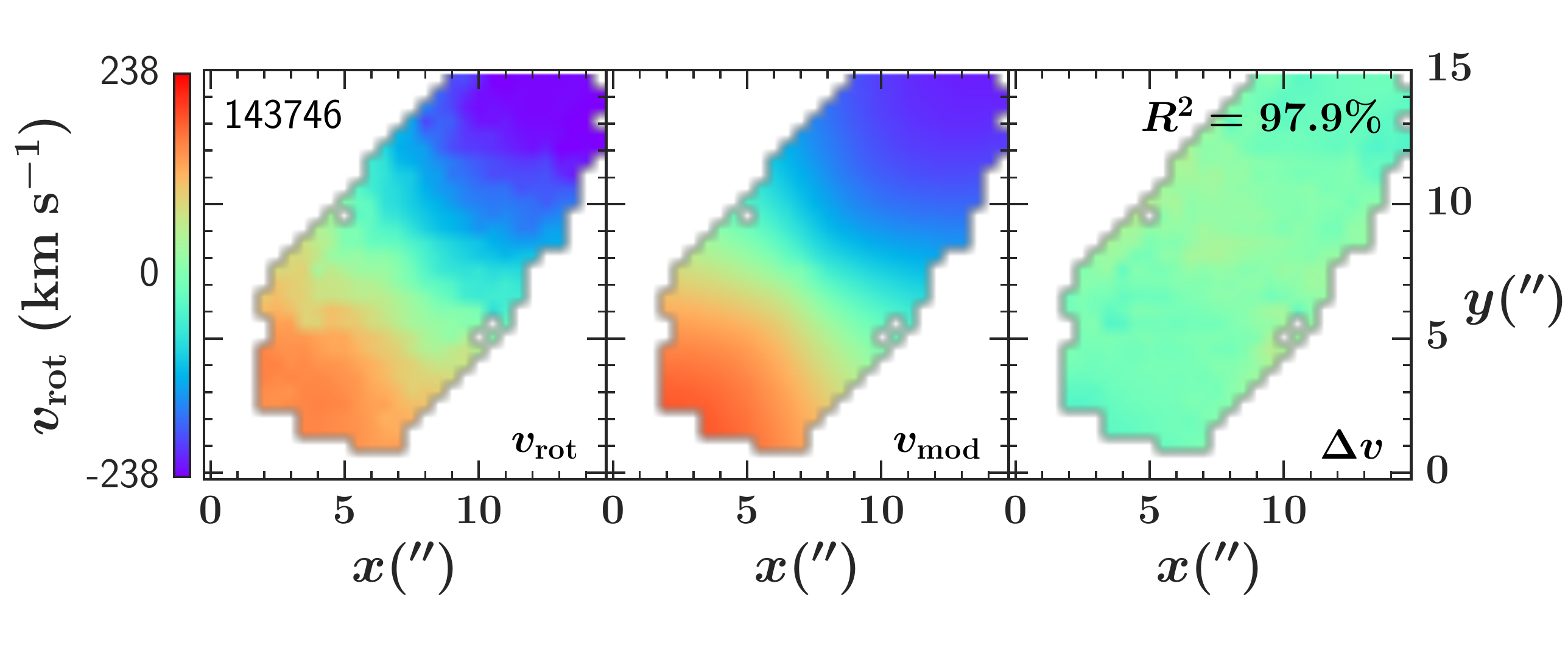}
\end{minipage}
\begin{minipage}[]{1.\textwidth}
\label{fig:DOGmap}
\includegraphics[width=0.485\textwidth,trim = 5 35 0 20, clip=True]{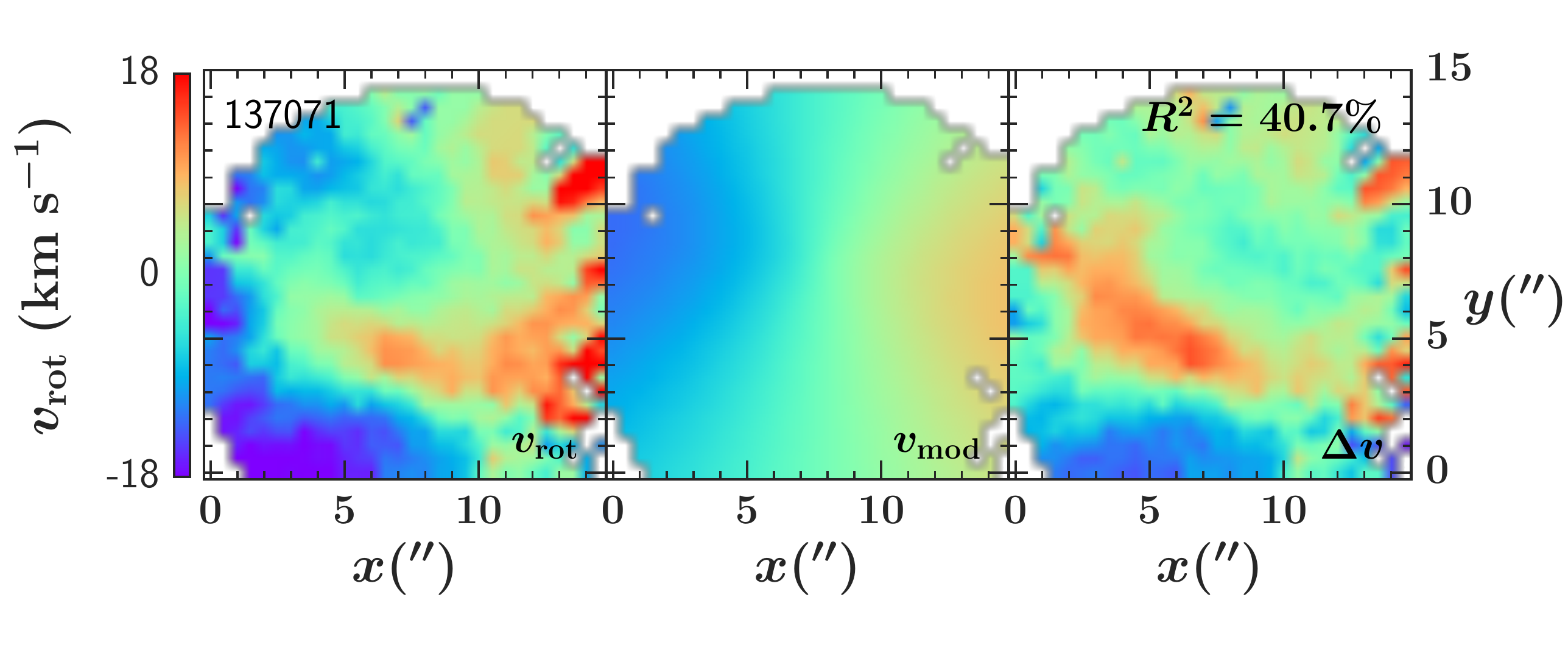}
\end{minipage}
\caption{Two example original SAMI velocity fields (left), along with their corresponding best fit two-dimensional exponential disk model (see Equation~\ref{eq:mod}) (middle), and the residual between the two (right). The upper panel is an example of a ``good'' fit ($R^{2}>80$\%), whilst the lower panel shows a ``bad'' fit ($R^{2} \leq 80$\%).}%
\label{fig:SAMIOGbadvels}
\end{figure}

Application of the various selection criteria results in {\it parent} sub-samples comprising 669, 490, and 530 galaxies for respectively the original SAMI, matched SAMI and KROSS data sets. The \textit{rot-dom} sub-samples contain respectively 309, 186, and 259 galaxies. Lastly, the \textit{disky} sub-sample comprises 134 galaxies for the original SAMI data, 70 galaxies for the matched SAMI data, and 112 galaxies for the KROSS data. The final \textit{disky} sub-sample for the original SAMI, matched SAMI, and KROSS data set then is respectively 20\%, 14\%, and 21\% the size of its corresponding {\it parent} sub-sample. A summary of the selection criteria of each sub-sample, along with the number of galaxies remaining after each selection criterion is applied, is provided in Table~\ref{tab:SAMIKROSSsamples} for each data set.

\section{Results}
\label{sec:KROSSvsSAMIresults}

\begin{figure}
\begin{minipage}[]{\textwidth}
\label{fig:sizemass}
\includegraphics[width=0.48\textwidth,trim= 25 86 50 30,clip=True]{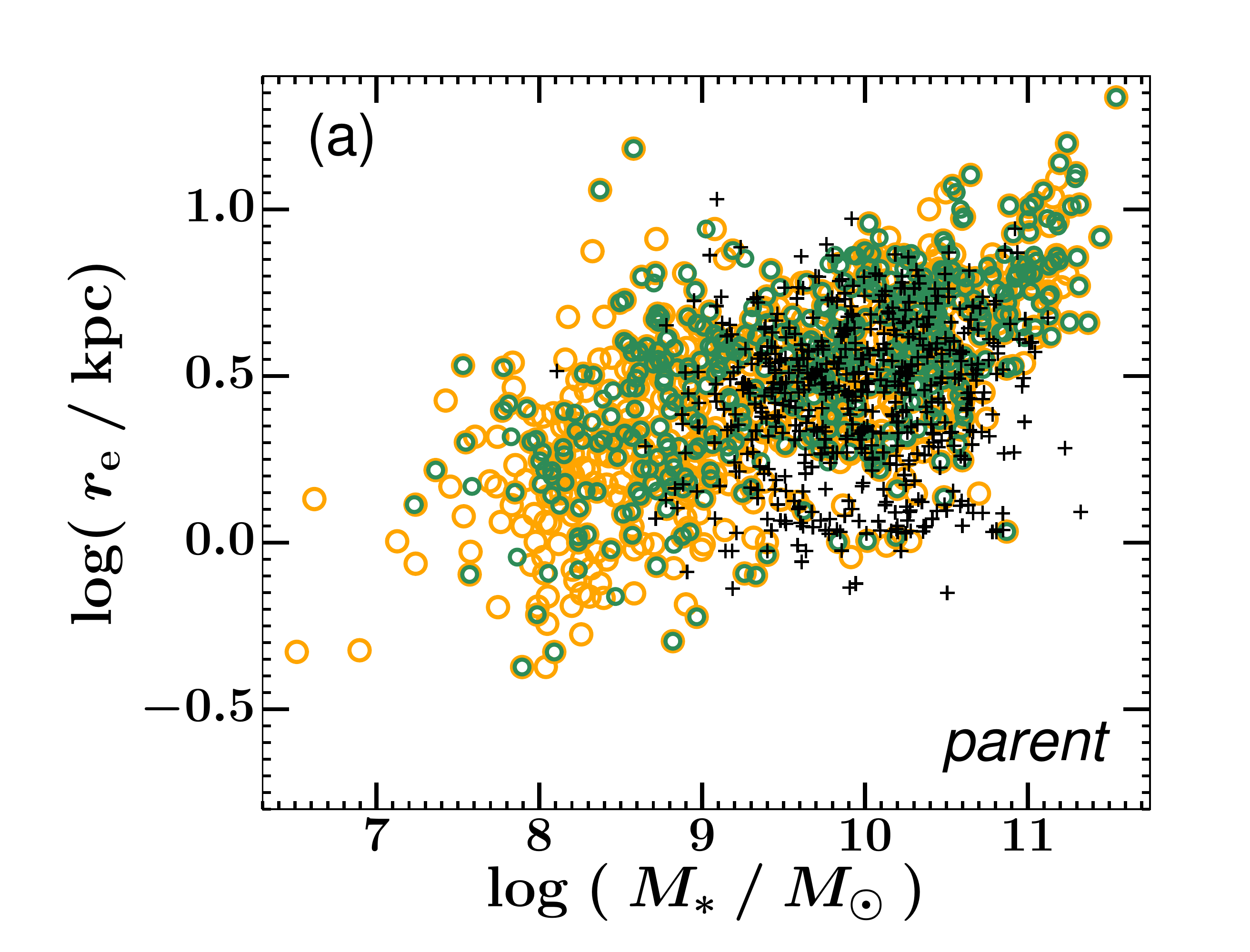}
\vfill
\includegraphics[width=0.48\textwidth,trim= 25 86 50 46,clip=True]{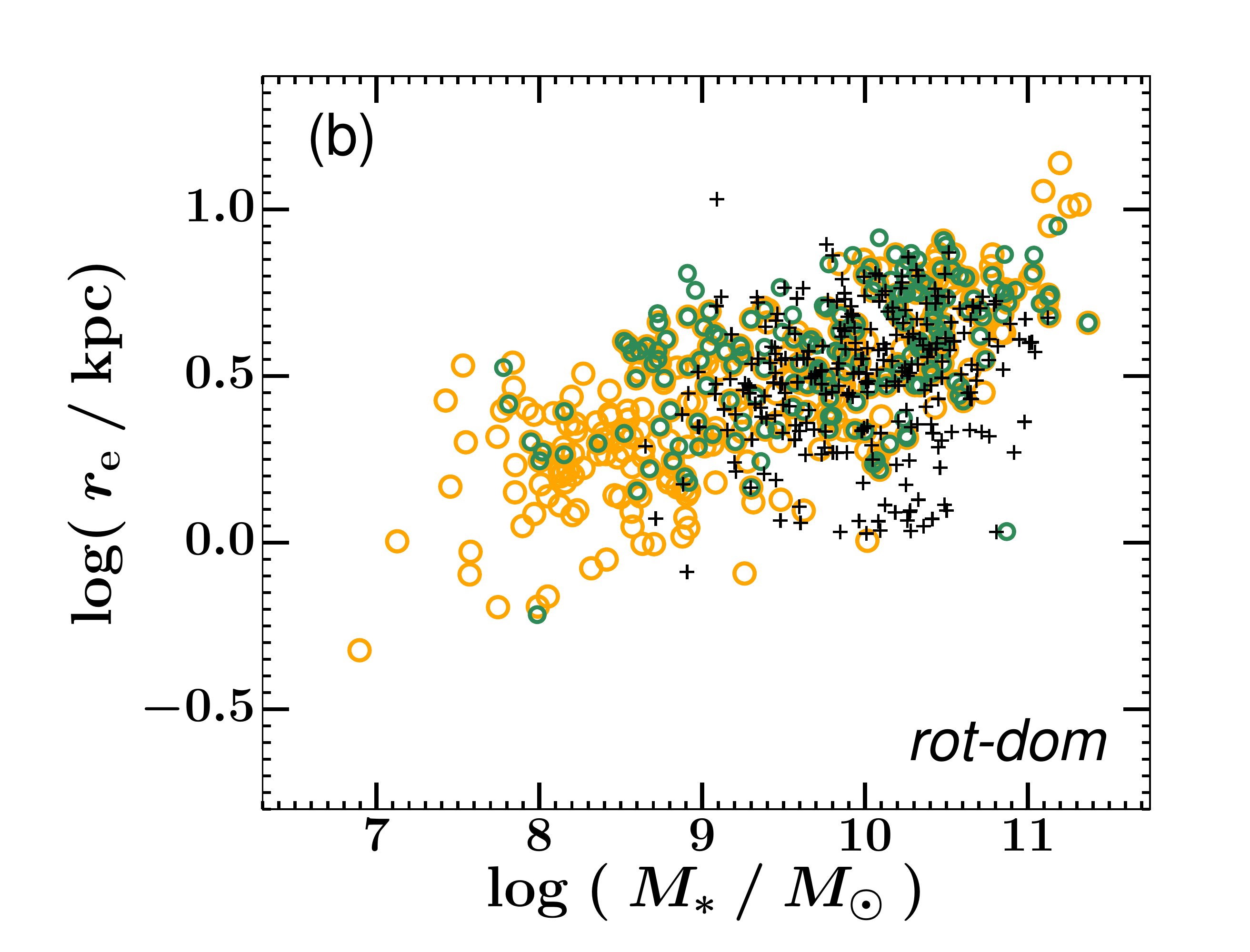}
\vfill
\includegraphics[width=0.48\textwidth,trim= 25 0 50 46,clip=True]{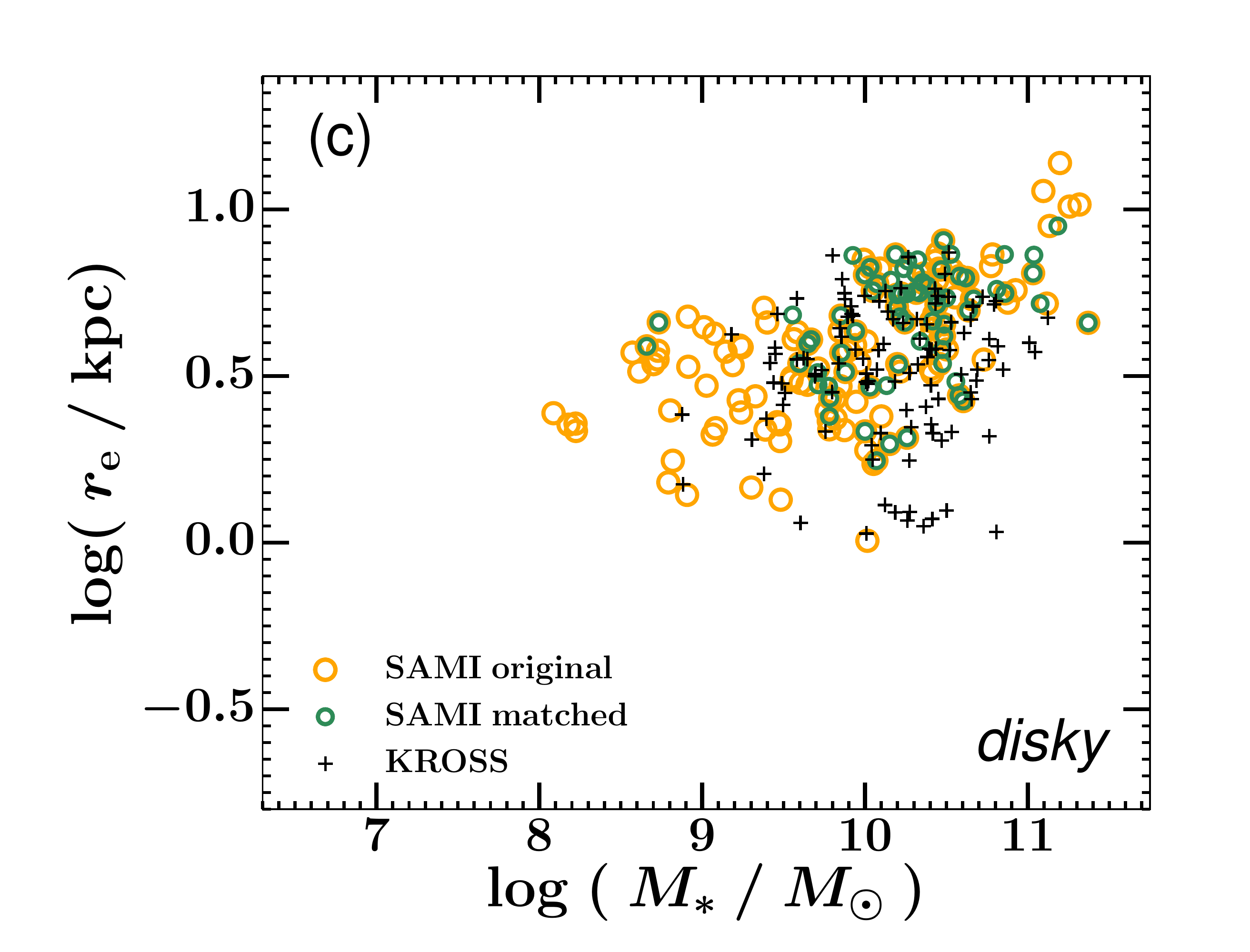}
\end{minipage}
\caption{%
Mass-size relation of the original SAMI, matched SAMI, and KROSS \textit{parent} (panel a), \textit{rot-dom} (panel b) and \textit{disky} (panel c) sub-samples. Sizes are measured from stellar continuum light in broadband images (in the restframe $r$-band ($\sim 0.6\mu$m) for SAMI galaxies, and restframe $i$-band ($\sim0.8\mu$m) for KROSS galaxies). Error bars are omitted for clarity. In each panel, the three data sets generally follow the same trend, more massive galaxies being larger (as expected). The KROSS sub-samples, however, are limited to higher stellar masses than the original SAMI and matched SAMI sub-samples.%
     }%
\label{fig:mass_vs_re_SAMIKROSS}
\end{figure}

In this section we present the $M_{K}$ and $M_{*}$ TFRs for the \textit{rot-dom} and \textit{disky} sub-samples of the original SAMI, matched SAMI and KROSS data sets. The values used to construct each of the TFRs presented here are tabulated in Table~\ref{tab:tfrvals}. In Figure~\ref{fig:mass_vs_re_SAMIKROSS} we present the mass-size relations of the original SAMI, matched SAMI and KROSS \textit{parent}, \textit{rot-dom} and \textit{disky} sub-samples, that show that the galaxies of all three data sets (and this across all three sub-samples) follow the same general trend of increasing size with increasing stellar mass \citep[as expected,][]{Shen:2003,Bernardi:2011}, although the SAMI sub-samples clearly extend to lower stellar masses (and smaller radii). 

The original SAMI and matched SAMI TFRs are compared in \S~\ref{subsec:HQvsLQ}, where the biases introduced via the matching process are explored. In \S~\ref {subsec:KROSSvsLQ} we compare the matched SAMI and KROSS TFRs, to measure the evolution of the relations since $z\approx1$.  

\subsection{Matched vs. Original SAMI}
\label{subsec:HQvsLQ}

In this sub-section we explore the extent to which the data matching process applied to the original SAMI data affects our ability to accurately recover key galaxy parameters needed for constructing the TFR. We also examine how the reduced data quality between the matched SAMI and original SAMI data can introduce biases between sub-samples selected from each using identical criteria. Finally, we compare TFRs constructed from both the original SAMI and matched SAMI data, highlighting significant differences between the two and determining the dominant factor that drives these differences.  

\subsubsection{Measurement Bias and Sample Statistics}
\label{subsubsec:measurebias}

To understand any differences between the matched SAMI and original SAMI TFRs, it is informative to first directly compare the measurements used to construct the relations, i.e.\ assess how the measurements of $v_{2.2}$ and $\sigma$ are affected by the degrading process and subsequent velocity field extraction and modelling. Similarly, we must quantify to what extent the truncation of the SAMI SEDs alters the derived $M_{*}$ of each galaxy ($M_{K}$ is nearly SED independent and thus unaffected). We thus compare the $v_{2.2}$, $\sigma$ and $M_{*}$ measurement of the galaxies in the matched SAMI \textit{parent} (and \textit{rot-dom}) sub-sample to the corresponding measurements made using the original SAMI data, for the same galaxies.

\begin{figure}
\begin{minipage}[]{1\textwidth}
\label{fig:SAMK}
\includegraphics[width=0.45\textwidth,trim= 0 0 47 15,clip=True]{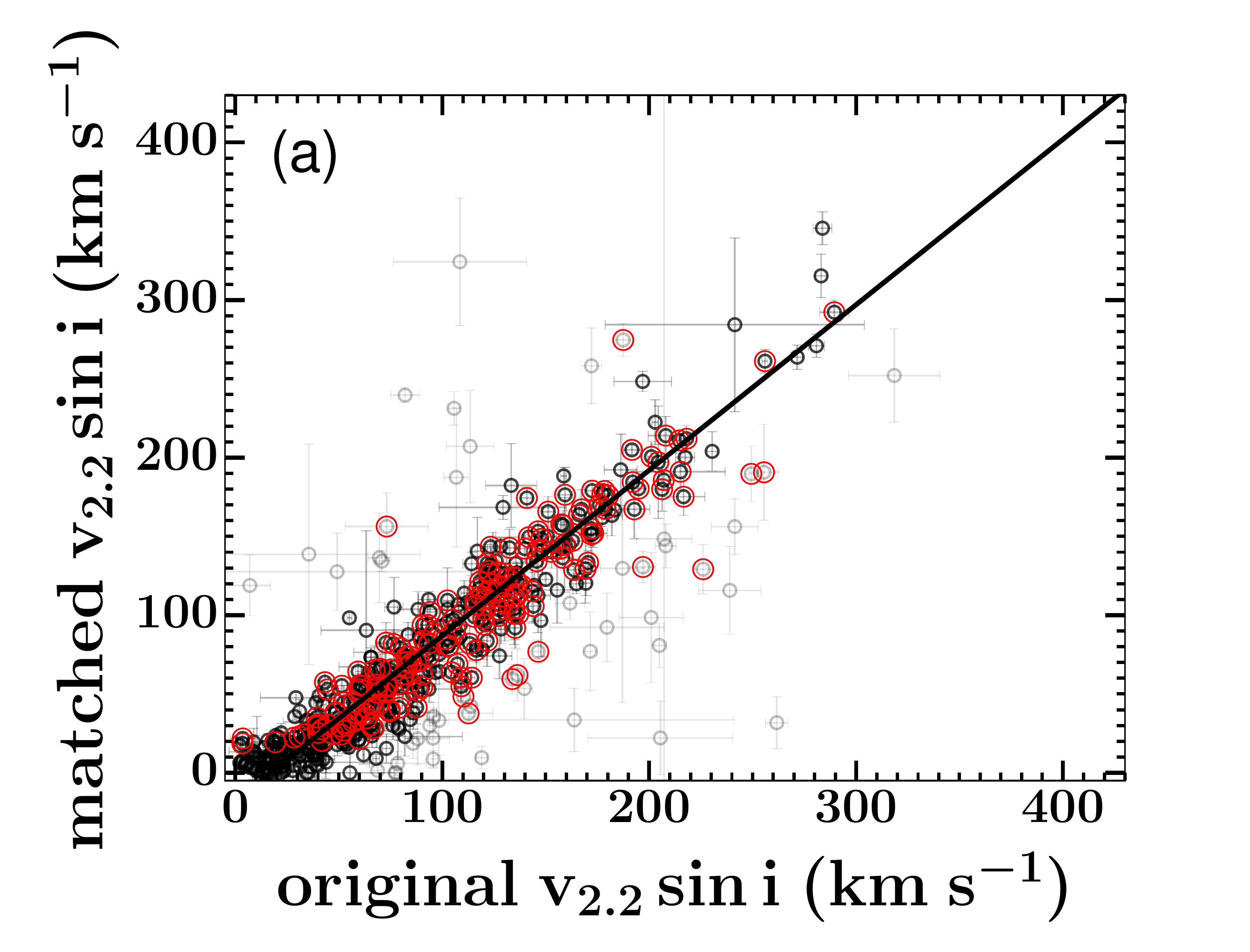}
\vfill
\includegraphics[width=0.45\textwidth,trim= 0 0 47 30,clip=True]{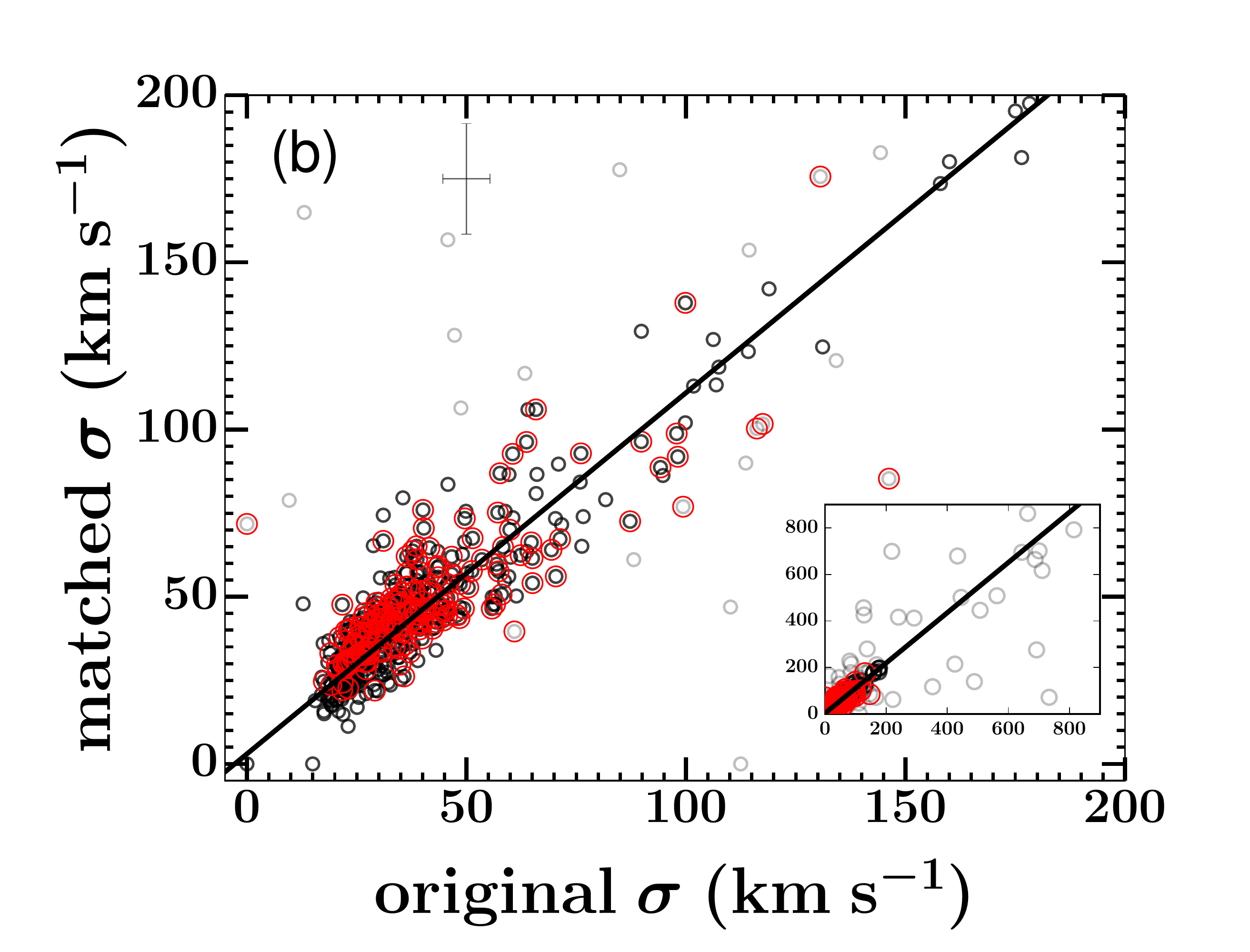}
\vfill
\includegraphics[width=0.45\textwidth,trim= 0 0 47 30,clip=True]{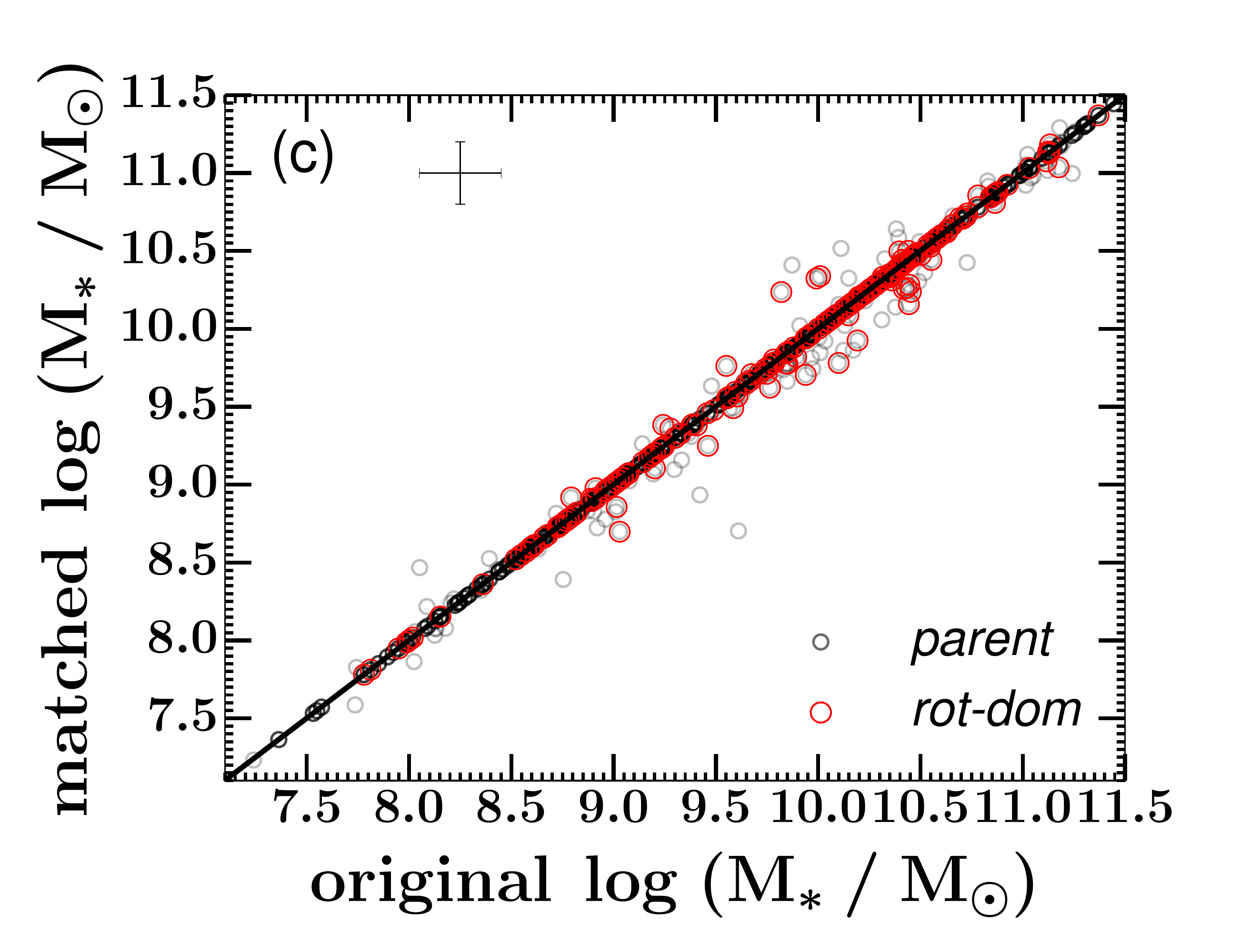}
\end{minipage}
\caption{%
Matched SAMI \textit{parent} and \textit{rot-dom} sub-sample measurements of $v_{2.2}\sin i$, $\sigma$ and $M_{*}$ versus the corresponding original SAMI measurements. The black solid line in each panel is the best (bisector) fit to the \textit{parent} data points. For clarity, in panels (b) and (c), the median uncertainty in both axes is indicated by a single point. We include an inset panel with increased axes limits in panel (b) to show outliers that are not displayed in the main plot. In all cases, the matched and original measurements generally agree, with the best fits nearly consistent with 1:1 relations, but with zero-points offset from zero to varying degrees and with varying scatters. Data points that are excluded via consecutive $2.5$-$\sigma$ clipping are shown as fainter points.%
     }%
\label{fig:HQLQ_directcomp}
\end{figure}

Figure~\ref{fig:HQLQ_directcomp} shows comparisons between the matched SAMI and original SAMI \textit{parent} and \textit{rot-dom} sub-sample measurements of $v_{2.2}\sin i$, $\sigma$ and $M_{*}$. The parameters of the best (bisector) fit straight line to each comparison between the \textit{parent} sub-samples are listed in Table~\ref{tab:HQLQ_directcomp}, along with measures of the total and intrinsic scatters along both axes. For each comparison we perform three consecutive fits, performing a $2.5$-$\sigma$ clip to the residuals between the data and the best fit line in each case. Those data points with residuals that are excluded via this clip are then excluded from the comparison and the next best fit found. 

The matched and original measurements for the \textit{parent} sub-samples generally agree with each other, being well correlated with varying total and intrinsic scatters. The stellar masses measured from the original SAMI and matched SAMI data follow a 1:1 relationship. This is also true of the $\sigma$ measurements, within uncertainties.  Similarly, the slope of the best fit to the $v_{2.2}\sin i$ comparison is close to unity. However the best-fit zero-point reveals a small systematic offset between the matched SAMI and original SAMI $v_{2.2}\sin i$ measurements. The measurements from the former are, on average, $18 \pm 2$ km s$^{-1}$ lower than the latter. This offset is seemingly driven mainly by galaxies with original SAMI measurements of $v_{2.2}\sin i \lesssim 100$ km s$^{-1}$. As discussed later, these are likely to be less massive, intrinsically smaller galaxies and therefore those most strongly affected by the data degrading process described in \S~\ref{subsec:SAMImatching}. 

A comparison between the same measurements but confined to only those galaxies in the \textit{rot-dom} sub-samples (the results of which we do not tabulate in Table~\ref{tab:HQLQ_directcomp}) reveals similar conclusions for the $v_{2.2}\sin i$ and $M_{*}$ comparisons but with less outliers than when considering the \textit{parent} sub-samples. Comparing the $\sigma$ values of the \textit{rot-dom} sub-samples for the matched SAMI and original SAMI data reveals a best fit slope consistent with unity ($0.88$ $\pm$ 0.05) but with a corresponding zero-point of $12 \pm 2$ km s$^{-1}$. We note that the outliers in the \textit{parent} sub-sample comparisons have low values of $R^{2}$, and typically $r_{\rm{H}\alpha,\rm{max}}/r_{\rm{e}} < 1$ for the matched SAMI data. Additionally, many of them exhibit non-disk-like kinematics in their original SAMI velocity maps. These outliers are therefore those galaxies for which the degrading process has most significantly reduced the accuracy with which we are able to recover measures of $v_{2.2}$ and $\sigma$.

Thus Figure~\ref{fig:HQLQ_directcomp} reveals that, after the application of the beam smearing correction from \citet{Johnson:2018}, the degradation of the original SAMI data to match the quality of KROSS observations results in measurements of $v_{2.2}\sin i$ and $\sigma$ that are only very slightly under-estimated and over-estimated, respectively. Whilst these biases are small, they are none-the-less important, given that accurate measurements of $v_{2.2}$ and $\sigma$ are essential to an accurate measure of the TFR. In particular, as will be discussed in \S~\ref{subsubsec:compTFRs}, a small systematic change to the rotation velocity has the potential to affect a large change in the TFR.

\begin{table*}
\begin{tabular}{ lllllll}
\hline
Measure & Slope  & \phantom{-1}Zero-point & $\sigma_{\rm{tot}}$ & $\sigma_{\rm{int}}$ & $\zeta_{\rm{tot}}$ & $\zeta_{\rm{int}}$ \\
 &    &  \phantom{-1}(km s$^{-1}$) & (km s$^{-1}$) & (km s$^{-1}$) & (km s$^{-1}$) & (km s$^{-1}$) \\

\hline
$v_{2.2}\sin i$ & $1.05$\phantom{00} $\pm$ 0.02	&	   	$-$18\phantom{.000} $\pm$ 2\phantom{.000} 		 & 16.51 $\pm$ 0.04  & 14.0 $\pm$ 0.2 					     & 15.64 $\pm$ 0.06 					   		& 13.50 $\pm$ 0.05 \\%
$\sigma$ & $1.08$\phantom{00} $\pm$ 0.04 & \phantom{$-$0}3\phantom{.000} $\pm$ 2\phantom{.700}  &  \phantom{0}8.2\phantom{0} $\pm$ 0.1\phantom{0}    &  \phantom{0}0\phantom{.0} \phantom{$\pm$ 0.2}  				& \phantom{1}7.6\phantom{4} $\pm$ 0.4\phantom{6}  & \phantom{1}0\phantom{.50} \phantom{$\pm$ 0.05}  \\%
\hline
 &    &  \phantom{-1}(dex) & (dex) & (dex) & (dex) & (dex) \\
\hline
$M_{*}$	& $1.0000$ $\pm$ 0.0001 	&	 \phantom{$-$0}0.003 $\pm$ 0.001    		 &  \phantom{0}0.00 $\pm$ 0.01  				   & \phantom{0}0\phantom{.0} \phantom{$\pm$ 0.2}  							 & \phantom{0}0.00 $\pm$ 0.01  			& \phantom{1}0\phantom{.50} \phantom{$\pm$ 0.05} \\%
\hline
\end{tabular}
\caption{Parameters of the best (bisector) straight line fits to the comparisons between the original and matched SAMI \textit{parent} sub-sample measurements of respectively $v_{2.2}$ sin$i$, $\sigma$ and $M_{*}$, as defined in the text. The total and intrinsic scatters are denoted respectively as $\sigma_{\text{tot}}$ and $\sigma_{\text{int}}$ along the ordinate, and $\zeta_{\text{tot}}$ and $\zeta_{\text{int}}$ along the abscissa. Uncertainties are quoted at the 1$\sigma$ level. We omit uncertainties for those entries for which the corresponding best fit has a reduced $\chi^{2} << 1$ and with zero intrinsic scatter i.e.\ best fits for which the uncertainties in the data more than compensate for the scatter around the fit. The best fits to the $M_{*}$ and the $\sigma$ comparisons are each consistent within uncertainties with a 1:1 relation. The best fit slope to the $v_{2.2}\sin i$ comparison differs only slightly from unity but there is a systematic offset between the matched SAMI and original SAMI measurements.}
\label{tab:HQLQ_directcomp}
\end{table*}

Before directly comparing the original SAMI and matched SAMI relations themselves, we also examine to what extent sub-samples of galaxies drawn from each data set using indentical selection criteria resemble one another. Figure~\ref{fig:SAMIvsKROSShists} shows, for each of the \textit{parent}, \textit{rot-dom}, and \textit{disky} sub-samples, comparisons of the distributions of key galaxy properties measured from the original SAMI and matched SAMI data, as well as those for the KROSS sample.  

Considering only the measurements for the original SAMI and matched SAMI sub-samples, panels (a--c), (d--f), and (g--i) of Figure~\ref{fig:SAMIvsKROSShists} demonstrate that the matching process described in \S~\ref{subsec:SAMImatching}, and the subsequent velocity field extraction, do not significantly bias the resultant $r_{e}$, $M_{*}$, and $M_{K}$ distributions for the \textit{parent} sub-samples. However, the distributions for the \textit{rot-dom} and \textit{disky} sub-samples for the matched SAMI galaxies {\it are} skewed to larger values in each of these parameters. In panels (j--l) we see that the distributions of $r_{\rm{H}\alpha,\rm{max}}$ for galaxies in the matched SAMI sub-samples are similar to those for galaxies in the corresponding original SAMI sub-samples but with a slightly lower mean in each case. Panels (m--o) reveal that the distribution of $v_{2.2}$ for the matched SAMI \textit{parent} sub-sample is skewed towards lower values than the corresponding original SAMI distribution. However, considering the \textit{rot-dom} sub-samples, the $v_{2.2}$ distributions of both data sets are similar. Further, the matched SAMI \textit{disky} sub-sample is biased to larger values of $v_{2.2}$ in comparison to the corresponding original SAMI distribution. Lastly, from panels (p--r) we see that each of the sub-samples for the matched SAMI galaxies are slightly skewed towards higher intrinsic velocity dispersions when compared to the corresponding original SAMI sub-samples. This may be a result of the selection criteria but could also partially be a reflection of the difficulty in recovering the intrinsic velocity dispersions from the matched SAMI data with complete accuracy with respect to the same measurement from the original SAMI galaxies, as discussed above.

\begin{figure*}
\begin{minipage}[]{1.\textwidth}
\centering
\includegraphics[width=0.24\textwidth,trim= 0 5 53 0,clip=True]{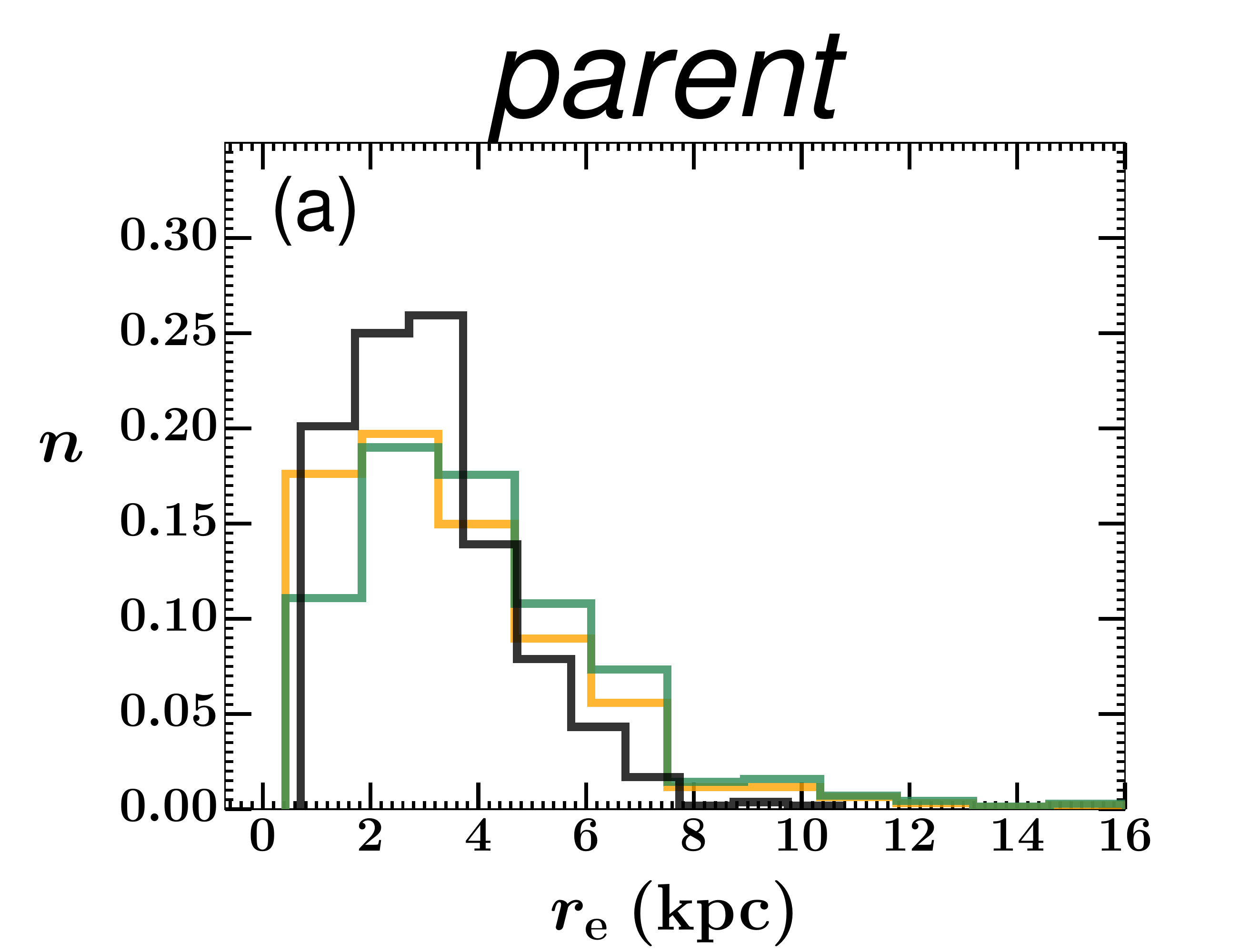}\includegraphics[width=0.24\textwidth,trim= 0 5 53 0,clip=True]{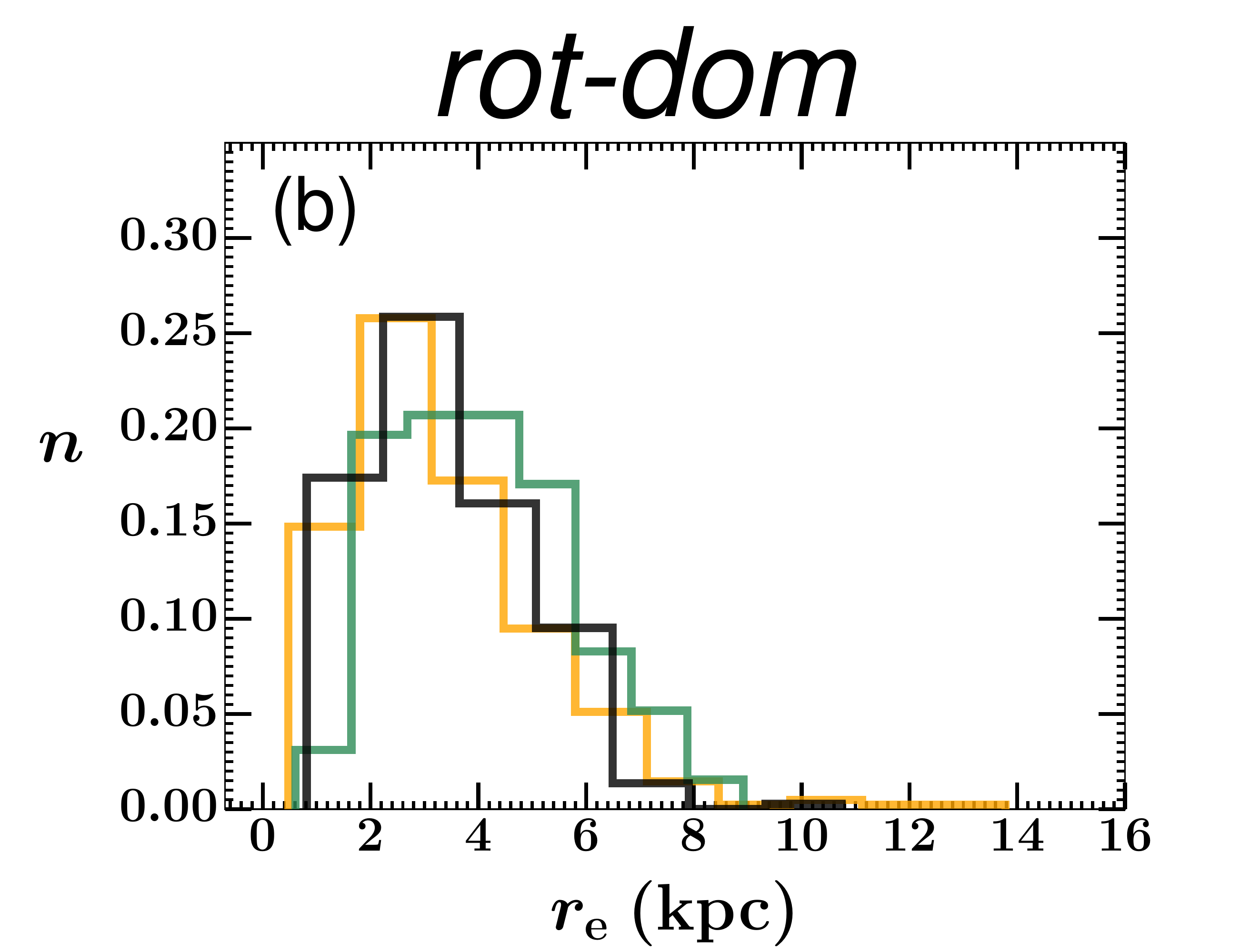}\includegraphics[width=0.24\textwidth,trim= 0 5 53 0,clip=True]{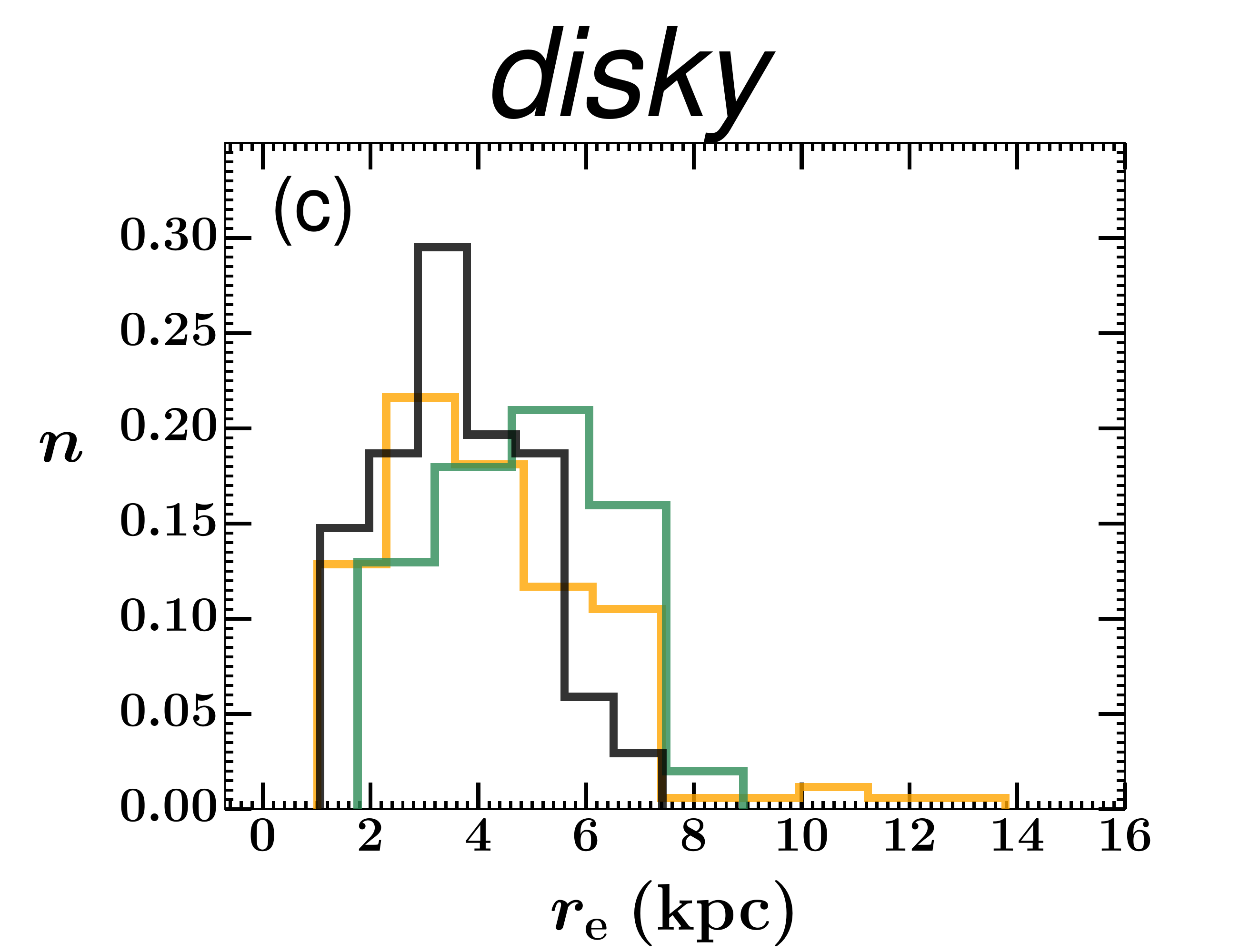}
\end{minipage}
\begin{minipage}[]{1.\textwidth}
\centering
\includegraphics[width=0.24\textwidth,trim= 0 5 53 40,clip=True]{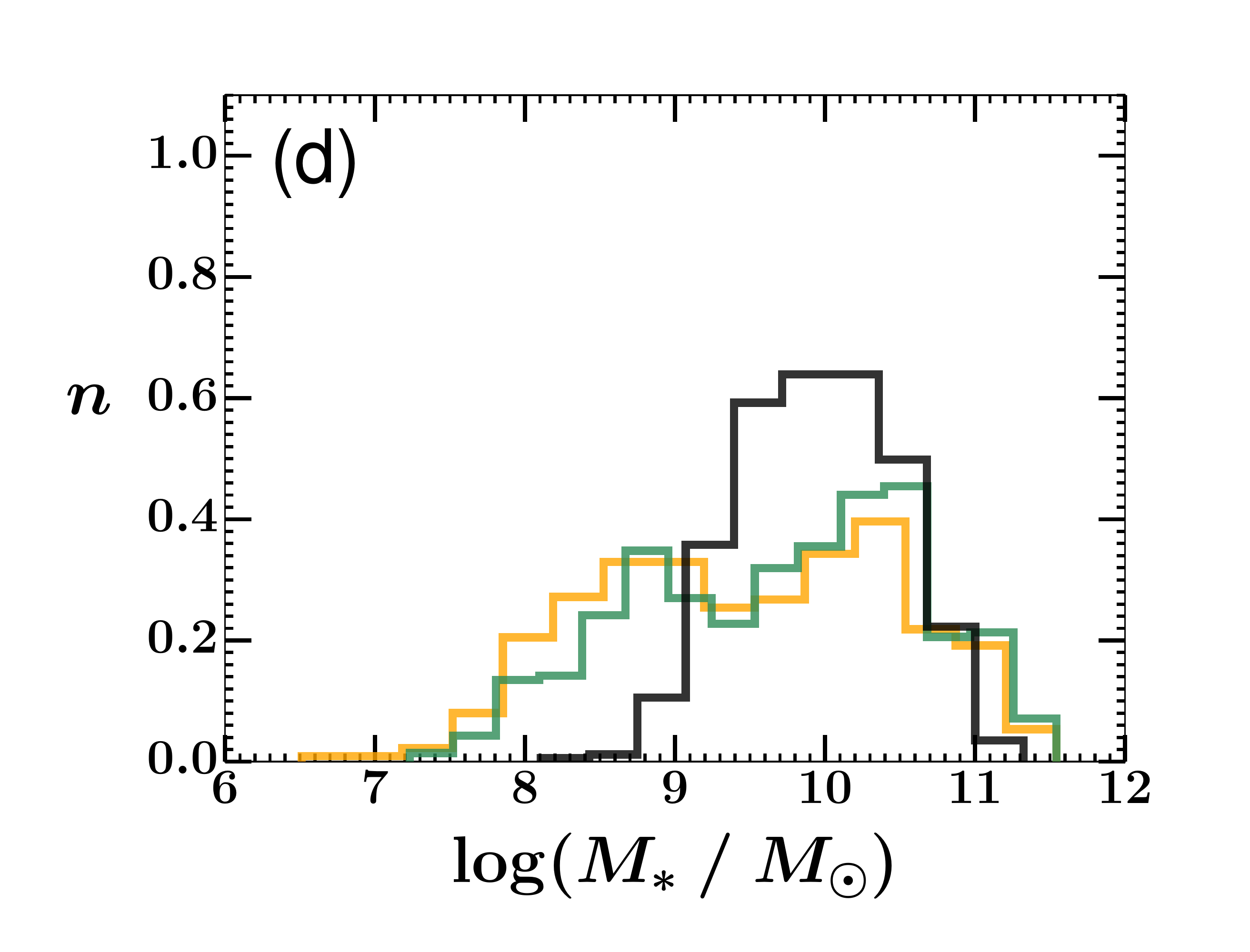}\includegraphics[width=0.24\textwidth,trim= 0 5 53 40,clip=True]{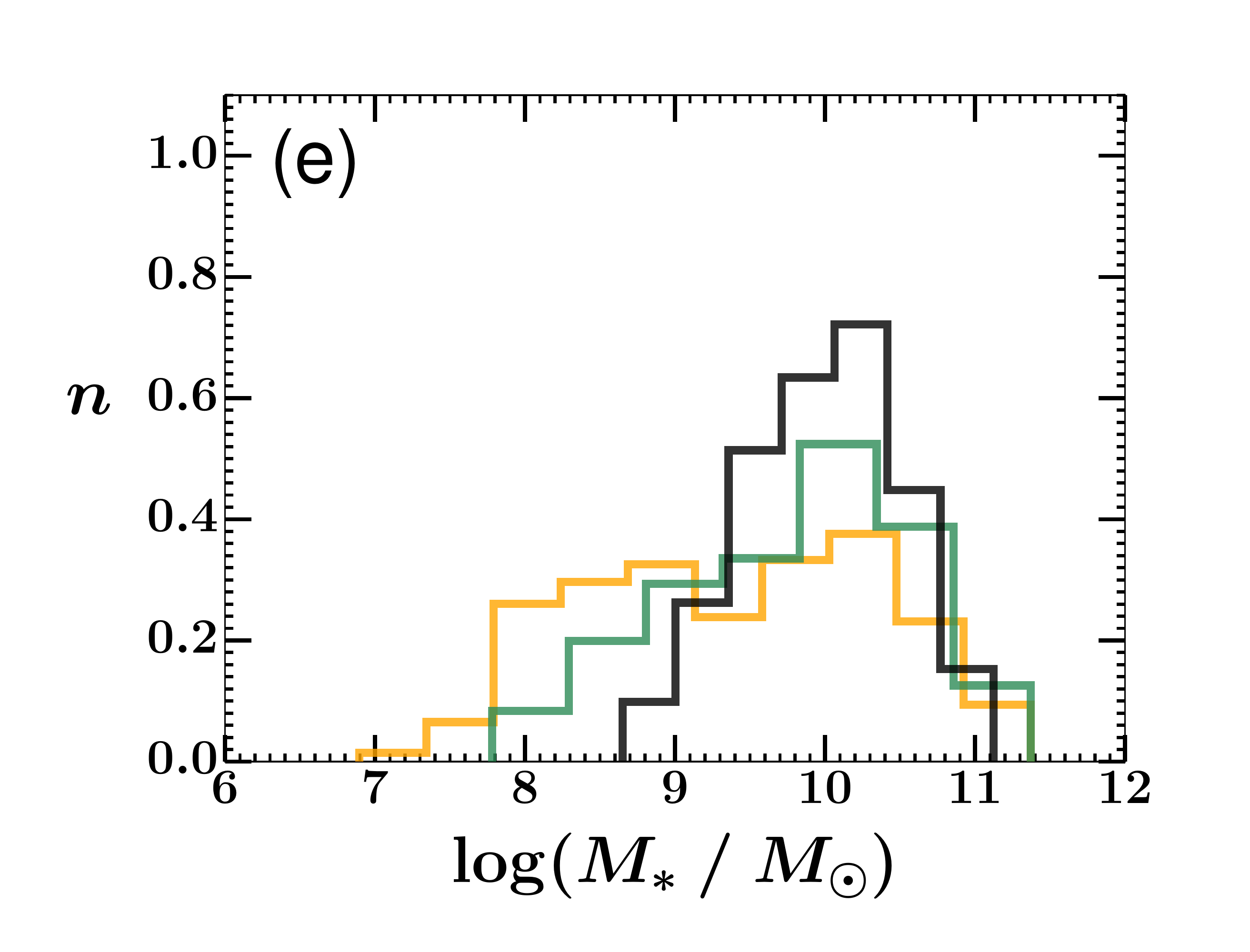}\includegraphics[width=0.24\textwidth,trim= 0 5 53 40,clip=True]{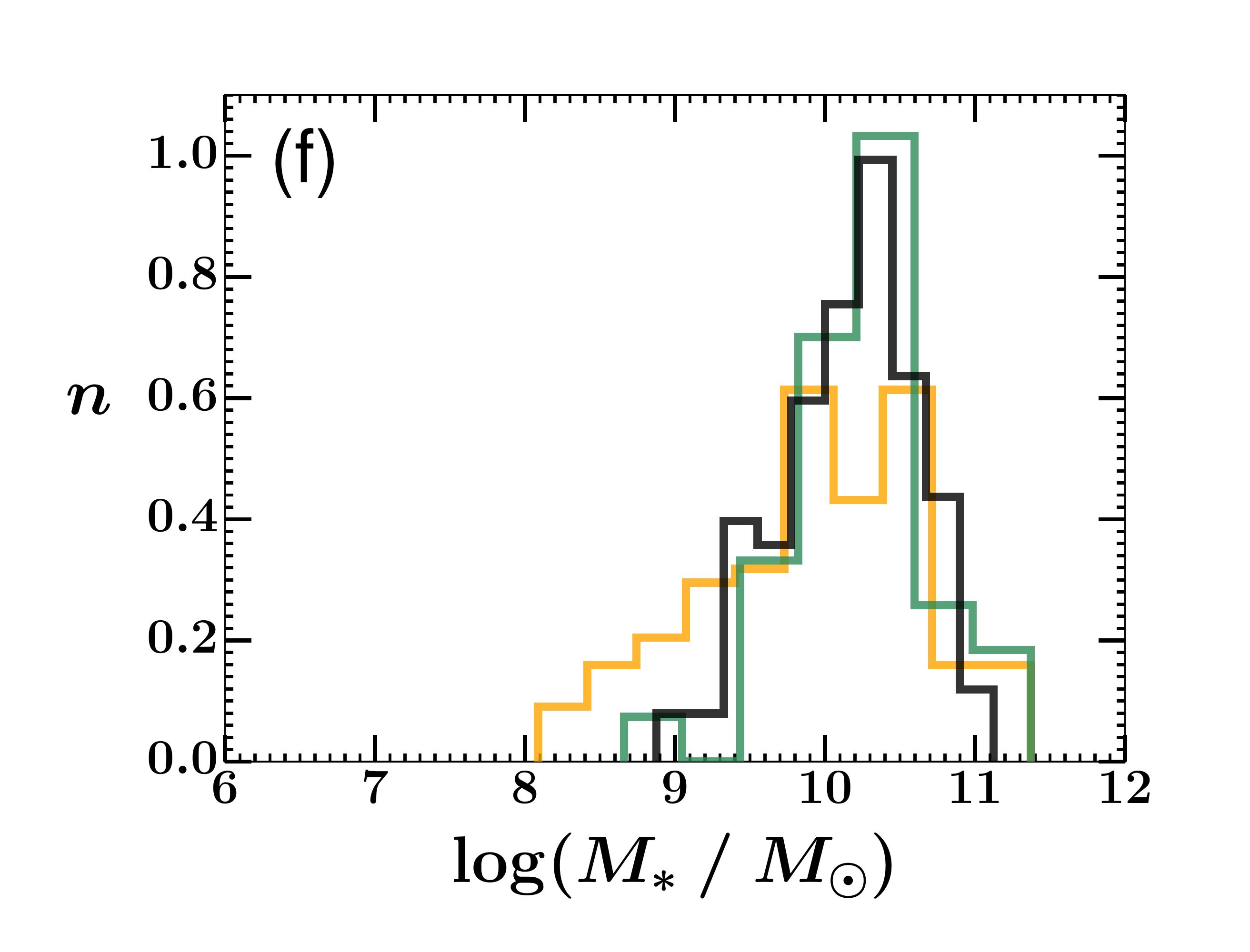}
\end{minipage}
\begin{minipage}[]{1.\textwidth}
\centering
\includegraphics[width=0.24\textwidth,trim= 0 5 53 40,clip=True]{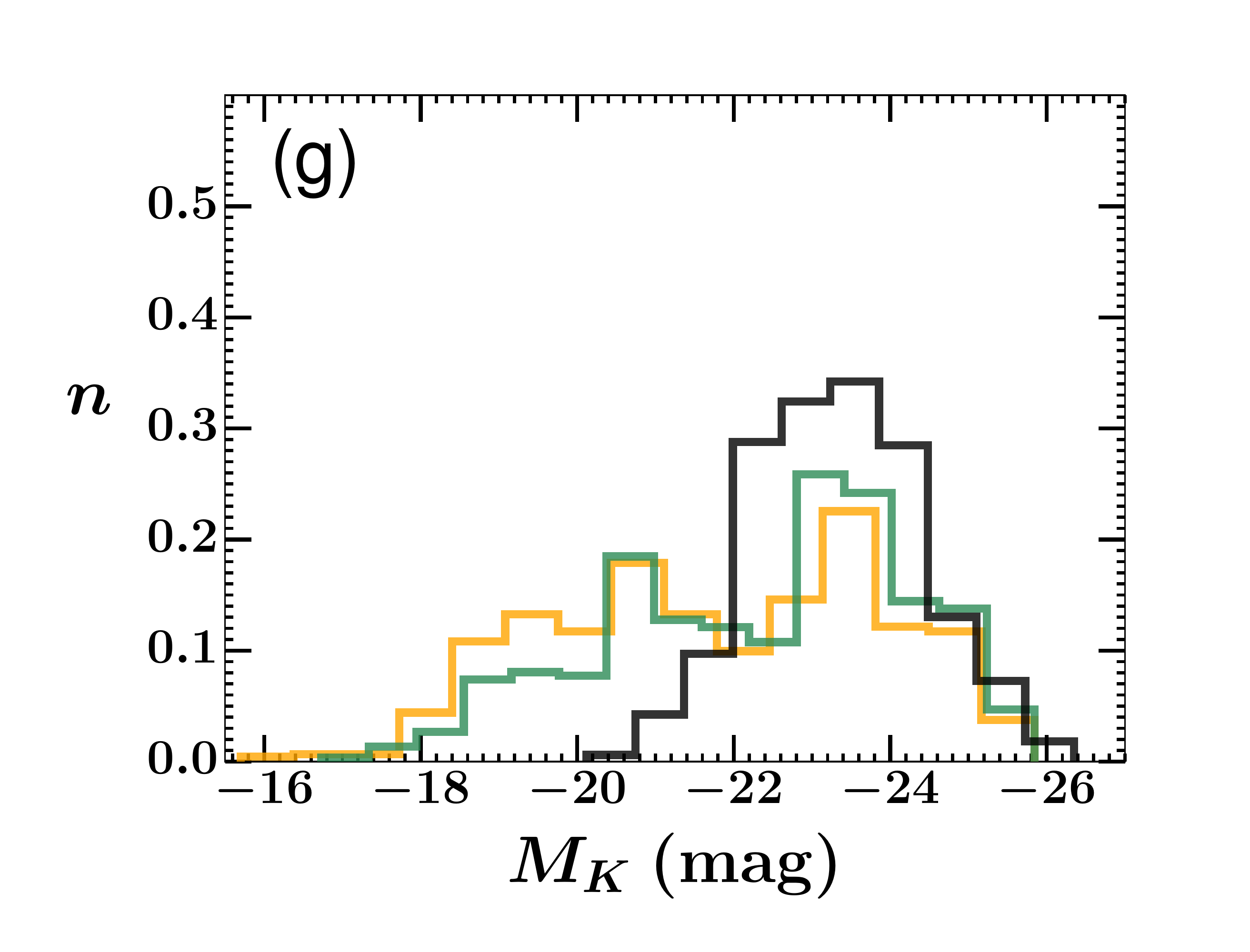}\includegraphics[width=0.24\textwidth,trim= 0 5 53 40,clip=True]{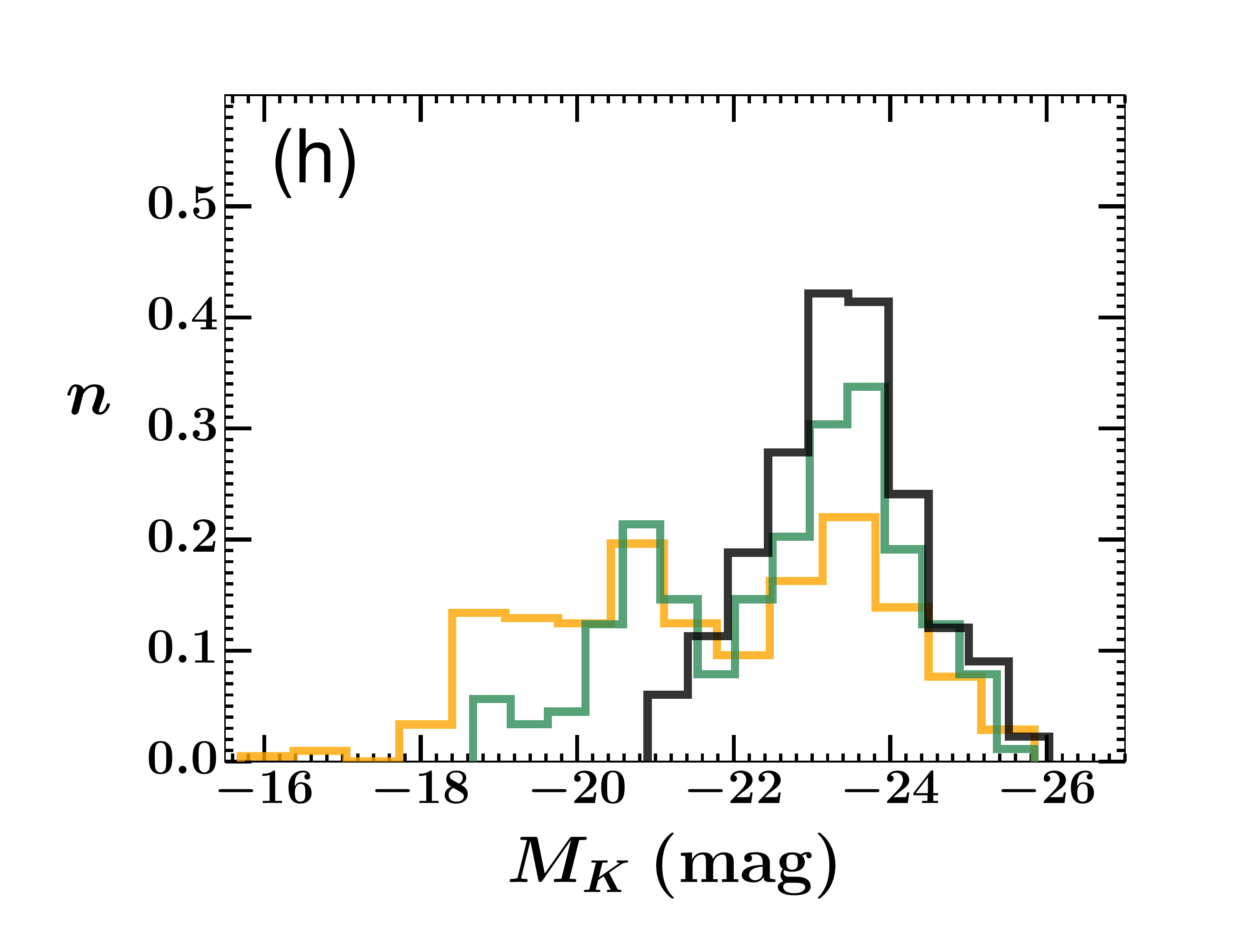}\includegraphics[width=0.24\textwidth,trim= 0 5 53 40,clip=True]{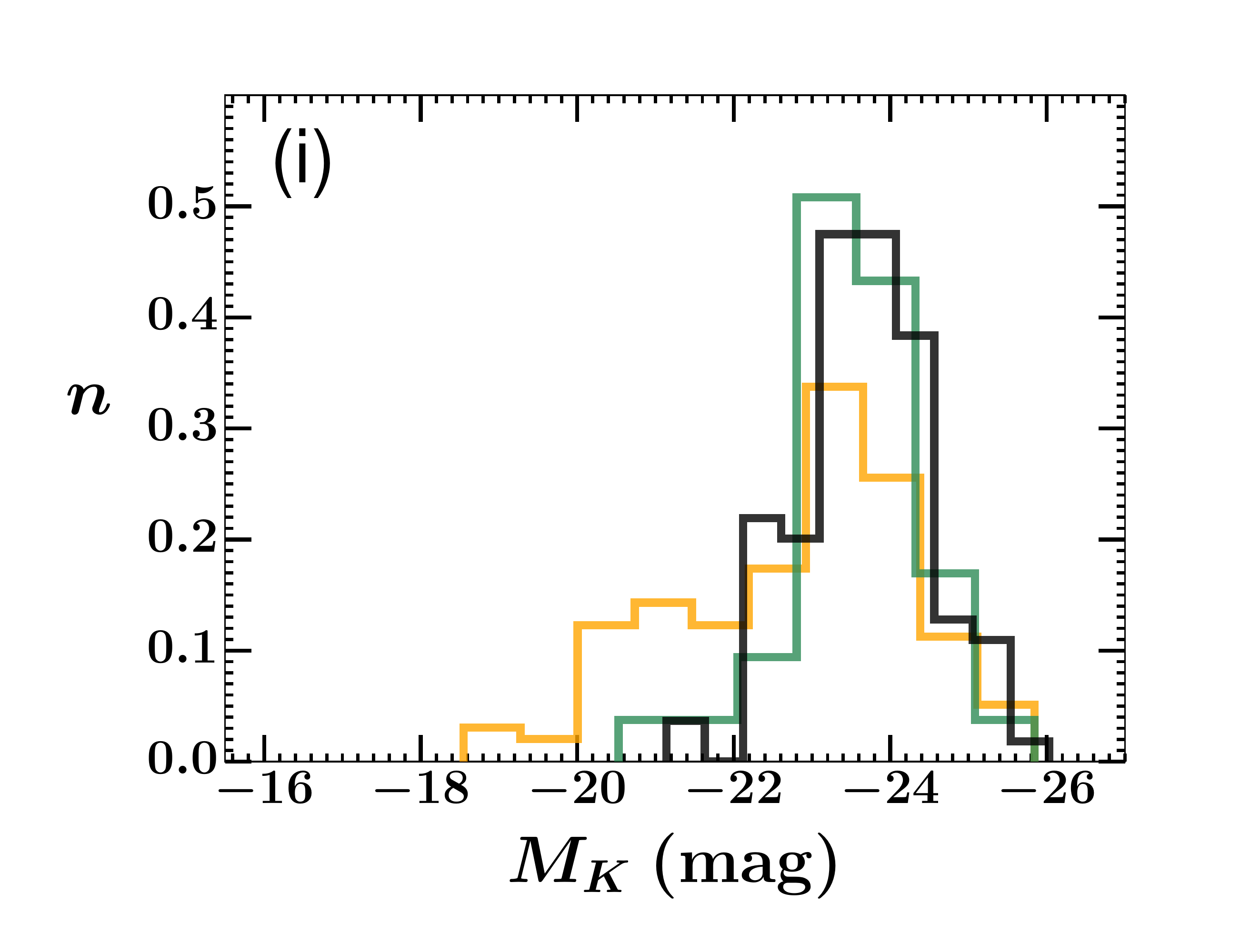}
\end{minipage}
\begin{minipage}[]{1.\textwidth}
\centering
\includegraphics[width=0.24\textwidth,trim= 0 5 53 40,clip=True]{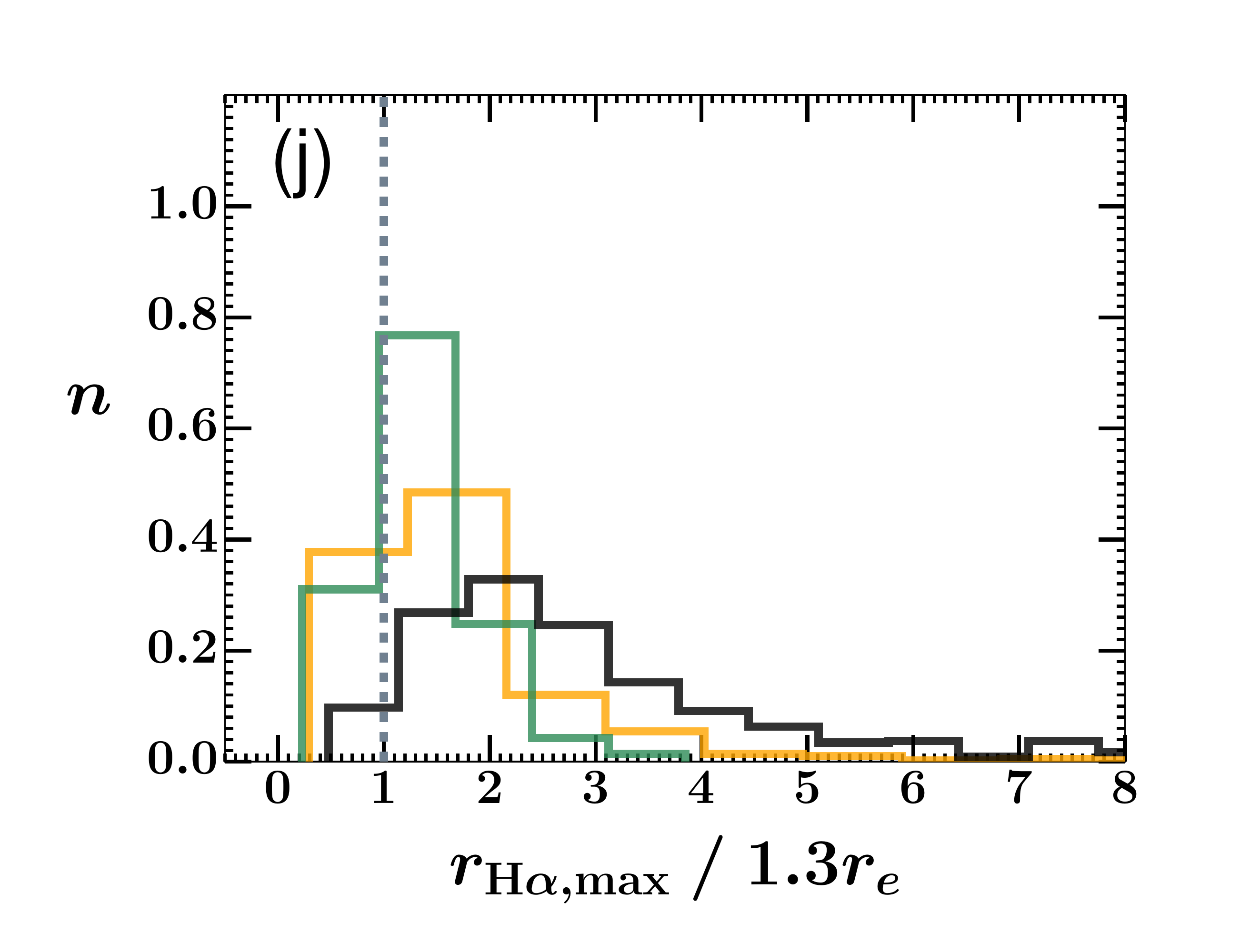}\includegraphics[width=0.24\textwidth,trim= 0 5 53 40,clip=True]{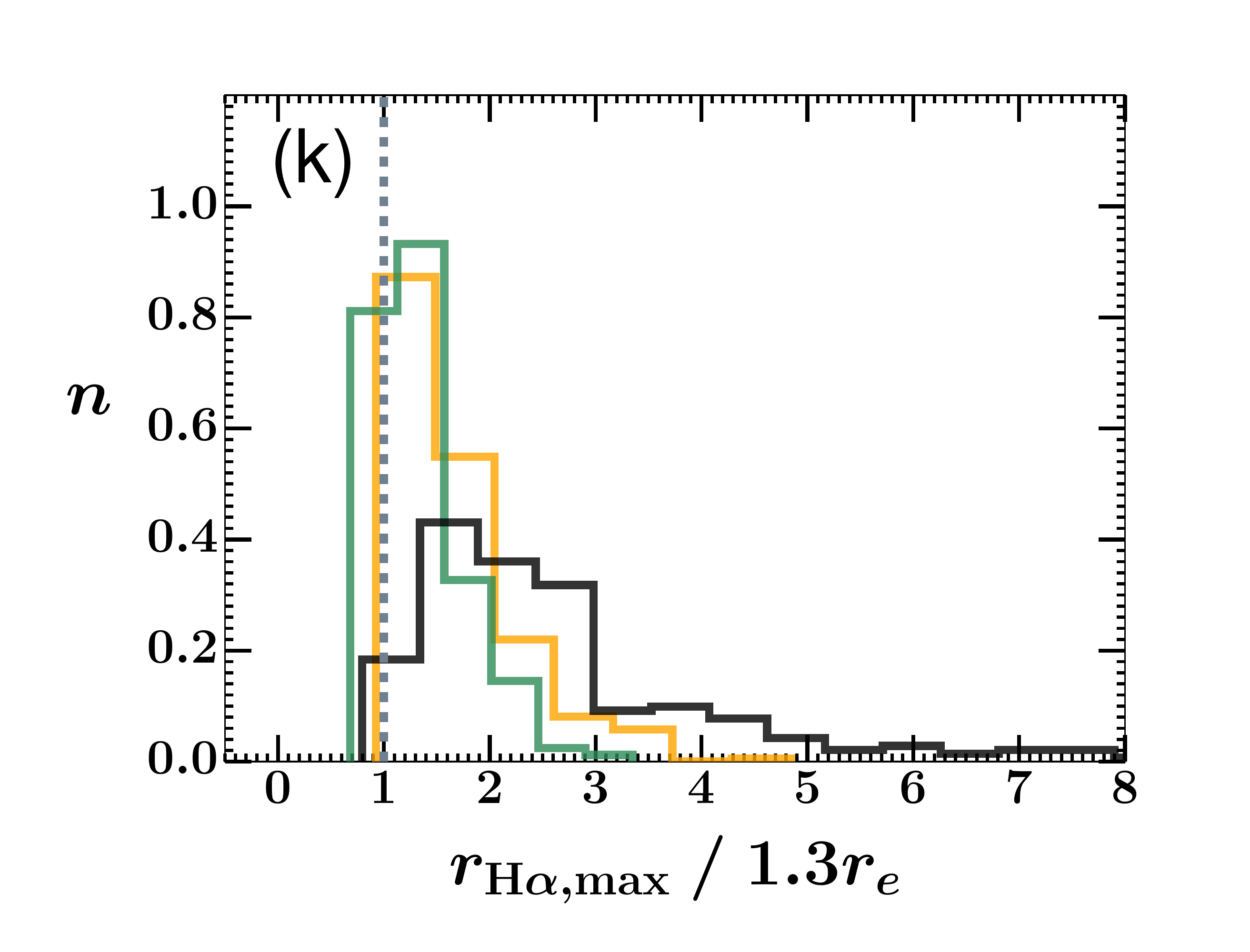}\includegraphics[width=0.24\textwidth,trim= 0 5 53 40,clip=True]{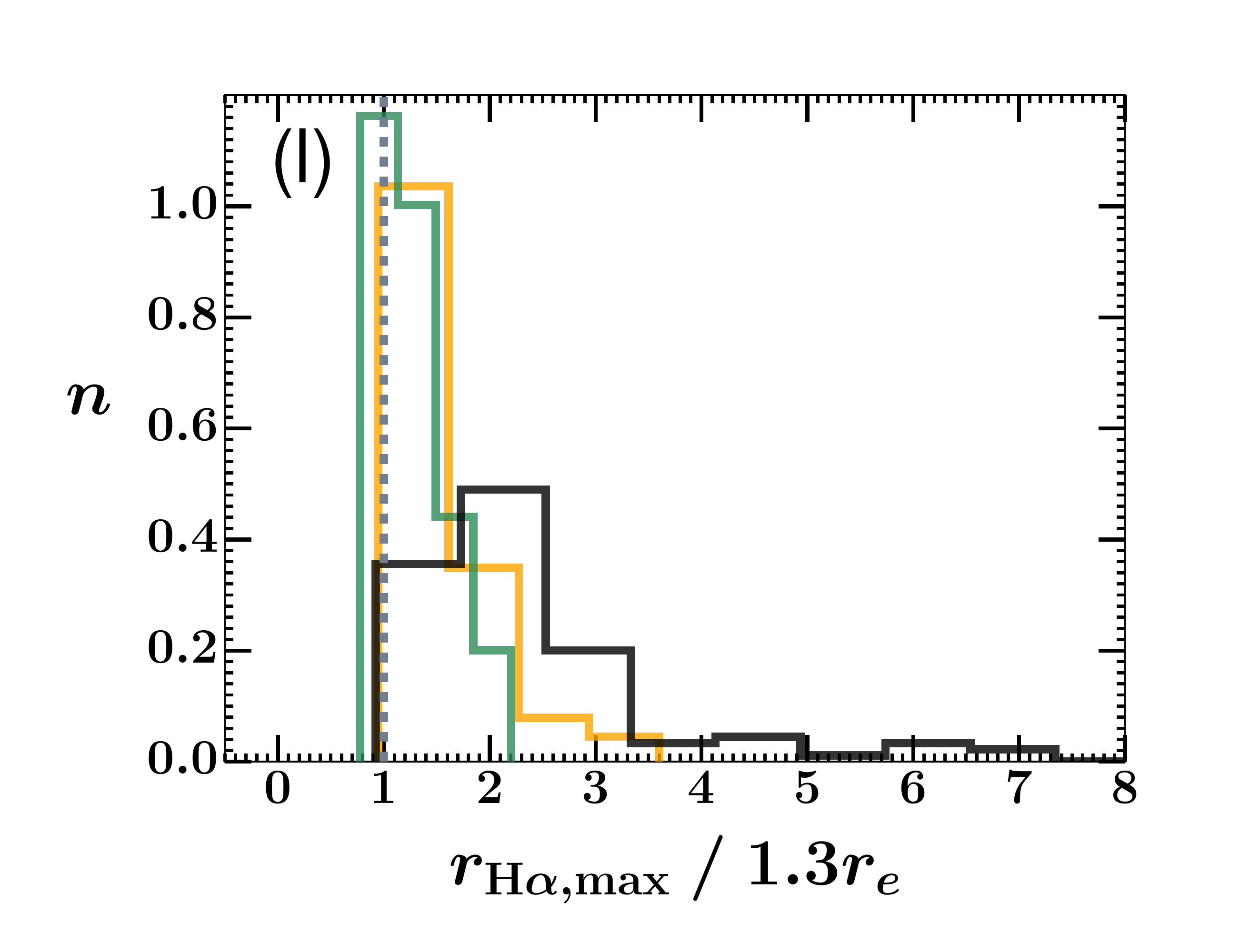}
\end{minipage}
\begin{minipage}[]{1.\textwidth}
\centering
\includegraphics[width=0.24\textwidth,trim= 0 5 53 40,clip=True]{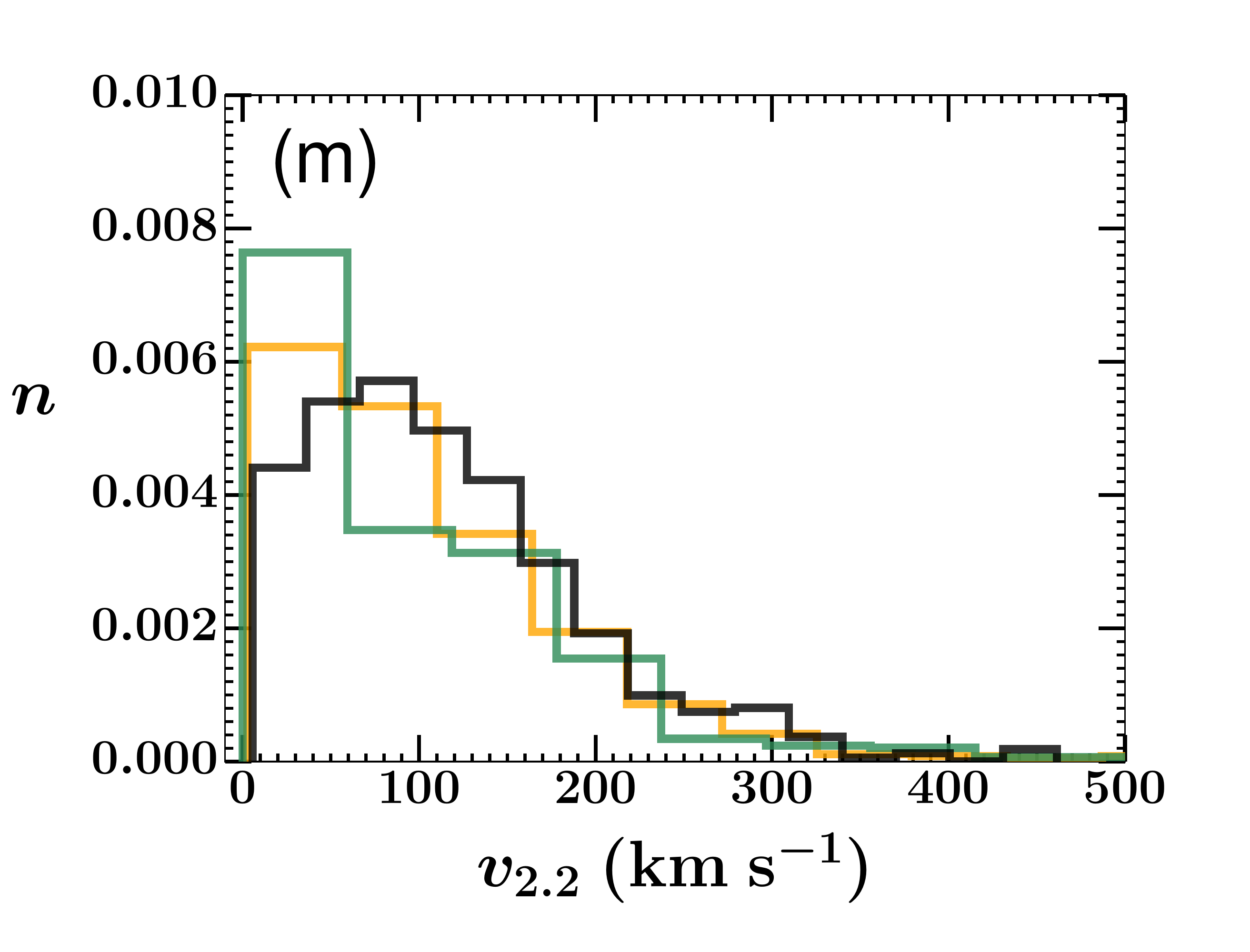}\includegraphics[width=0.24\textwidth,trim= 0 5 53 40,clip=True]{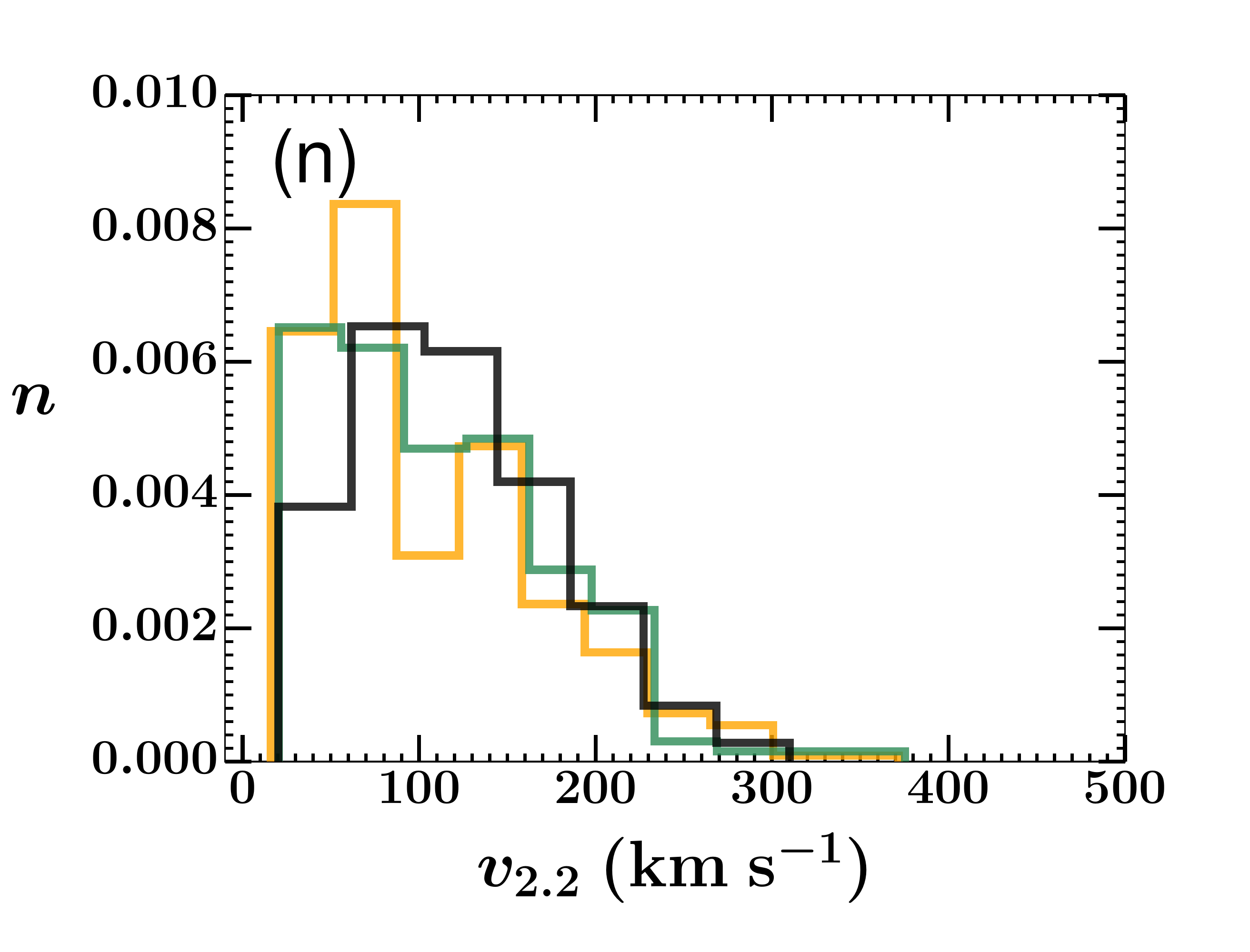}\includegraphics[width=0.24\textwidth,trim= 0 5 53 40,clip=True]{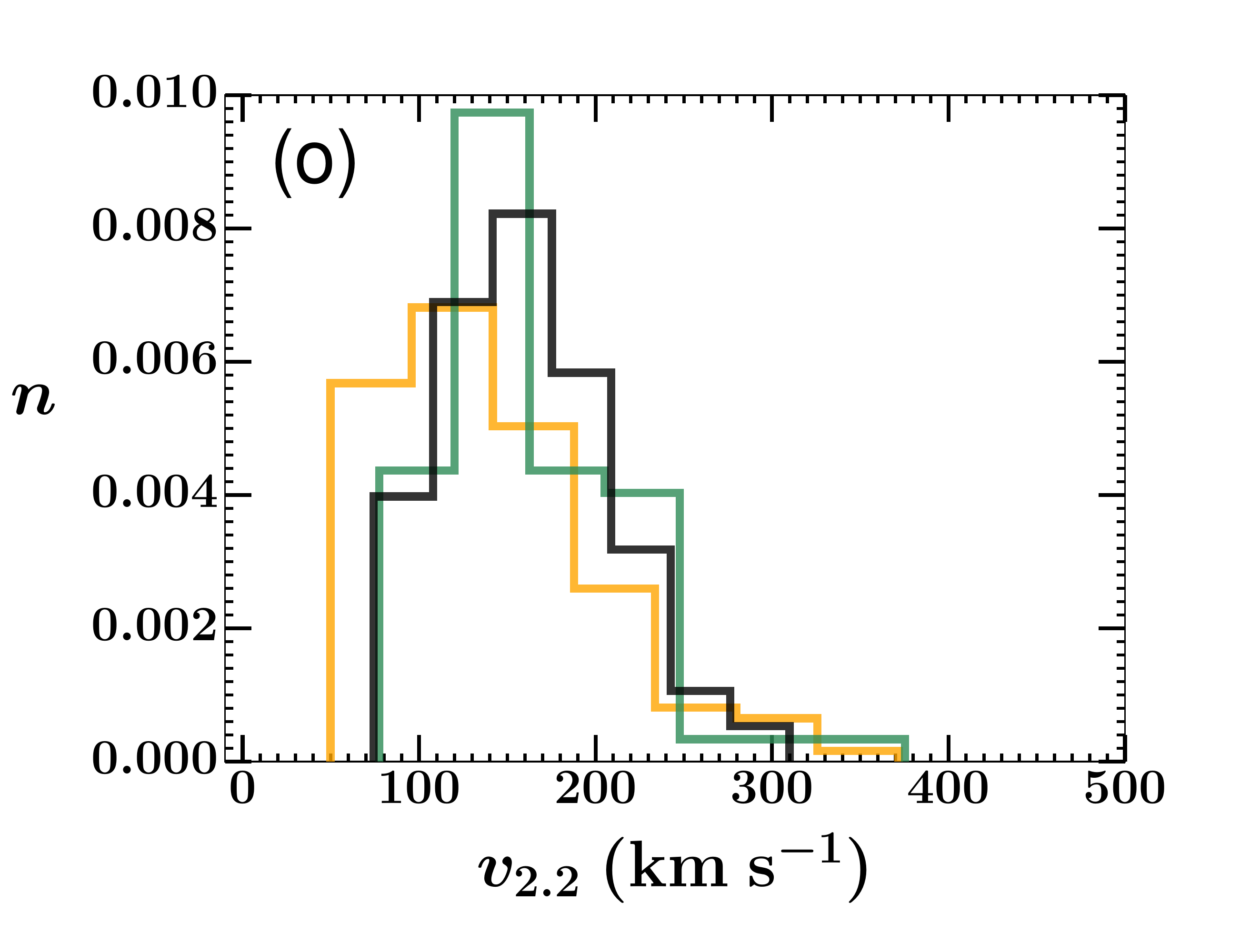}
\end{minipage}
\begin{minipage}[]{1.\textwidth}
\centering
\includegraphics[width=0.24\textwidth,trim= 0 5 53 40,clip=True]{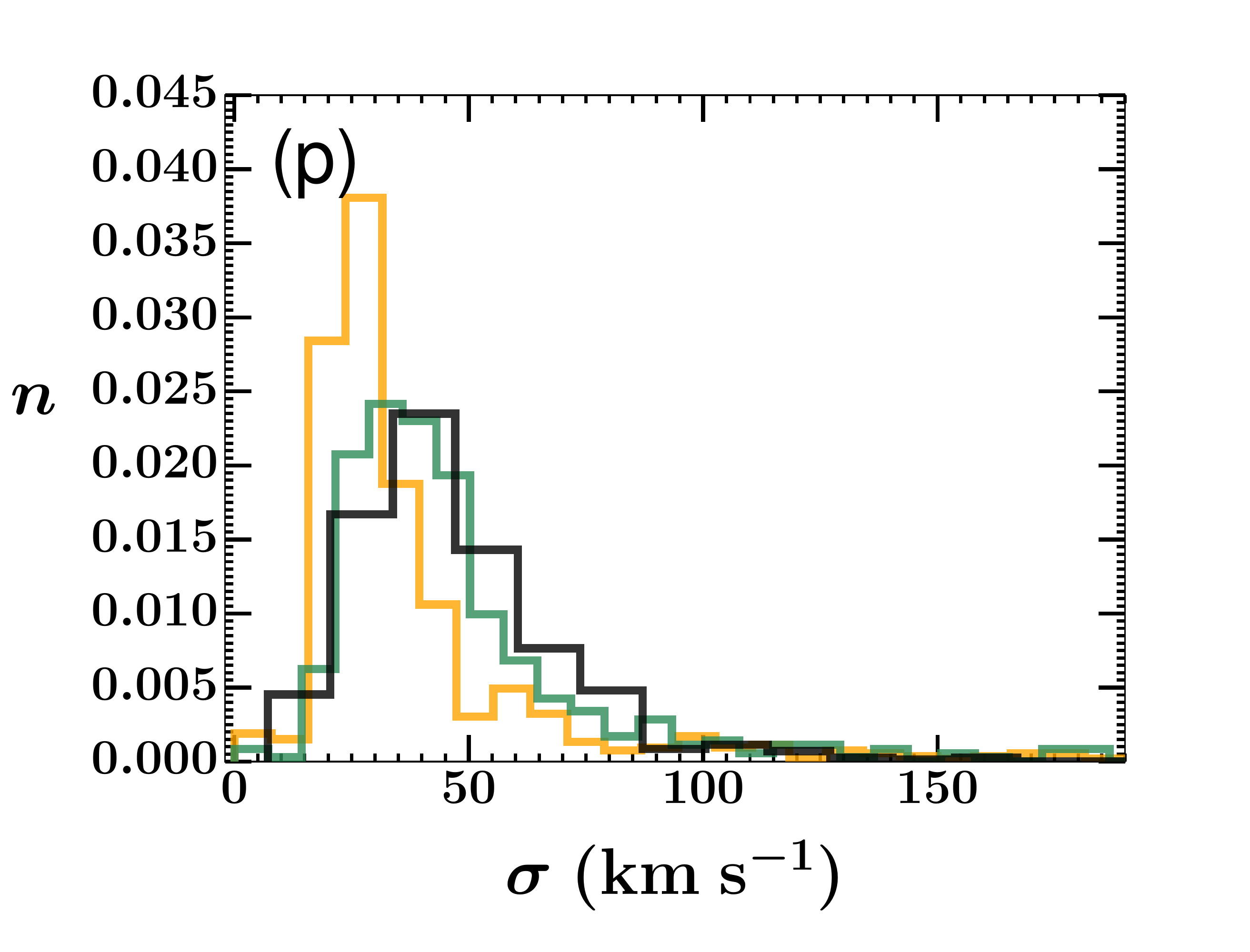}\includegraphics[width=0.24\textwidth,trim= 0 5 53 40,clip=True]{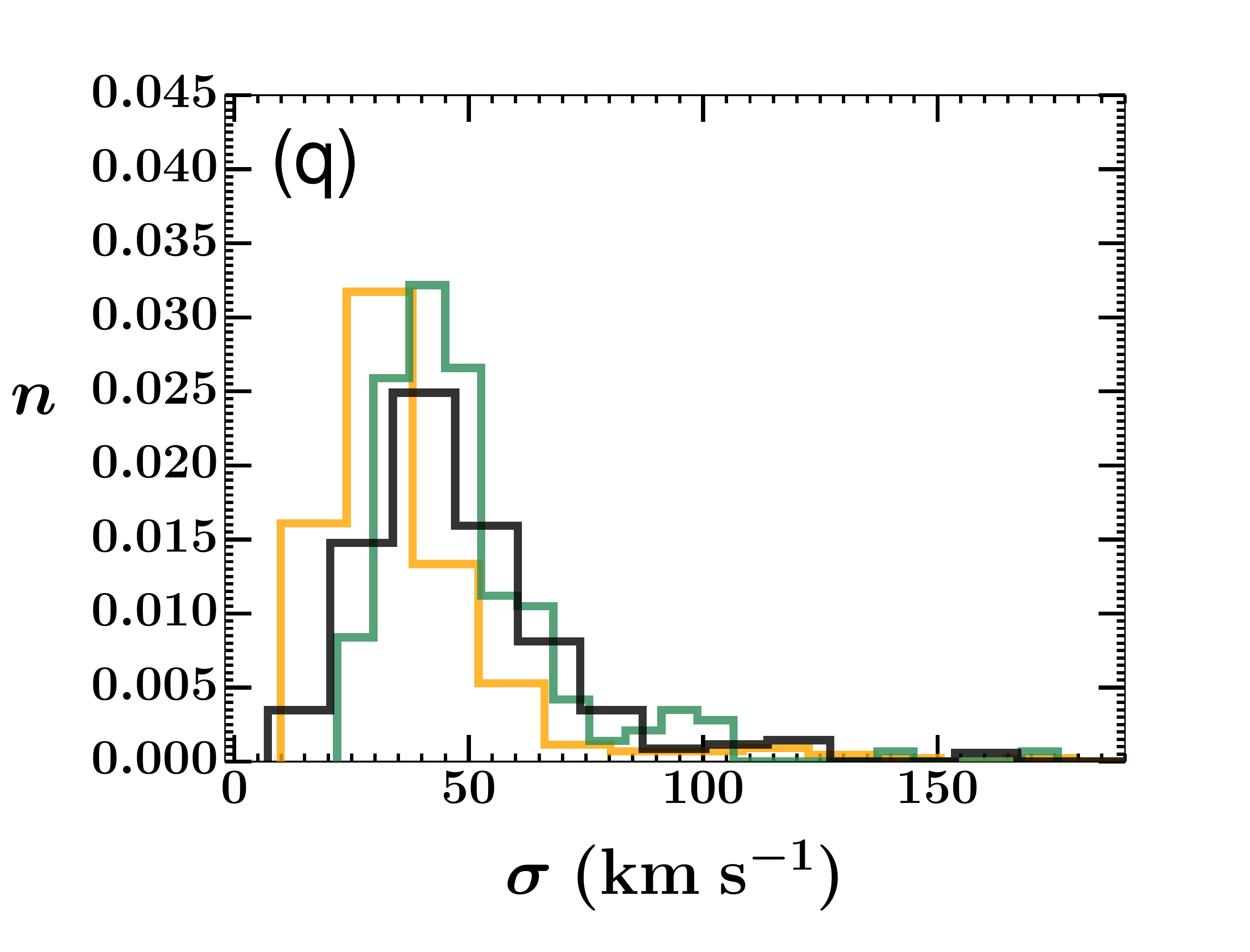}\includegraphics[width=0.24\textwidth,trim= 0 5 53 40,clip=True]{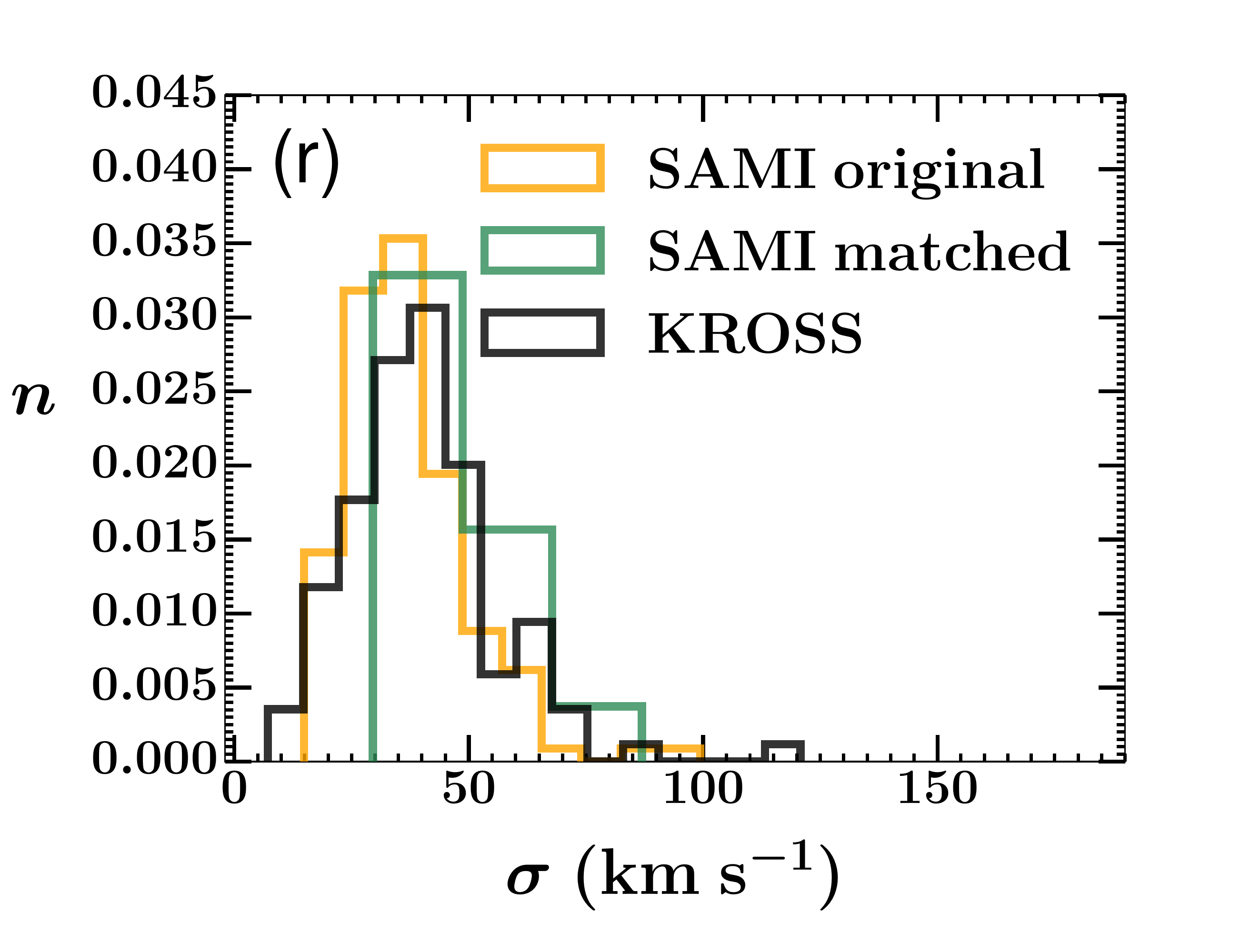}
\end{minipage}
\caption{%
Distributions of $r_{\text{e}}$, $M_{*}$, $M_{K}$, $r_{\text{H}\alpha,\text{max}}/1.3r_{\text{e}}$, $v_{2.2}$ and $\sigma$, as defined in the text, for respectively the original SAMI, matched SAMI and KROSS \textit{parent} (left column), \textit{rot-dom} (middle column) and \textit{disky} (right column) sub-samples (defined in \S~\ref{sec:SAMIvsKROSSsample}). The dashed line in panels (j--l) indicates the radius at which we measure the rotation velocity for the TFRs shown in \S~\ref{subsec:HQvsLQ} and \S~\ref{subsec:KROSSvsLQ}. By design, $r_{\text{H}\alpha,\text{max}} \gtrsim 1.3r_{\text{e}}$ for the \textit{rot-dom} and \textit{disky} sub-samples of all three data sets. The same sub-sample selection criteria, when successively applied to both the original SAMI and matched SAMI data, tend to select larger, more massive, and more rapidly rotating galaxies from the latter data set than from the former. On average, the majority of KROSS galaxies are more massive, more rapidly rotating, and have more spatially extended H$\alpha$ emission (relative to their size) than the majority of SAMI galaxies. %
     }%
\label{fig:SAMIvsKROSShists}
\end{figure*}

The key point from Figure~\ref{fig:SAMIvsKROSShists} is that the same sub-sample selection criteria, when successively applied to both the original SAMI and matched SAMI data, tend to select larger, more massive, and more rapidly rotating galaxies from the latter data set than from the former. This can be understood by considering how the matching process disproportionately affects those galaxies that are intrinsically compact. In these cases, decreasing the spatial resolution and sampling make it harder to measure a velocity gradient across the H$\alpha$ emission. In this respect it is not suprising that the primary effect of the matching process is to exclude those galaxies that are more compact. Since in general a galaxy's size, mass and rotation are coupled \citep[e.g.][]{Ferguson:1994,Shen:2003,Trujillo:2004,Bernardi:2011}, it also follows that those excluded galaxies will also tend to be more slowly rotating and less massive. 

This premise is further evidenced by the fact that the matched SAMI and KROSS distributions become increasingly well-matched in the majority of the key galaxy properties ($M_{*}$, $M_{K}$, $v_{2.2}$ and $\sigma$, as well as $r_{\rm{e}}$ to a lesser extent) as successively stricter selection criteria are applied. Indeed the distributions of these parameters are very similar for the matched SAMI and KROSS \textit{disky} sub-samples, again suggesting that the comparatively decreased data quality of these two data sets, in conspiracy with neccessarily strict selection criteria, results in the preferential exclusion of the smallest, least massive, and most slowly rotating galaxies from both samples. Of course, the KROSS galaxies themselves are also subject to an effective lower stellar mass limit as a result of the limiting magnitude for the survey (see \S~\ref{subsec:KROSS}). The differences between the matched SAMI and KROSS distributions in $r_{\rm{H}\alpha,\rm{max}}/1.3r_{\rm{e}}$ (panels (j--l)) can be explained by {\it intrinsic} differences between the properties of the high- and low-redshift samples; star-forming galaxies at $z\approx1$ tend to have more spatially extended star-forming (and thus H$\alpha$-emitting) regions \citep{Stott:2016}. 

We thus proceed to compare the original SAMI and matched SAMI TFRs of the \textit{rot-dom} and \textit{disky} sub-samples in the knowledge that differences between the relations may be introduced by either the data degradation process or the resulting sample selection (or potentially via a conspiracy between the two).

\subsubsection{TFRs}
\label{subsubsec:compTFRs}

\begin{figure*}
\centering
\begin{minipage}[]{1\textwidth}
\label{fig:SAMK}
\centering
\includegraphics[width=.9\textwidth,trim= 0 0 0 0,clip=True]{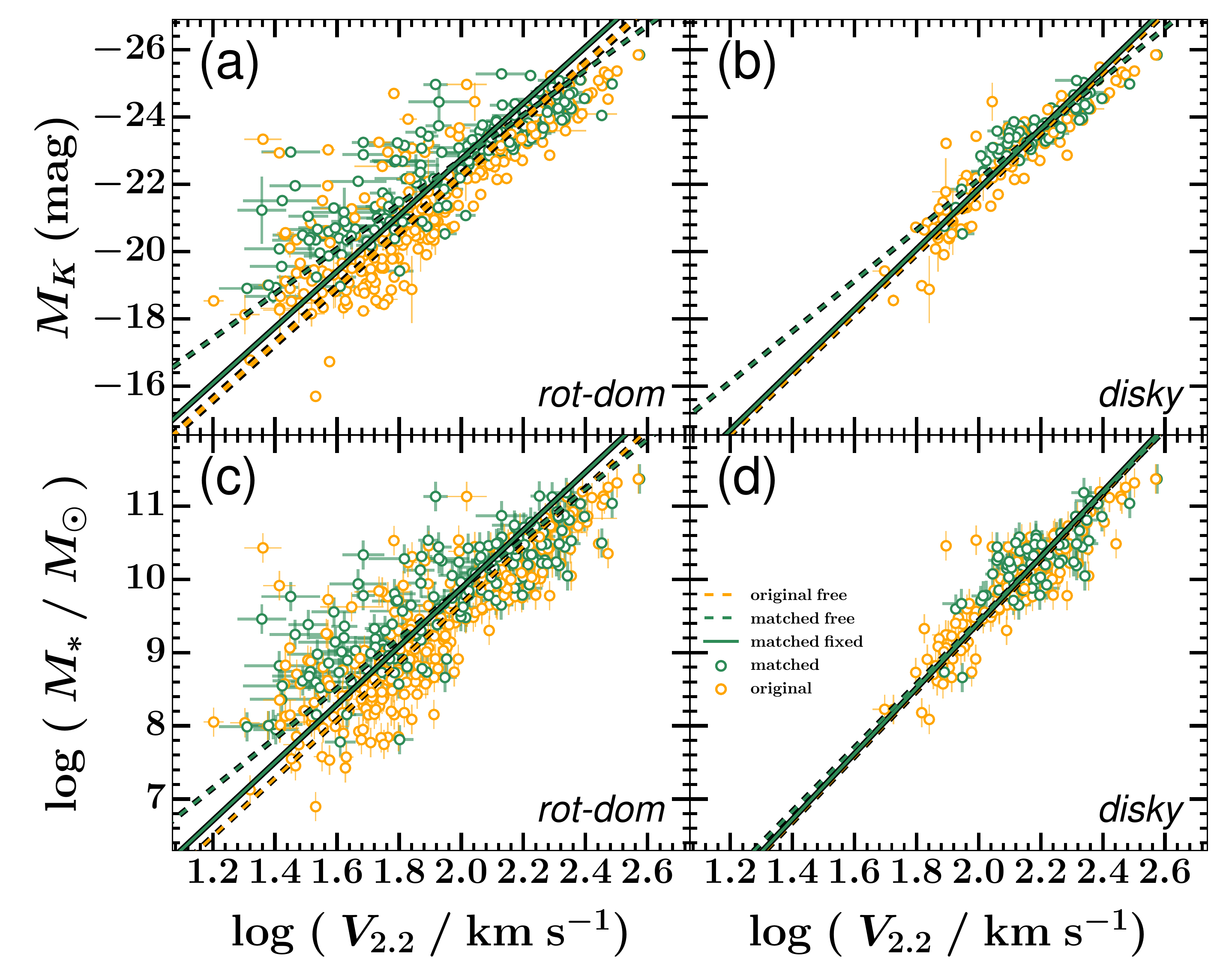}
\end{minipage}
\caption{%
The $M_{K}$ (top) and $M_{*}$ (bottom) TFRs for the original SAMI and matched SAMI \textit{rot-dom} (left) and \textit{disky} (right) sub-samples, as described in \S~\ref{sec:SAMIvsKROSSsample}. In each panel the dashed lines are the best fits with free slopes to each data set. The solid line is the best straight line fit to the matched SAMI data with the slope fixed to that of the best free fit slope to the corresponding original data. In all cases the best free fit slope for the matched SAMI TFR is shallower than the best fit slope for the corresponding original SAMI relation. For a fixed slope the zero-point of the TFR in each cases differs between the original SAMI and matched SAMI data, although this difference is not significant between the \textit{disky} sub-samples. %
     }%
\label{fig:TFR_comp_SAMI}
\end{figure*}

Here we present the matched SAMI and original SAMI TFRs and examine the differences between the two.

Figure~\ref{fig:TFR_comp_SAMI} shows the $M_{K}$ TFRs (upper panels) and $M_{*}$ TFRs (lower panels) of the original SAMI and matched SAMI \textit{rot-dom} (left panels) and \textit{disky} (right panels) sub-samples. To find the best fit straight line to each relation we employ the {\sc hyper-fit} package \citep[][written for the {\sc r} statistical language\footnote{\url{github.com/asgr/hyper.fit}}]{Robotham:2015} via the web-based interface\footnote{\url{http://hyperfit.icrar.org}}. The fitting routine relies on the basis that for a set of $N$-dimensional data with uncertainties that vary between data points (and are potentially covariant), provided the uncertainties are accurate, there exists a single, unique best-fit ($N-1$)-dimensional plane with intrinsic scatter that describes the data. This is contrary to the approach of many previous TFR studies \citep[including][]{Tiley:2016} that employ either one or the other, or some form of average, of two unique best fit straight lines to the TFR i.e.\ the now well-known forward and reverse best fits \citep[e.g.][]{Willick:1994aa} that effectively treat as the independent variable the galaxy rotation velocity or the absolute magnitude (or stellar mass), respectively. Such an approach tends to result in two best fits that differ significantly in slope and zero-point, and thus too the total and intrinsic scatter. \citet{Robotham:2015}, however, derive the general likelihood function to be maximised to recover the single best fit model, with single values of associated intrinsic and total scatter (that may be decomposed into components either orthogonal to the best fit or parallel to the ordinate).

\begin{table*}
\centering
\begin{tabular}{ llllllllll}
\hline
Data Set & Sample & Fit &Slope & $M_{K\ v_{2.2}=100}$ & $\sigma_{\rm{int,orth.}}$ & $\sigma_{\rm{tot,orth.}}$ & $\sigma_{\rm{int,vert.}}$ & $\sigma_{\rm{tot,vert.}}$ \\
		 & 		  &     & (mag dex$^{-1}$)     & 	 (mag)      & (dex mag$^{-1}$)               & (dex mag$^{-1}$)              & (mag)              & (mag)              \\
\hline
SAMI original  & \textit{rot-dom}   & free  & $-8.3$ $\pm$ 0.3 		   			 & $-$22.26 $\pm$ 0.07 & 0.134 $\pm$ 0.006  					   & 0.139 $\pm$ 0.007 & 1.13 $\pm$ 0.06					   & 1.17\phantom{0} $\pm$ 0.06 \\%
 		& \textit{disky} & free  & $-9.0$ $\pm$ 0.3 		   			 & $-$21.71 $\pm$ 0.05 & 0.063 $\pm$ 0.004  					   & 0.066 $\pm$ 0.005 & 0.57 $\pm$ 0.04 & 0.59\phantom{0} $\pm$ 0.04 \\%
SAMI matched  & \textit{rot-dom}   & free  & $-6.6$ $\pm$ 0.3 		   			 & $-$22.73 $\pm$ 0.06 & 0.112 $\pm$ 0.007  					   & 0.126 $\pm$ 0.007 & 0.75 $\pm$ 0.06 & 0.85\phantom{0} $\pm$ 0.05 \\%
	    & 		      & fixed & $-8.3$ \phantom{$\pm$} \phantom{0.2} 		     & $-$22.74 $\pm$ 0.08				        & 0.125 $\pm$ 0.008  					   & 0.136 $\pm$ 0.001 & 1.05 $\pm$ 0.07					   & 1.142 $\pm$ 0.004 \\%
	    & \textit{disky} & free  & $-7.5$ $\pm$ 0.5 		   			 & $-$22.14 $\pm$ 0.05 & 0.052 $\pm$ 0.006  					   & 0.059 $\pm$ 0.006 & 0.40 $\pm$ 0.05 & 0.44\phantom{0} $\pm$ 0.04 \\%
	    & 			  & fixed & $-9.0$ \phantom{$\pm$} \phantom{0.2} 		   	 & $-$21.86 $\pm$ 0.07  & 0.055 $\pm$ 0.006  					   & 0.061 $\pm$ 0.001 & 0.50 $\pm$ 0.05				       & 0.554 $\pm$ 0.006 \\%
KROSS   & \textit{rot-dom}   & free  & $-8.3$ $\pm$ 0.9 		   			 & $-$23.1\phantom{0} $\pm$ 0.1\phantom{0} 					   & 0.188 $\pm$ 0.009  					   & 0.20\phantom{0} $\pm$ 0.02\phantom{0} & 1.6\phantom{0} $\pm$ 0.2 					   & 1.6\phantom{00} $\pm$ 0.2 \\%
	    & 		      & fixed & $-6.6$ \phantom{$\pm$} \phantom{0.2}			 & $-$23.16			  $\pm$ 0.08 		   & 0.194 $\pm$ 0.009  					   & 0.198 $\pm$ 0.001 & 1.30 $\pm$ 0.06 					   & 1.334 $\pm$ 0.004 \\%
	    & \textit{disky} & free  & $-11$\phantom{.} $\pm$ 1\phantom{.2} & $-$21.6\phantom{0} $\pm$ 0.1\phantom{0} & 0.089 $\pm$ 0.008  & 0.11\phantom{0} $\pm$ 0.01\phantom{0} & 1.0\phantom{0} $\pm$ 0.2\phantom{0} & 1.1\phantom{00} $\pm$ 0.1\phantom{0} \\%
	    & 			  & fixed & $-7.5$ \phantom{$\pm$} \phantom{0.2} 			 & $-$22.21 $\pm$ 0.08  & 0.098 $\pm$ 0.009  & 0.111 $\pm$ 0.001 & 0.74 $\pm$ 0.07\phantom{0} & 0.844 $\pm$ 0.005 \\%
\hline
\end{tabular}
\caption{Parameters of the best-fit $M_{K}$ TFRs for the \textit{rot-dom} and \textit{disky} sub-samples. Uncertainties are quoted at the 1$\sigma$ level.}
\label{tab:SAMIvsKROSSMKTFR}
\end{table*}

\begin{table*}
\centering
\begin{tabular}{ llllllllll}
\hline
Data Set & Sample & Fit &Slope & $\log (\frac{M_{*}}{M_{\odot}})_{v_{2.2}=100}$& $\sigma_{\rm{int, orth.}}$ & $\sigma_{\rm{tot, orth.}}$ & $\sigma_{\rm{int, vert.}}$ & $\sigma_{\rm{tot, vert.}}$ \\
		 & 		  &     &      & 	(dex)       &                &                & (dex)              & (dex)              \\
\hline
SAMI original  & \textit{rot-dom}     & free  & $4.0$\phantom{0} $\pm$ 0.1\phantom{0} 			 & \phantom{1}9.66 $\pm$ 0.03 & 0.129 $\pm$ 0.006  					   & 0.141 $\pm$ 0.007 					   & 0.53 $\pm$ 0.03 					   & 0.58\phantom{0} $\pm$ 0.03 \\%
 			   & \textit{disky} & free  & $4.5$\phantom{0} $\pm$ 0.2\phantom{0}  		 & \phantom{1}9.37 $\pm$ 0.03 & 0.065 $\pm$ 0.006  					   & 0.079 $\pm$ 0.005 					   & 0.30 $\pm$ 0.03 					   & 0.36\phantom{0} $\pm$ 0.02 \\%
SAMI matched   & \textit{rot-dom}     & free  & $3.4$\phantom{0} $\pm$ 0.2\phantom{0} 			 & \phantom{1}9.87 $\pm$ 0.04 & 0.117 $\pm$ 0.009    					   & 0.141 $\pm$ 0.009			           & 0.42 $\pm$ 0.04 					   & 0.50\phantom{0} $\pm$ 0.03 \\%
	    	   & 	           & fixed & $4.0$ \phantom{$\pm$} \phantom{0.09} 			 & \phantom{1}9.88 $\pm$ 0.04	& 0.127 $\pm$ 0.009					       & 0.145 $\pm$ 0.001 					   & 0.52 $\pm$ 0.04					   & 0.591			 $\pm$ 0.002 \\%
	           & \textit{disky} & free  & $4.3$\phantom{0} $\pm$ 0.4\phantom{0}  		 & \phantom{1}9.44 $\pm$ 0.04 & 0.059 $\pm$ 0.009    & 0.078 $\pm$ 0.009 				       & 0.26 $\pm$ 0.05 					   & 0.34\phantom{0} $\pm$ 0.04 \\%
	    	   & 			   & fixed & $4.5$\phantom{0} \phantom{$\pm$} \phantom{0.10} & \phantom{1}9.41 $\pm$ 0.04	& 0.061 $\pm$ 0.008					       & 0.078 $\pm$ 0.001 					   & 0.28 $\pm$ 0.04 				       & 0.360 $\pm$ 0.004 \\%
KROSS  	 	   & \textit{rot-dom}     & free  & $3.7$\phantom{0} $\pm$ 0.3\phantom{0}			 & \phantom{1}9.88 $\pm$ 0.04 & 0.172 $\pm$ 0.009  					   & 0.19\phantom{0} $\pm$ 0.02\phantom{0} & 0.66 $\pm$ 0.07 					   & 0.72\phantom{0} $\pm$ 0.06 \\%
	    	   & 		       & fixed & $3.4$ \phantom{$\pm$} \phantom{0.09} 	     	 & \phantom{1}9.89 $\pm$ 0.04 & 0.172 $\pm$ 0.009  					   & 0.189 $\pm$ 0.001 					   & 0.61 $\pm$ 0.03 					   & 0.670 $\pm$ 0.002 \\%
	    	   & \textit{disky} & free  & $5.2$\phantom{0} $\pm$ 0.6\phantom{0}			 & \phantom{1}9.19 $\pm$ 0.05           & 0.075 $\pm$ 0.009  					   & 0.10\phantom{0} $\pm$ 0.01\phantom{0} & 0.39 $\pm$ 0.07 					   & 0.52\phantom{0} $\pm$ 0.05\phantom{0} \\%
	    	   & 			   & fixed & $4.3$\phantom{0} \phantom{$\pm$} \phantom{0.10} & \phantom{1}9.35 $\pm$ 0.04 & 0.074 $\pm$ 0.009  					   & 0.100 $\pm$ 0.001 					   & 0.33 $\pm$ 0.04 					   & 0.447 $\pm$ 0.003 \\%
\hline
\end{tabular}
\caption{Parameters of the best-fit $M_{*}$ TFR for the \textit{rot-dom} and \textit{disky} sub-samples. Uncertainties are quoted at the 1$\sigma$ level.}
\label{tab:SAMIvsKROSSmassTFR}
\end{table*}

\begin{table}
\centering
\begin{tabular}{ llll}
\hline
TFR & Sample & matched-original  & KROSS-matched  \\
\hline
$M_{K}$ 	 & \textit{rot-dom}    & $-$0.5\phantom{0} $\pm$ 0.1\phantom{0}  mag & $-$0.4\phantom{0} $\pm$ 0.1\phantom{0} mag\\%
 			 & \textit{disky}  & $-$0.15 $\pm$ 0.08 mag & $-$0.08 $\pm$ 0.09 mag\\%
$M_{*}$ 	 & \textit{rot-dom}    & \phantom{$-$}0.21 $\pm$ 0.06 dex			& \phantom{$-$}0.02 $\pm$ 0.06 dex\\%
	    	 & \textit{disky}  & \phantom{$-$}0.04 $\pm$ 0.05 dex	& $-$0.09 $\pm$ 0.06 dex\\%
\hline
\end{tabular}
\caption{Zero-point offsets between respectively the original SAMI and matched SAMI TFRs and the KROSS and matched SAMI TFRs, measured with a fixed slope at a given rotation velocity. Uncertainties are quoted at the 1$\sigma$ level.}
\label{tab:SAMIvsKROSSoffsets}
\end{table}

The parameters of the best straight line fits (with free slopes) with {\sc hyper-fit} to each relation are listed in Tables~\ref{tab:SAMIvsKROSSMKTFR} and \ref{tab:SAMIvsKROSSmassTFR}, along with measures of the total and intrinsic scatters both orthogonal to the best fit and along the ordinate. The slopes of the TFRs from the matched SAMI data are, in every case, shallower than the corresponding relations from the original SAMI data. 

To measure the offset between the zero-points of the original SAMI and matched SAMI TFRs, we fix the slopes of the matched SAMI relations to those of the corresponding original relations. The resulting best-fit TFR parameters are listed in Tables~\ref{tab:SAMIvsKROSSMKTFR} and \ref{tab:SAMIvsKROSSmassTFR} and the matched SAMI-original SAMI zero-point offsets are listed in Table~\ref{tab:SAMIvsKROSSoffsets}. There is a significant offset (greater than three times its standard error) between the zero-point of the matched SAMI TFR and the corresponding original SAMI TFR when considering either the $M_{K}$ or $M_{*}$ relation for the \textit{rot-dom} sub-samples, in the sense that the matched SAMI galaxies are on average respectively brighter, or more massive at fixed rotation velocity. However, there is no significant offset between either TFR for the \textit{disky} sub-samples for the same data sets.

Considering the scatter of the relations, within each data set (original and matched) and for both the $M_{K}$ and $M_{*}$ TFRs the total and intrinsic scatters in both the orthogonal and vertical directions are reduced in the \textit{disky} sub-sample relations compared to the \textit{rot-dom} sub-sample relations, as a result of the tight selection for rotation-dominated systems. Whilst these scatters may not be driven purely by the inclusion (or exclusion) of systems with small $v_{2.2}/\sigma$, it is apparent that given the typically low velocity dispersions of the SAMI galaxies and indeed local late-type galaxies, a $v_{2.2}/\sigma + \Delta v_{2.2}/\sigma >1$ cut is not stringent enough to select for galaxies that obey well the assumption of circular motions implicit in the TFR. One should thus take caution in considering the rotation velocities of galaxies in a TFR sample as a tracer of their total (dynamical) mass unless one is certain that the sample is one of strictly rotationally-dominated systems. This is particularly relevant when comparing the TFRs of galaxies at increasing redshift to those of galaxies in the local Universe; \citet{Turner:2017b} show that with increasing lookback time, a galaxy's velocity is decreasingly representative of its dynamical mass. 

Comparing the TFR scatters between the original and matched data sets, for both the \textit{rot-dom} and \textit{disky} sub-samples the TFRs of both data sets in most cases exhibit intrinsic scatters that are consistent within uncertainties. However in one case (for the \textit{rot-dom} $M_{K}$ relations), somewhat suprisingly, the TFR from the matched SAMI data actually exhibits significantly {\it lower} intrinsic scatter than the corresponding TFR from the original SAMI data. 

\subsubsection{The Effect of Data Quality on the TFR}
\label{subsubsec:dataqualTFRs}

\begin{figure*}
\centering
\begin{minipage}[]{1\textwidth}
\label{fig:SAMK}
\centering
\includegraphics[width=.9\textwidth,trim= 0 0 0 0,clip=True]{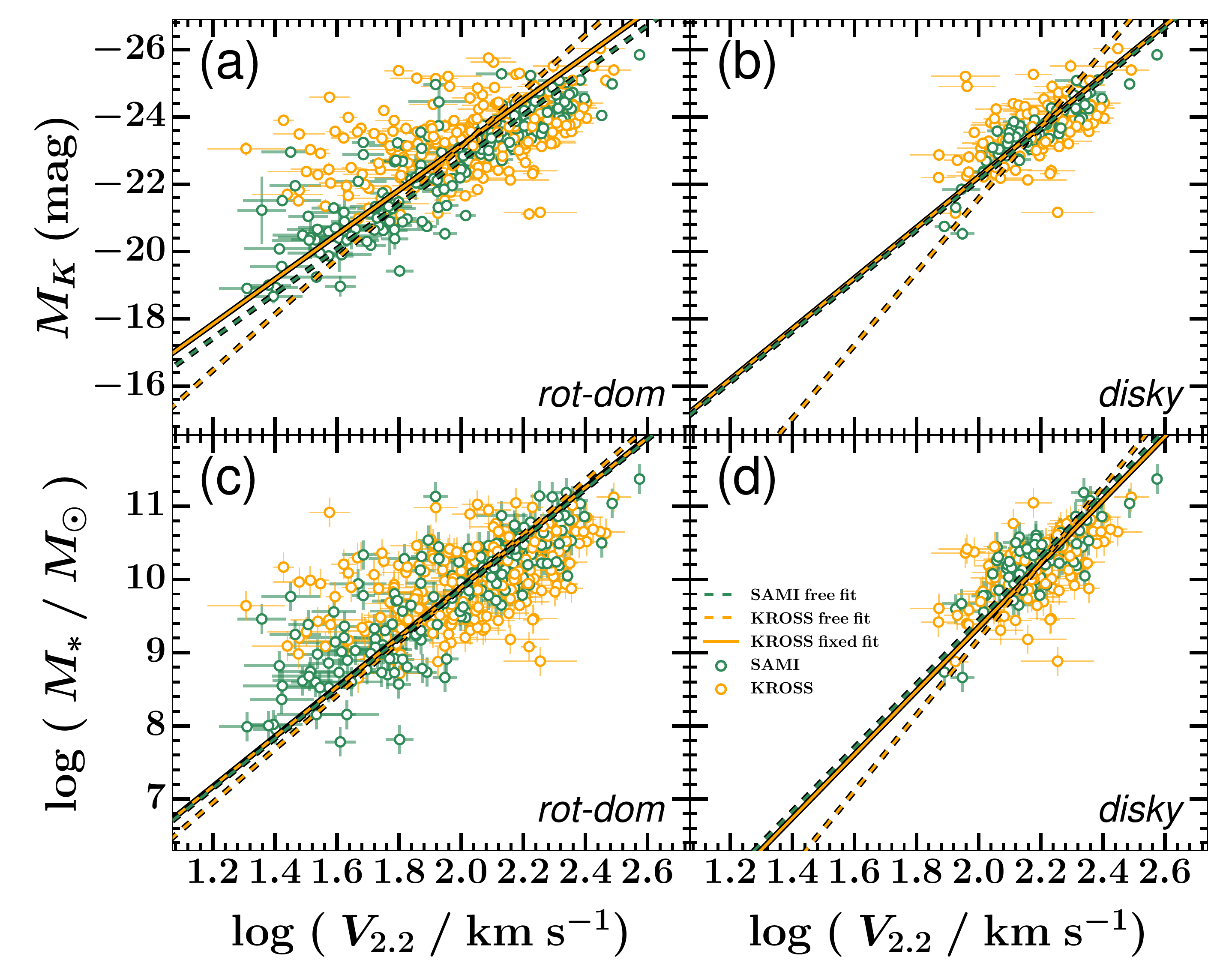}
\end{minipage}
\caption{%
The $M_{K}$ (top) and $M_{*}$ (bottom) TFRs of the matched SAMI and KROSS \textit{rot-dom} (left) and \textit{disky} (right) sub-samples, as described in \S~\ref{sec:SAMIvsKROSSsample}. The dashed lines are the best fits with free slopes to each data set. The solid line in each panel is the best fit the KROSS TFR with the slope fixed to that of the best (free) fit to the corresponding matched SAMI TFR. There is no significant offset between the matched SAMI and KROSS $M_{K}$ or $M_{*}$ TFRs zero-points for the \textit{disky} sub-samples that comprise disk-like galaxies that obey well the assumption of circular motion inherent in the TFR.%
     }%
\label{fig:TFR_MK_KROSSLQ}
\end{figure*}

Here we examine the dominant factor(s) causing the slopes, zero-points, and scatters of the TFRs for the \textit{rot-dom} and \textit{disky} sub-samples to differ between the original SAMI and matched SAMI data sets. 

We have so far shown that the matched SAMI relations have shallower slopes than the corresponding original SAMI relations and zero-points (for fixed slope fits) that are offset towards brighter magnitude or larger stellar mass in the ordinate (equivalent to lower velocities in the abscissa). We also found that the scatter in the matched SAMI TFRs is equivalent to, or sometimes even less than, the corresponding original SAMI relations. 

To explain these differences we recall our findings from \S~\ref{subsubsec:measurebias} that we are able to recover key galaxy parameters from the matched SAMI data with considerable accuracy but that, nonetheless, these matched SAMI measurements still suffered from small systematic biases. We also showed that the sub-sample selection criteria described in \S~\ref{sec:SAMIvsKROSSsample} tended to select larger, more massive, and more rapidly rotating galaxies from the matched SAMI data than the original SAMI data. However, it is not immediately clear which of these effects is most important in explaining the differences between the matched SAMI and original SAMI TFRs. 

To determine whether the measurements bias or the sample bias is the dominant factor we take the stellar mass TFRs for the \textit{rot-dom} and \textit{disky} sub-samples of the matched SAMI data. For each galaxy in each relation we then swap the matched SAMI values in the ordinate and abscissa for the corresponding measurements from the original SAMI data for the same galaxy. For clarity we refer to these relations as the ``swapped'' TFRs. We find significant differences between the slopes, zero-points and scatters of the swapped and matched SAMI TFRs that are in each case equal in size, within uncertainties, and opposite in effect to the differences we measured previously between the matched SAMI and original SAMI relations. We can therefore state with certainty that the matched TFRs (for both the \textit{rot-dom} and \textit{disky} sub-sample) do not differ from the corresponding original SAMI TFRs as a result of differing selection biases between the two. Instead they differ purely as a result of biases in the matched SAMI measurements.

It is intuitively obvious how the bias in the matched SAMI measurements can affect the changes between the slopes and zero-points of the matched SAMI TFRs and the original SAMI TFRs. It is clear that the systematic offset between the matched SAMI and original SAMI $v_{2.2}$ measurements, although small (only $18 \pm 2$ km s$^{-1}$), is amplified in the log-space of the TFR plane leading to a significant flattening of the resultant TFR slope as the fractional difference in $v_{2.2}$ becomes large for galaxies with low rotation velocities. As discussed in \S~\ref{subsubsec:measurebias}, this systematic offset is driven mainly by those galaxies with original SAMI measurement of $v_{2.2} \lesssim 100$ km s$^{-1}$, meaning galaxies with lower velocities are worse affected again,  exacerbating the effect. This change in slope also drives a corresponding offset in the TFR zero-point, ``dragging'' the TFR towards lower velocities, corresponding to a shift in zero-point toward higher stellar mass, or brighter magnitudes.

Reassuringly, the sub-sample selection criteria can help alleviate this effect. For example, there is no significant zero-point offset between the matched SAMI and original SAMI TFRs (either $M_{*}$ or $M_{K}$) for the \textit{disky} sub-samples. However, as is shown in Figure~\ref{fig:SAMIvsKROSShists} and discussed in \S~\ref{subsubsec:measurebias}, the corresponding cost is an increasingly biased sample that is correspondingly reduced in size. 

Finally, the matched SAMI measurements lead to TFRs with intrinsic scatters that are in most cases consistent within uncertainties with those of the corresponding original SAMI relation. However, we note that the matched SAMI intrinsic scatters are in every case {\it formally} lower (and in one case significantly lower) than those of the original SAMI TFRs. To understand this we consider a combination of two factors. First, and most importantly, the fractional uncertainty in the matched SAMI $v_{2.2}$ measurements is, on (median) average, inflated by a factor of $3 \pm 2$ compared to the original SAMI measurements. This alone, in the majority of cases, accounts for the difference in the intrinsic scatter between the matched SAMI and original SAMI TFRs. However, for the case in which the intrinsic scatter in the matched SAMI and original SAMI TFRs significantly differs, it only accounts for a maximum of 40 percent of the difference. It is therefore possible that here we are also witnessing a more subtle impact of the degrading process used to produce the matched SAMI sample. The matching process preferentially selects larger, more massive, and faster-rotating galaxies. These also tend to be the systems that most strictly obey the assumptions of the TFR (see \S~\ref{sec:SAMImotivation}), resulting in a matched SAMI TFR with a \textit{real} reduction in its intrinsic scatter in comparison to the corresponding original SAMI relation.

To make a fair comparison between the KROSS $z\approx1$ TFRs and the SAMI $z\approx0$ TFRs, we now proceed to compare only the matched SAMI relations to the KROSS relations, that should be equally biased in their galaxy measurements and sample selection, and do not discuss further the original SAMI relations.   

\subsection{KROSS vs. Matched SAMI TFRs}
\label{subsec:KROSSvsLQ}

Figure~\ref{fig:TFR_MK_KROSSLQ} shows the $M_{K}$ and $M_{*}$ TFRs of the matched SAMI and KROSS \textit{rot-dom} and \textit{disky} sub-samples. The corresponding best-fit parameters (with free slopes) are listed in Tables~\ref{tab:SAMIvsKROSSMKTFR} and \ref{tab:SAMIvsKROSSmassTFR}, along with the parameters of the best-fits to the KROSS TFRs when the slope is fixed to that of the corresponding matched SAMI relation.

We note here that whether we consider either the $M_{*}$ or $M_{K}$ TFRs, we observe similar trends in the slope and scatter between the \textit{rot-dom} and \textit{disky} relations for either the matched SAMI or KROSS data sets. As for the SAMI TFRs, we find the KROSS TFRs to exhibit steeper slopes and reduced scatter for the \textit{disky} sub-sample in comparison to the corresponding relations for the \textit{rot-dom} sub-samples. We now discuss in further detail the differences in the TFRs between the KROSS and matched SAMI data sets.    

Firstly, considering the scatters of the matched SAMI and KROSS TFRs, the KROSS relations exhibit larger total and intrinsic scatters (in the ordinate and orthogonal to the best fit line) in comparison to the corresponding matched SAMI relations in every case. Since every effort has been taken to match the data quality, analysis methods, and sample selection, this suggests that the KROSS TFRs exhibit {\it intrinsically} larger scatter than the corresponding matched SAMI relations.   

Considering the slopes of the matched SAMI and KROSS TFRs, the KROSS relations for both the \textit{rot-dom} and the \textit{disky} sub-samples are, in each case, steeper than the corresponding matched SAMI relations. The steeper slopes of the KROSS TFRs, which are not well constrained at the lower mass, lower-velocity end of the relations, are most likely strongly affected by incompleteness in stellar mass and velocity as a result of the initial KROSS survey limiting magnitude and the combination of the reduced data quality with the strict \textit{disky} selection criteria. Evidence of the latter effect can be seen strongly in the matched SAMI data; application of the \textit{disky} selection criteria effectively result in a lower mass (or magnitude) cut with respect to the \textit{rot-dom} sub-sample, with a corresponding cut in velocity. At the same time, the best fit free slope for the \textit{disky} matched SAMI TFR is, in each case, steeper than the slope for the corresponding \textit{rot-dom} relation. 

Due to the potential for systematic effects in the measured TFR slopes, to measure the offset between the zero-points of the matched SAMI and KROSS TFRs we fix the slopes of the KROSS relations to those of the corresponding matched SAMI relations. The resulting KROSS-matched SAMI zero-point offsets are listed in Table~\ref{tab:SAMIvsKROSSoffsets}.  The $M_{K}$ TFR zero-point for the KROSS \textit{rot-dom} galaxies is $0.4\pm0.1$ mag brighter than the zero-point for the corresponding matched SAMI relation. However, there is no significant zero-point offset ($-0.08 \pm 0.09$ mag) when considering the \textit{disky} sub-samples instead. The KROSS and matched SAMI $M_{*}$ TFRs zero-points are consistent within uncertainties when comparing either the \textit{rot-dom} or the \textit{disky} sub-samples.

\section{Discussion}
\label{sec:SAMIvsKROSSdiscussion}

In this section we summarise and discuss our results in the context of previous studies in the literature and theoretical expectations for the redshift evolution of the TFR. We also discuss their importance in the physical context of galaxy evolution over the last $\approx$8 Gyr.

\subsection{Data quality, sample selection, and measurement effects}

The comparisons of the original SAMI and matched SAMI data sets presented in \S~\ref{subsec:HQvsLQ} constitute a direct measure of the observational biases that IFS studies of $z\approx1$ galaxies must account for. We find the matched SAMI sub-samples to be biased against those systems that are more compact, less massive, and that more slowly rotate compared to the original SAMI sub-samples constructed using the same selection criteria. Further, we find that, after the application of a correction factor to account for the increased beam smearing in the matched SAMI cubes, we are able to recover measurements of the intrinsic rotation velocity and velocity dispersion with considerable accuracy. However, on average, the matched SAMI measurements remain slightly biased respectively high and low in comparison to measurements of the same galaxies from the corresponding original SAMI cubes. 

Importantly, we find that a small systematic difference between the matched SAMI and original SAMI measurements of $v_{2.2}$ is amplified in the log-space of the TFR plane, leading to significant differences between the slopes and zero-points of the matched SAMI TFRs and those of the corresponding  original SAMI relations. Most of the difference between the intrinsic scatters of the matched SAMI and original SAMI TFRs can be attributed to an inflation of the fractional uncertainty in measurements of $v_{2.2}$ from the former as compared to the latter. We found that the application of the \textit{disky} selection criteria helped to reduce differences between the TFRs for the two data sets, but at the expense of smaller and more biased galaxy sub-samples.

That TFRs constructed with identical selection criteria differ as the result of data quality alone is important. Accurately determining the intrinsic slope, zero-point and scatter in the TFRs at $z\approx1$ and $z\approx0$ is a vital step in understanding any difference between the relations at the two epochs (and thus the physical processes driving galaxy evolution over this period). In this work, even if we cannot absolutely remove all potential biases from our measures of the relations at each redshift, we can at least be sure that our careful matching of data quality, analysis methods, and sample selection allows an accurate measure of the {\em relative} differences between the two. Worryingly, we find that the differences between the original SAMI and matched SAMI TFRs are, in all cases, as large (within uncertainties) or larger than the differences between the KROSS and matched SAMI TFRs. This has potentially wide ranging implications for our interpretation of previous reports of evolution in the TFR as a function of redshift \citep[including those of][]{Tiley:2016}. Future IFS studies that wish to {\it directly} compare the TFRs of galaxies at different epochs therefore must nullify the potential difference in biases between observations at each redshift, either by similarly matching the data quality, analysis methods, and sample selection as we have done in this work or otherwise.

Whilst we have quantified the effects of matching $z\approx0$ IFS data to the quality of typical (KROSS) $z\approx1$ observations, we must bear in mind that this is not the same as asking what one might observe were the $z\approx0$ galaxy population to be figuratively placed at $z\approx1$ and observed in the same manner as the actual galaxy population at that epoch. Since only the less common systems at $z \approx 0$ have star formation rates comparable to typical $z\approx1$ galaxies, and given the sensitivity limits of current detectors and with similar exposure times as KROSS, we would detect very few $z\approx0$ star-forming galaxies were they observed with current telescopes over the same distances as we observe $z\approx1$ galaxies. Only the brightest, most prolifically star-forming $z\approx0$ systems would be observed. Comparing these systems to typical $z\approx1$ star-forming galaxies asks a very different question than comparing ``main sequence" star forming galaxies at each epoch as in the current work. Moreover, such a comparison would require very large numbers of IFS observations of $z\approx0$ galaxies, to then select a sub-sample of comparable size to KROSS. So, whilst both are worthwhile comparisons, we have concentrated on the latter here, as this is what the current $z\approx0$ IFS surveys are most appropriate for (but see \citealt{Green:2014} and \citealt{Fisher:2016}).  

\subsection{The Tully-Fisher relation at $z\mathbf{\approx}1$ and $z\mathbf{\approx}0$}

In this sub-section we discuss the differences between the KROSS TFRs at $z\approx1$ and the corresponding matched SAMI relations at $z\approx0$.

\subsubsection{Scatter}
\label{subsubsec:scatterdiscussion}

After accounting for the ratio of rotation velocity-to-velocity dispersion that is evidently a large source of scatter in the TFRs of both data sets (i.e. considering only the \textit{disky} sub-samples), the KROSS TFRs still display larger intrinsic scatter than the corresponding matched SAMI relations. We note also that the TFRs residuals (i.e.\ the perpendicular distance from the best fit line of the galaxies in the TFR plane) for both the matched SAMI and KROSS \textit{disky} sub-samples do not strongly correlate with any of the key galaxy properties we measure in this work. We do find a {\it moderate} correlation (a correlation coefficient of $r=0.5$) between the matched SAMI \textit{disky} TFR residuals and $v_{2.2}/\sigma$ of the galaxies, however this correlation is weak for the KROSS \textit{disky} galaxies ($r=0.2$). This therefore implies a larger intrinsic variation in $\Sigma(M/L)$ across the KROSS galaxies at $z\approx1$ than the matched SAMI galaxies at $z\approx0$. This most simply supports a secular evolution scenario of mass assembly at $z\approx1$, with in-situ gas accretion \citep[and perhaps minor mergers, e.g.][]{McLure:2013} and star formation in galaxies that allows for variation in the $M/L$ ratio between systems. If major mergers played a dominant role in stellar mass assembly at this epoch, one might instead expect minimal variation in the $M/L$ across the galaxy population as heirarchical assembly maintained this ratio. Of course this assumption ignores any sharp increases in star formation rates as a result of the mergers themselves \citep[e.g.][]{Joseph:1984,Hernquist:1989,Barnes:1991,Teyssier:2010}, and that the surface mass density may also change via the same process. There is also the possibility for secular changes to the surface mass density such as bulge formation \citep[e.g.][]{Kormendy:2004}.

\subsubsection{Slope}

A careful and homogeneous comparison between the $z\approx1$ and $z\approx0$ TFRs in this work reveals that the KROSS TFRs in general have steeper best fit slopes than those of the corresponding matched SAMI relations. However, given that the KROSS relations exhibit large scatters, and are not well constrained at low stellar mass and rotation velocity, it is likely that the {\it intrinsic} slope of the $M_{*}$ TFR for star-forming galaxies at $z\approx1$ is poorly constrained by the KROSS sample. We therefore avoid any physical interpretation of the difference in slopes between the KROSS and matched SAMI relations, preferring instead to perform fixed slope fits to the KROSS relations (where the slope is fixed to that of the corresponding matched SAMI relation in each case). 

\subsubsection{Zero-point}

Since the comparison of the SAMI and KROSS TFR zero-points of the \textit{rot-dom} sub-samples is clearly affected by increased scatter in the abscissas of the relations (as a result of the inclusion of galaxies with velocity dispersions comparable to their rotation velocities), in the remainder of this section we safely favour the comparison between the matched SAMI and KROSS TFRs of the \textit{disky} sub-samples and concentrate on these measurements in our discussion. As discussed in \S~\ref{subsec:diskysample}, for these systems the rotation velocity term in the collisionless Boltzmann equation should dominate, accounting for $\gtrsim$90\% of their dynamical mass under reasonable assumptions. In other words, they obey well the assumption of circular motion inherent in the derivation of the TFR.

In Appendix~\ref{sec:samplevsoffset}, however, we explore how the TFR zero-point offset (in both stellar mass and velocity) between $z\approx1$ and $z\approx0$ differs for each possible comparison between the \textit{rot-dom} and \textit{disky} sub-samples of the matched SAMI and KROSS $M_{*}$ relations. There we demonstrate that the size (and sign) of the TFR zero-point offset between the KROSS TFR at $z\approx1$ and the matched SAMI TFR at $z\approx0$ depends strongly on the difference in average $v_{2.2}/\sigma$ and stellar mass between the samples used to construct the relation at each epoch i.e.\ the larger the average stellar mass and $v_{2.2}/\sigma$ of the $z\approx1$ sub-sample (the more disk-like the sub-sample is) in comparison to the $z\approx0$ sub-sample, the more negative (with respect to $z\approx0$) the stellar mass TFR zero-point and the more positive the TFR zero-point offset in velocity \citep[see also][]{Turner:2017b}. In other words, perhaps unsurprisingly, the measured TFR evolution changes as a function of which types of galaxies you compare between epochs. Appendix~\ref{sec:samplevsoffset} therefore underlines the importance of controlling for (or at least awareness of) sample selection when comparing TFRs in order to avoid inadvertently biasing the physical interpretation of any measured differences between the relations.

Considering only \textit{disky} galaxies, we find no significant difference between the zero-points of the $M_{K}$ or $M_{*}$ TFRs between the two epochs (at fixed slope), implying both relations are in place by $z\approx1$ for the most rotation-dominated, star-forming systems. This is in agreement with the results of \citet{Miller:2011,Miller:2012} but in disagreement with other IFS studies of the TFR at similar redshifts that do measure a significant negative offset of the $M_{*}$ TFR zero-point with respect to $z\approx0$ \citep[e.g.][]{Puech:2008}. Most importantly our result disagrees even with other KMOS TFR studies of star-forming galaxies at the same redshift including these authors' previous work (\citet{Tiley:2016}) as well as the work of \citet{Ubler:2017}, both of which report a large negative offset in the $M_{*}$ TFR zero-point at $z\approx1$ with respect to $z\approx0$.

The difference between this work and that of \citet{Tiley:2016} is driven by a combination of factors. Firstly, different $z\approx0$ TFR baselines are used for comparison in each work. Secondly, a different measure of velocity is used in this work ($v_{2.2}$) compared to the previous work ($V_{80}$ in \citealt{Tiley:2016}). Lastly, different sample selection criteria are applied in each study - in particular this work employs a ``soft'' cut (i.e.\ incorporating uncertainty) in the galaxy rotation-to-dispersion ratio whilst in \citet{Tiley:2016} a ``hard'' cut is applied. The last two factors have the dominant effect. Correcting for them reduces the difference in TFR zero-point offset to within twice its standard error making it statistically insignificant. 

\citet{Ubler:2017} also use a different $z\approx0$ comparison relation, a measure of rotation velocity (the peak rotation velocity) different to $v_{2.2}$, and employ a hard cut in galaxy rotation-to-dispersion ratio during their sample selection. Their measured TFR zero-point evolution also agrees with that of \citet{Tiley:2016}. As such we invoke a similar explanation for the difference between their result and ours in the current work. We further note that such issues (i.e.\ differences in sample selection criteria and the definitions of variables) also entirely explain the discrepancy between the reported zero-point offset between the TFRs at $z\approx1$ and at $z\approx0$ in \citet{Turner:2017b}, and that reported in this work - both of which use KROSS data to construct the $z\approx1$ relation. After accounting for each, the two studies are in good agreement. Such discrepancies clearly highlight the importance of matched sample selection criteria and uniformly defined variables, applied across each epoch under consideration, in comparing the difference between TFRs.

\begin{figure}
\centering
\begin{minipage}[]{.5\textwidth}
\label{fig:SAMK}
\centering
\includegraphics[width=1.02\textwidth,trim= 10 10 0 0,clip=True]{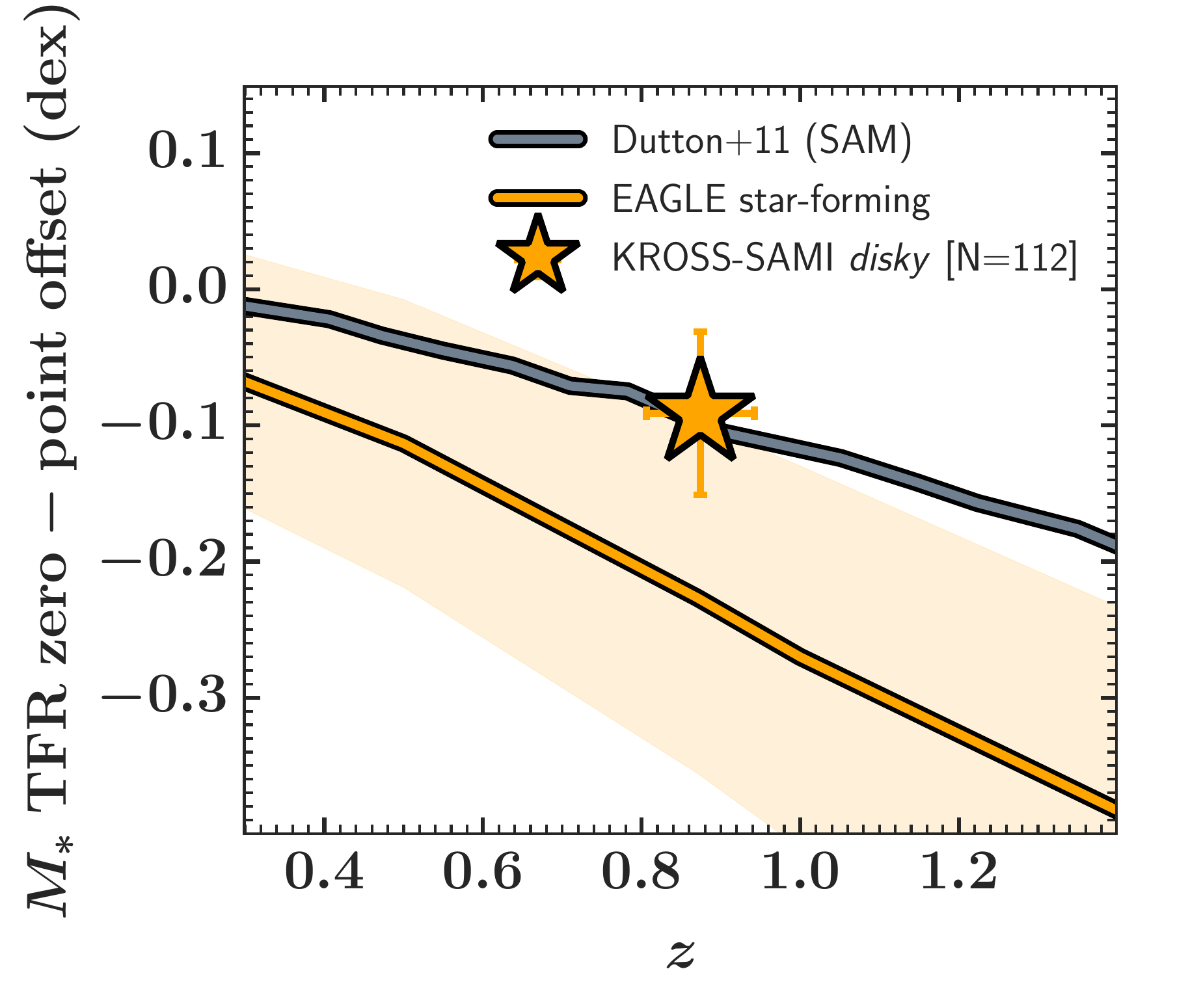}
\end{minipage}
\caption{%
Evolution of the stellar mass TFR zero-point offset as a function of redshift (with respect to $z\approx0$). We compare our measure of the zero-point offset between $z\approx1$ and $z\approx0$ (from the KROSS-matched SAMI fixed slope comparison) to the zero-point evolution predicted by the semi-analytical modelling of \citet{Dutton:2011} and star-forming model galaxies in the EAGLE simulation \citep{Schaye:2015,Crain:2015,McAlpine:2015}. We linearly interpolate between each point to better highlight the predicted trends as a function of redshift. The shaded orange region indicates the root-mean-square scatter of the star-forming model EAGLE galaxies at each redshift, again linearly interpolated to highlight the trend with redshift. Our measure of the TFR zero-point evolution since $z\approx1$ for \textit{disky} galaxies agrees well with the prediction from \citet{Dutton:2011}, but is also consistent with the zero-point evolution seen for the model EAGLE galaxies.  %
     }%
\label{fig:TFR_vs_z_SAMIKROSS}
\end{figure}

In Figure~\ref{fig:TFR_vs_z_SAMIKROSS}, we set our measure of the $M_{*}$ TFR zero-point offset between the KROSS and matched SAMI \textit{disky} sub-samples in context by comparing it with the TFR zero-point evolution of star-forming ($>1M_{\odot}$ yr$^{-1}$) galaxies from the Evolution and Assembly of GaLaxies and their Environments (EAGLE) simulation \citep{Schaye:2015,Crain:2015,McAlpine:2015}. To calculate this we first found the best fit to the TFR of $z = 0$ star-forming EAGLE galaxies and then the best fit to the TFRs in successively higher redshift slices ($z = 0, 0.5, 0.87, 1.0, 1.5, 2, 3$), with the slope fixed to that of the $z = 0$ relation. We also include the predicted zero-point evolution from \citet{Dutton:2011}, based on $N$-body simulations of baryonic disks growing in dark matter haloes over time. The stellar mass TFR zero-point evolution seen for the model EAGLE galaxies in the period $0.3 \lesssim z \lesssim 1.4$  is stronger than that predicted from the semi-analytic modelling of \citet{Dutton:2011}. Figure~\ref{fig:TFR_vs_z_SAMIKROSS}, shows that our measurement of the TFR zero-point evolution for disk-like, star-forming galaxies between $z\approx1$ and $z\approx0$ agrees well with the predictions of \citet{Dutton:2011}. However, accounting for uncertainties, the same measurement does not significantly deviate from the zero-point evolution seen for the model EAGLE galaxies (an offset of $-0.23$ dex at $z = 0.87$ with respect to $z = 0$). We therefore avoid any further detailed interpretation of Figure~\ref{fig:TFR_vs_z_SAMIKROSS} since we are not able to safely discard either model with certainty. Rather we highlight that whilst we measure only a small offset in the TFR zero-point between $z\approx1$ and $z\approx0$ (consistent with zero), this is in line with theoretical expectations from both semi-analytic modelling, and a full hydrodynamical $\Lambda$CDM cosmological simulation.

That neither the $M_{*}$ nor $M_{K}$ TFR zero-point for \textit{disky} galaxies significantly evolves between $z\approx1$ and $z\approx0$, combined with the fact that the \textit{disky} KROSS and matched SAMI galaxies occupy a similar region in the $r_{\rm{e}}$--$M_{*}$ plane (Figure~\ref{fig:mass_vs_re_SAMIKROSS}), implies that within uncertainties disk-like, star-forming galaxies at $z\approx1$ have equal amounts of stellar mass, and emit similar amounts of $K$-band light, per dark mass as those at $z\approx0$. Formally, assuming a constant surface mass density, our measurement of the $M_{*}$ TFR zero-point offset formally implies a remarkably modest increase (a factor $1.23^{+0.18}_{-0.16}$) of the stellar mass-to-total mass ratio of disk-like, star-forming galaxies since $z\approx1$. Combined with the measured offset between the $M_{K}$ TFR zero-points for the same galaxies, this implies an increase  by a factor $1.48^{+0.61}_{-0.43}$ of the $K$-band stellar mass-to-light ratio over the same period. In other words, at fixed velocity (which here we may tentatively view as a proxy for total mass) there is a formal but insignificant increase in the amount of stellar mass within disk-like, star-forming galaxies between $z\approx1$ and $z\approx0$ and a slight reduction in how luminous that stellar mass is in the $K$-band over the same period.   

The total stellar mass of the KROSS galaxies is toward the upper end of the range for late-type galaxies at $z\approx0$, primarily the result of an absolute magnitude cut in the KROSS target selection skewing the KROSS sample towards the brighter, more massive star-forming systems at $z\approx1$. Thus, that we measure no significant offset between the $M_{*}$ TFR zero-points for the KROSS and matched SAMI \textit{disky} galaxies naively implies that stellar mass assembly is nearly complete by $z\approx1$ in at least the most massive disk-like star-forming galaxies at this epoch. However, this overly simplistic conclusion must be reconciled with the large gas fractions of KROSS galaxies ($35\pm7$ percent gas to baryonic mass fraction, \citealt{Stott:2016}, that accounts for a maximum 0.24 dex offset in stellar mass - see also \citealt{Wuyts:2016} for an independent measure of baryonic fractions of star forming galaxies at $z\gtrsim 1$), their short gas depletion times and the evidence for further accretion of large amounts of gas onto galaxies since then - previous studies observe high specific baryonic accretion rates at $z\approx1$ ($\approx 0.6$--$0.8$ Gyr$^{-1}$) for galaxies of moderate stellar mass ($\log(M_{*}/M_{\odot} = 9.3$--$10.7$), that decline (to $\approx0.1$--$0.2$ Gyr$^{-1}$) to the present day \citep{Elbaz:2007,Salim:2007,Dutton:2010}. These imply that significant stellar mass growth {\em must} have taken place since $z\approx1$. Of course, one might worry whether the \textit{disky} sub-sample for the KROSS galaxies is representative of the star-forming population as a whole at $z\approx1$ in terms of its average specific star formation rate. However, we find that, although formally lower, the median specific star formation rate of the \textit{disky} KROSS sample ($\log \frac{\rm{sSFR}_{\rm{med}}}{\rm{yr}^{-1}}=-9.3 \pm 0.3$) is consistent, within uncertainties, with those of the \textit{parent} and \textit{rot-dom} KROSS galaxies ($\log \frac{\rm{sSFR}_{\rm{med}}}{\rm{yr}^{-1}}=-9.1 \pm 0.3$, for each respectively). 

These apparently contradictory conclusions then seemingly require that dark and stellar mass growth have been intimately linked from the epoch of peak star formation to the present day, with matched levels of accretion of both dark and baryonic matter, subsequent star formation being fueled by the latter. In this manner galaxies would only evolve {\it along} the TFR over this period, and not parallel to it. This is more easily understood if we also bear in mind that we are not ``following" a single population of star-forming galaxies and charting their evolution since $z\approx1$ but rather comparing ``snapshots'' of galaxies at different epochs. At both epochs we have directly selected for galaxies that are star-forming at rates typical for the epoch at which they reside. In a secular evolution scenario, such star-forming galaxies have the ability to maintain their stellar mass-to-total mass ratios by continually building stellar mass to match the rate at which they accrete dark mass. At such a point that a star-forming galaxy is quenched but continues to accrete dark mass, its stellar mass-to-total mass ratio will of course begin to decrease and it will move off the TFR. However, that same galaxy should then also drop out of our analysis. Nevertheless it is remarkable that the stellar mass-to-total mass ratio of star-forming galaxies compared between epochs $\approx$8 Gyr apart do not significantly differ; in a secular evolution scenario this suggests that the star-formation rates of typical star-forming galaxies at a given epoch are tightly linked to their mass accretion rates and therefore that star-formation efficiency is likely a function of galaxy mass. 

Previous studies, complimentary to the TFR, also support such a view. For example, \citet{Hudson:2015} use weak lensing measurements to show that the stellar-to-halo mass ratio of galaxies evolves between $z\approx0.7$ and $z\approx0.3$ but that the evolutionary trends are dominated by {\it red} galaxies. Contrastingly, the stellar-to-halo mass ratio of {\it blue} (i.e.\ star-forming) galaxies over this period can be described by a power law with no redshift evolution. This too suggests that star-forming galaxies form stars over this period at such a rate as to balance the rate at which they accrete dark matter. Galaxy formation modelling can also lend plausibility to such a scenario. For example, \citet{Mitchell:2016} show that the semi-analytic galaxy formation model {\sc galform} predicts that the stellar mass-halo mass relation for star-forming galaxies evolves very little between $z\approx2$ and $z\approx0$ as most star-forming galaxies only evolve {\it along} the relation with corresponding mass evolution in their star-formation efficiency. 

Of course, a lack of very strong evolution of the TFR zero-point may instead simply support the hypothesis that disk-like galaxies have formed in a predominantly hierarchical manner since $z\approx1$, maintaining an approximately constant dynamical (total) mass-to-light ratio as they grow. However, this hierarchical scenario is more difficult to reconcile with the evidence discussed above for large gas fractions, short depletion timescales, and high baryonic accretion rates for star-forming galaxies in the past.

\section{Conclusions}
\label{sec:SAMIvsKROSSconclusions}

We have presented a detailed comparison of the $M_{K}$ and $M_{*}$ TFRs at $z\approx1$ and $z\approx0$, derived using IFS observations of H$\alpha$ emission from respectively $z\approx1$ star-forming galaxies from KROSS \citep{Stott:2016,Harrison:2017} and local galaxies from the SAMI Galaxy Survey \citep{Croom:2012,Bryant:2015}. To minimise potentially different biases in the relations resulting from differing data quality and analysis methods, we matched the spectral and spatial resolution, sampling and H$\alpha$ signal-to-noise ratios of the SAMI data to those typical of KROSS observations.

We compared the TFRs derived from the pre- and post-matched SAMI data for carefully selected sub-samples of galaxies with associated $v_{2.2}$, $M_{K}$ and $M_{*}$ measurements reliable enough for inclusion in our TFR analysis and ratios of rotation-to-velocity greater than 1 ($v_{2.2}/\sigma + \Delta v_{2.2}/\sigma > 1$), and referred to as the \textit{rot-dom} sub-samples. We defined a further \textit{disky} sub-sample for each data set, containing those galaxies of the \textit{rot-dom} sub-sample that have an even higher ratio of rotation velocity to velocity dispersion, and with observed velocity maps that were well enough fit by a two-dimensional exponential disk model to suggest they are disk-like ($v_{2.2}/\sigma + \Delta v_{2.2}/\sigma > 3$; $R^{2} > 80$\%). \\

\smallskip

\noindent In this work we found that:

\smallskip 

\begin{itemize}
\item \noindent Degrading the SAMI data quality so that it matched that of typical KROSS observations did not grossly affect our ability to recover key galaxy parameters (rotation velocity $v_{2.2}\sin i$, intrinsic velocity dispersion $\sigma$, and stellar mass $M_{*}$); after the application of a correction factor \citep{Johnson:2018} designed to account for the effects of beam smearing (as also applied to the KROSS galaxies), the matching process only tended to slightly bias measurements of galaxies' intrinsic rotation velocity and intrinsic velocity dispersion to respectively lower and higher values. 
\newline 
\item \noindent Nevertheless we found that, although small, those biases in the matched SAMI measurements - in particular the measurement of $v_{2.2}$ - are amplified in the log-space of the TFR plane, leading to significant differences between the slopes and zero-points of the TFRs constructed using the original SAMI and matched SAMI data. The inferred intrinsic scatter was also lower than expected in the matched SAMI relations as a result of inflated fractional uncertainties in the measurements of $v_{2.2}$ from that data set. Most concerningly, we found that the differences between the matched SAMI and original SAMI relations are, in every case, as large or larger than the effect we are most interested in measuring i.e. the intrinsic differences between the $z\approx1$ and $z\approx0$ TFRs. We did find that the application of our strict {\it disky} selection criteria helped alleviate these differences, but at the cost of smaller and more biased galaxy sub-samples.
\newline
\item \noindent For a fairer measure of the differences between the $z\approx1$ and $z\approx0$ TFRs, we compared the KROSS $z\approx1$ relations to those constructed from the matched SAMI data at $z\approx0$, that are each equally biased in their measurements and sample selection. We found that the slope of the TFR is, in general, apparently higher at $z \approx 1$ in comparison to $z \approx 0$. However we avoided any physical intepretation of this result given that the KROSS TFRs exhibit large scatter and their slopes are therefore unlikely to accurately reflect the {\it intrinsic} slope of the TFR at $z\approx1$.
\newline
\item \noindent The intrinsic scatter of the $z \approx 1$ TFRs is, in all cases, larger than that of the corresponding $z \approx 0$ relations. Since every effort was made to control for systematic biases in our comparison, this suggests a real reduction in the scatter of the TFR between $z\approx1$ and $z\approx0$.
\newline
\item \noindent At fixed slope, for disk-like, star-forming galaxies (i.e. for the \textit{disky} sub-samples) there is no significant evolution in either the $M_{K}$ or $M_{*}$ TFR zero-point between $z\approx1$ and $z\approx0$. The non-evolution of the $M_{*}$ TFR zero-point is consistent, within uncertainties, with both the prediction from the semi-analytical modelling of \citealt{Dutton:2011}, as well as the zero-point evolution for star-forming model galaxies in EAGLE over the same redshift range. Assuming constant surface mass density, our results imply that the stellar mass-to-total mass ratio of the rotation-dominated, disk-like star-forming galaxy population has only modestly increased (by a factor of $1.23^{+0.18}_{-0.16}$) over the last $\approx8$ Gyr, with a corresponding, and more uncertain, increase in the $K$-band stellar mass-to-light ratio (by a factor of $1.48^{+0.61}_{-0.43}$).
\end{itemize}

Our results highlight how observational data quality can strongly bias the slope, scatter and zero-point of the TFR. Whilst these differences may be somewhat alleviated with the application of strict selection criteria, this comes at the cost of a smaller and more biased sample. Given that there are distinct differences in data quality between IFS observations at intermediate redshifts ($1 \lesssim z \lesssim 3$) and those conducted in the local Universe ($z\approx0$), this work therefore underlines the requirement to first account for these differences (as well as differences in measurement definitions and selection criteria) before being able to reliably measure differences in the TFR between epochs. Indeed, that the measured TFR changes as much as a function of data quality than as a function of redshift should serve as a reminder of the dangers of directly comparing heterogeneous data sets, and is particularly relevant for future studies that aim to measure the evolution of galaxy scaling relations as a function of redshift.

After matching these biases between our $z\approx1$ and $z\approx0$ samples, one main conclusion may be drawn from our results. Given the large rates of star formation and the high gas fractions of $z\approx1$ (KROSS) galaxies (and thus their short gas depletion times), at least moderate amounts of stellar mass growth {\em must} have occured in galaxies since $z\approx1$. Therefore, that the stellar mass TFR for disk-like, star-forming galaxies has apparently not evolved since $z\approx1$ (and that these systems do not have significantly reduced specific star formation rates in comparison to the larger KROSS sample of star-forming galaxies at the same redshift) suggests that the stellar mass growth of such galaxies must be closely matched with an equal amount of growth in dark matter over the $\approx8$ Gyr between $z\approx1$ and $z\approx0$. This must be reconciled not only with the expectation that these galaxies will convert their already-present gas reservoirs to stars during this time but also that these reservoirs should be continually replenished via further baryonic accretion over the same period (albeit at a decreasing rate with decreasing redshift). Therefore, any accretion of gas onto these galaxies (and its subsequent conversion to stars) since $z\approx1$ must be closely matched with similar levels of dark matter accretion to keep the stellar mass-to-total mass ratio constant and equal to that of $z\approx0$ galaxies.

The conclusion drawn here is based on ``snapshots" of the TFR for star-forming galaxies at only two epochs in the history of the Universe. We cannot rule out the possibility that the mass-to-light and stellar mass-to-total mass ratios of star-forming galaxies have varied more chaotically during the intervening period since $z\approx1$. Whilst other studies from the literature do not seem to support such a scenario, we must take caution in drawing over-arching conclusions from multiple studies with heterogeneous analyses. Thus, given that we know stark changes in the properties of galaxies must have occured in the $\approx8$ Gyr since $z\approx1$, we can gain further insights by extending our carefully matched comparisons of the $z\approx1$ and $z\approx0$ TFRs to other redshifts.  Programmes analagous to KROSS and SAMI to observe similar samples of galaxies at $z\approx0.4$ and $z\approx1.5$ with KMOS are already complete. The analysis of galaxies at each of these epochs will be the subject of future work.

\section*{Acknowledgments}

ALT, AMS, IRS, and CMH acknowledge support from STFC (ST/L00075X/1 and ST/P000541/1). ALT also acknowledges support from the ASTRO 3D Visitor program. IRS also acknowledges support from the ERC Advanced Grant DUSTYGAL (321334) and a Royal Society/Wolfson Merit Award. MSO acknowledges the funding support from the Australian Research Council through a Future Fellowship (FT140100255). KG acknowledges support from Australian Research Council Discovery Project DP160102235. Support for AMM is provided by NASA through Hubble Fellowship grant \#HST-HF2-51377 awarded by the Space Telescope Science Institute, which is operated by the Association of Universities for Research in Astronomy, Inc., for NASA, under contract NAS5-26555. Parts of this research were supported by the Australian Research Council Centre of Excellence for All Sky Astrophysics in 3 Dimensions (ASTRO 3D), through project number CE170100013.

The SAMI Galaxy Survey is based on observations made at the Anglo-Australian Telescope. The Sydney-AAO Multi-object Integral field spectrograph (SAMI) was developed jointly by the University of Sydney and the Australian Astronomical Observatory. The SAMI input catalogue is based on data taken from the Sloan Digital Sky Survey, the GAMA Survey and the VST ATLAS Survey. The SAMI Galaxy Survey is funded by the Australian Research Council Centre of Excellence for All-sky Astrophysics (CAASTRO), through project number CE110001020, and other participating institutions. The SAMI Galaxy Survey website is http://sami-survey.org/ .

Based on observations made with ESO Telescopes at the La Silla Paranal Observatory under the programme IDs 60.A-9460, 092.B- 0538, 093.B-0106, 094.B-0061 and 095.B-0035. This research uses data from the VIMOS VLT Deep Survey, obtained from the VVDS data base operated by Cesam, Laboratoire d’Astrophysique de Mar- seille, France. This paper uses data from the VIMOS Public Extra- galactic Redshift Survey (VIPERS). VIPERS has been performed using the ESO VLT, under the ‘Large Programme’ 182.A-0886. The participating institutions and funding agencies are listed at \url{http://vipers.inaf.it}. This paper uses data from zCOSMOS which is based on observations made with ESO Telescopes at the La Silla or Paranal Observatories under programme ID 175.A-0839. We acknowledge the Virgo Consortium for making their simulation data available. The EAGLE simulations were performed using the DiRAC-2 facility at Durham, managed by the ICC, and the PRACE facility Curie based in France at TGCC, CEA, Bruyres-le-Chtel. This work is based in part on data obtained as part of the UKIRT Infrared Deep Sky Survey. This work is based on observations taken by the CANDELS Multi-Cycle Treasury Program with the NASA/ESA HST, which is operated by the Association of Uni- versities for Research in Astronomy, Inc., under NASA contract NAS5-26555. HST data was also obtained from the data archive at the Space Telescope Science Institute.




\bibliographystyle{mnras}
\bibliography{TILEY_KROSSvsSAMI.bib} 




\appendix

\section{Table of Values}

Table~\ref{tab:tfrvals} presents examples of the derived values from the KROSS, original SAMI and matched SAMI \textit{rot-dom} and \textit{disky} galaxy sub-samples, that were used to construct the TFRs in Figures~\ref{fig:TFR_comp_SAMI} and \ref{fig:TFR_MK_KROSSLQ}. Upon publication this table will be available in full, in machine readable format, at \url{http://astro.dur.ac.uk/KROSS/data.html}. As stated in the main text, the KROSS measurements of $v_{2.2}$ (and $\sigma$) are taken directly from \citet{Harrison:2017}.

\begin{table*}
\centering
\begin{tabular}{ llllll}
\hline
Survey & ID  & \textit{disky} flag & $\log(v_{2.2}/\rm{km\ s}^{-1})$ & $\log(M_{*}/M_{\odot})$ & $M_{K}$ \\
(1) & (2)  & (3) & (4) & (5) & (6) \\
\hline
SAMI (original) & 8353  & 0 & 1.79 $\pm$ 0.01 & \phantom{0}9.0 $\pm$ 0.2 & $-21.19$\phantom{0} $\pm$ 0.09 \\
SAMI (original) & 16026  & 1 & 2.237 $\pm$ 0.009 & 10.0 $\pm$ 0.2 & $-23.438$ $\pm$ 0.009 \\
. & .  & . & . &  \phantom{0}. & . \\
SAMI (matched) & 22633  & 0 & 1.68\phantom{0} $\pm$ 0.04\phantom{0} &  \phantom{0}9.8 $\pm$ 0.2 & $-23.2$\phantom{00} $\pm$ 0.1\phantom{0} \\
SAMI (matched) & 16026  & 1 & 2.13\phantom{0} $\pm$ 0.02\phantom{0} & 10.0 $\pm$ 0.2 & $-23.438$ $\pm$ 0.009 \\
. & .  & . & . &  \phantom{0}. & . \\
KROSS & 15  & 0 & 1.58 $\pm$ 0.04\phantom{0} &  \phantom{0}9.4 $\pm$ 0.2 & $-22.04$ $\pm$ 0.08 \\
KROSS & 20  & 1 & 2.19 $\pm$ 0.03 &  \phantom{0}9.6 $\pm$ 0.2 & $-23.15$ $\pm$ 0.02 \\
. & .  & . & . & \phantom{0}. & . \\
. & .  & . & . &  \phantom{0}. & . \\
. & .  & . & . &  \phantom{0}. & . \\
\hline
\end{tabular}

\caption{The derived properties used to construct the KROSS, original SAMI and matched SAMI TFRs for the \textit{rot-dom} and \textit{disky} sub-samples. {\bf (1)} The source survey for the galaxy.  {\bf (2)} The object ID for the corresponding survey.  {\bf (3)} A flag indicating to which sub-sample(s) the galaxy belongs. If equal to $0$, the galaxy only belongs to the \textit{rot-dom} sub-sample. If equal to $1$, the galaxy belongs to both the \textit{rot-dom} and \textit{disky} sub-samples. {\bf (4)} Log of the intrinsic rotation velocity, $v_{2.2}$ {\bf (5)} The stellar mass derived via SED fitting with {\sc LePhare}. {\bf (6)} The absolute $K$-band magnitude in the Vega system.}
\label{tab:tfrvals}
\end{table*}

\section{TFR zero-point vs. Sample Selection}
\label{sec:samplevsoffset}

\begin{figure*}
\begin{minipage}[]{1.\textwidth}
\centering
\label{fig:SAMK}
\hspace{-0.2cm}\includegraphics[width=.85\textwidth,trim= 10 15 20 10,clip=True]{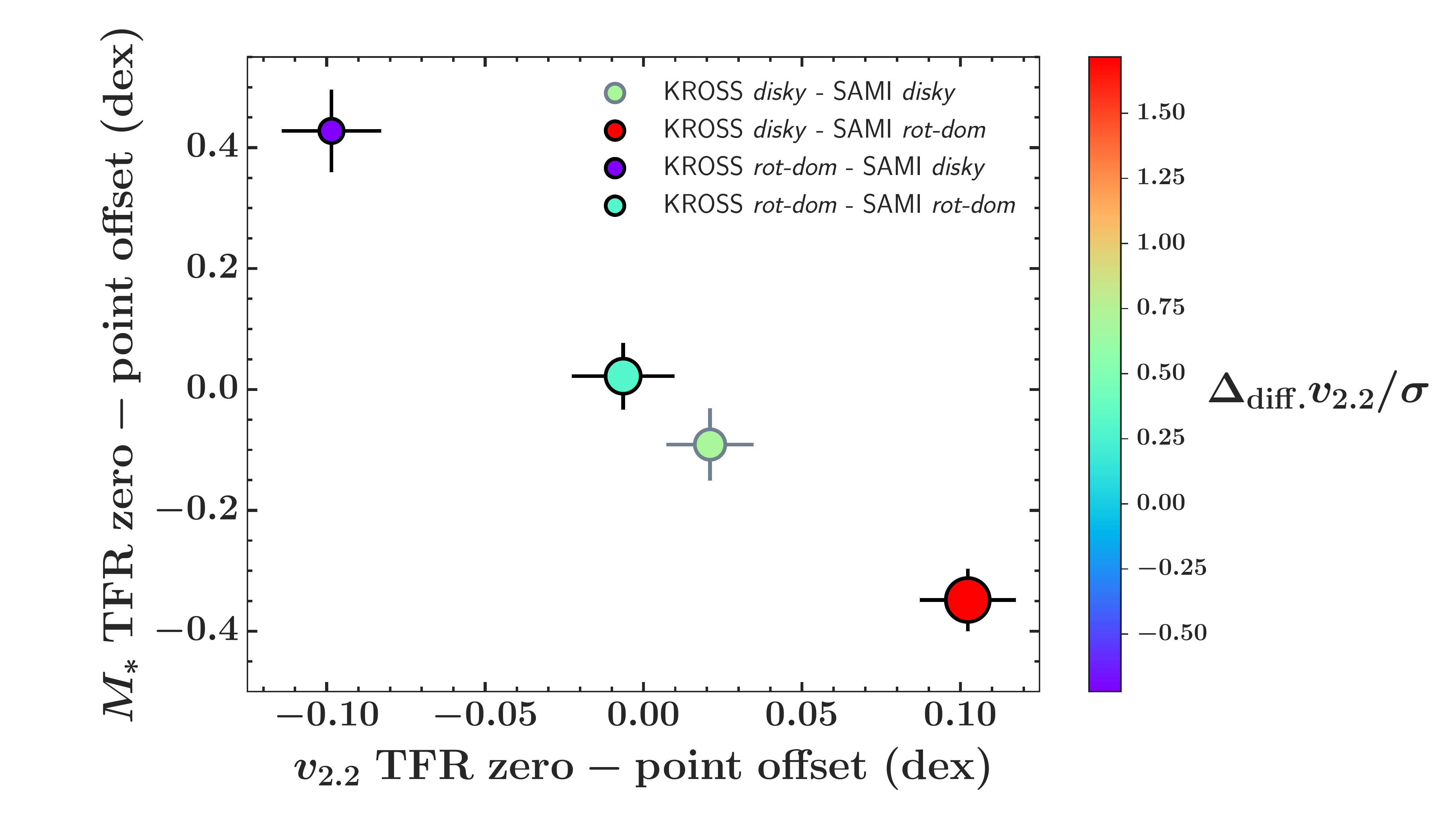}
\end{minipage}
\caption{%
The stellar mass and velocity ($v_{2.2}$) TFR zero-point offsets for each possible comparison between the KROSS and matched SAMI relations. For each KROSS sub-sample (\textit{rot-dom} and \textit{disky}) we compare the TFR zero-point to that of the TFR for each of the matched SAMI sub-samples (\textit{rot-dom} and \textit{disky}). In each case we fix the slope to that of the best free fit to whichever SAMI relation is considered. For each comparison between TFRs we calculate the difference between the best fit stellar mass zero-point to each relation. We convert this to a corresponding offset in the abscissa (i.e.\ the change in $\log(v_{2.2}/\rm{km\ s}^{-1}$)) by multiplication with a scaling factor $-1/m$, where $m$ is the (fixed) slope of the relation. Each point (corresponding to each comparison) is colour coded by the difference in the median $v_{2.2}/\sigma$ between the samples ($\Delta_{\rm{diff.}} v_{2.2}/\sigma$). Similarly, the size of each point corresponds to the difference in the median stellar mass between the samples ($\Delta_{\rm{diff.}} M_{*}$, with the smallest point representing $\Delta_{\rm{diff.}} M_{*}=-0.25$ dex, and the largest $\Delta_{\rm{diff.}} M_{*}=0.26$ dex). The zero-point offsets correlate with both $\Delta_{\rm{diff.}} M_{*}$ and $\Delta_{\rm{diff.}} v_{2.2}/\sigma$, with the largest positive values of both corresponding to respectively the largest, most negative stellar mass TFR zero-point offsets and the largest, most positive $v_{2.2}$ TFR zero-point offsets. %
     }%
\label{fig:sample_offsets}
\end{figure*}

In this work we stress the importance of matching the selection criteria of different sub-samples before drawing comparisons between the TFRs constructed from each. In Figure~\ref{fig:sample_offsets}, we show how the stellar mass TFR zero-point offset (both in stellar mass itself, and in rotation velocity) between the KROSS TFR at $z\approx1$ and the matched SAMI TFR at $z\approx0$ varies as a function of the sample selection criteria applied to each data set. It shows large differences in the magnitude and sign of the zero-point offsets depending on whether the sub-samples compared are selected with matched or differing criteria. The former results in modest offsets in zero-point with the same sign and (comparitively) similar sizes, whether we compare the \textit{rot-dom} or \textit{disky} sub-samples. The latter results in (comparitively) much larger zero-point offsets that differ in sign depending on whether we compare \textit{disky} galaxies at $z\approx1$ to \textit{rot-dom} galaxies at $z\approx0$, or vice versa. 

The trend in Figure~\ref{fig:sample_offsets} reflects the underlying difference (or similarity) in the average $v_{2.2}/\sigma$ and $M_{*}$ between sub-samples i.e. the TFR zero-point offsets correlate with both the difference in stellar mass and the difference in $v_{2.2}/\sigma$ between the sub-samples considered. Figure~\ref{fig:sample_offsets} therefore confirms the importance of matched selection criteria for a true measure of the evolution in the TFR zero-point as a function of redshift and also highlights the danger of simply comparing TFRs constructed at higher-$z$ to previously established comparison relations in the literature constructed at $z\approx0$.

\bsp	
\label{lastpage}
\end{document}